\documentclass{llncs}

\usepackage{times}
\usepackage{amsfonts,amssymb,amsmath}
\usepackage{graphicx}

\newcommand{\cA}{{\cal A}}

\newcommand{\eps}{\epsilon}

\newcommand{\floor}[1]{\left \lfloor #1 \right\rfloor}

\newtheorem{fact}[theorem]{Fact}

\begin{document}

\title{The Complexity of Scheduling for p-norms of Flow and Stretch }
\titlerunning{The Complexity of Scheduling for p-norms of Flow and Stretch }

\author{
 Benjamin Moseley\inst{1}\fnmsep
\and
Kirk Pruhs\inst{2}\fnmsep\thanks{ Supported in part by NSF grants CCF-0830558, CCF-1115575, CNS-1253218, and an IBM Faculty Award.  }
\and
 Cliff Stein\inst{3}\fnmsep \thanks{Research partially supported by NSF grant
CCF-0915681.}
   }

\institute{
Toyota Technological Institute, Chicago IL, 60637, USA. \email{moseley@ttic.edu}
\and
Computer Science Department, University of Pittsburgh, Pittsburgh, PA 15260, USA. \email{kirk@cs.pitt.edu}.
\and
 Department of Industrial Engineering \& Operations Research, Columbia University, Mudd 326,500W 120th Street, New York, NY 10027, USA. \email{cliff@ieor.columbia.edu}.
}
 
 \authorrunning{Moseley, Pruhs and Stein}
\maketitle

\begin{abstract}
We consider computing optimal $k$-norm preemptive schedules of jobs that arrive over time. 
In particular, we show that computing the optimal $k$-norm of flow schedule,
$1 \mid r_j, pmtn \mid \sum_j (C_j - r_j)^k$ in standard 3-field scheduling notation, 
is strongly NP-hard for
$k \in (0, 1)$ and integers $k \in (1, \infty)$.  Further we show that computing the  optimal $k$-norm of stretch schedule,
$1 \mid r_j, pmtn \mid \sum_j ((C_j - r_j)/p_j)^k$ in standard 3-field scheduling notation, 
is strongly NP-hard for
$k \in (0, 1)$ and integers $k \in \cup (1, \infty)$. 
 \end{abstract}

\section{Introduction}
\label{sec:introduction}

In the ubiquitous client-server computing model, multiple clients issue requests over time,  and
a request specifies a job for the server to perform.
When the requested jobs 
have widely varying processing times --- as is the case for compute servers, database 
servers, web servers, etc. --- the server system generally must allow (presumably long) jobs
to be preempted for waiting (presumably smaller) jobs in order to provide a reasonable 
quality of service to the clients. The most commonly considered and most natural quality of
service measure for a job $j$  is the flow/waiting/response time, which is $C_j - r_j$ the
duration of time between  time $r_j$ when the request is issued, and time $C_j$ when the
job is completed. Another commonly considered and natural quality of service measure 
for a job $j$ is the stretch/slowdown, $(C_j - r_j)/p_j$, 
the flow time divided by the processing time requirement $p_j$ of the job. The stretch of a job measures
how much time the job took relative to how long the job would have taken on a dedicated
server. Flow time is probably more
appropriate when the client has little idea of the time required for this requested job, as
might be the case for a non-expert database client. Stretch is probably more appropriate
when the client has at least an approximate idea of the time required for the job, as would
be the case when clients are requesting static content from a web server (e.g. when requesting
large video files clients will expect/tolerate a longer response than requesting small text files).

The server must have some scheduling policy to determine which requests to prioritize in the case
that there are multiple outstanding requests. 
To measure the quality of service of the schedule produced by the server's scheduling policy,
one needs to combine the quality of service measures of the individual requests. 
In the computer systems literature, the most commonly considered quality of service measure for a schedule is the $1$-norm, or equivalently average or total,
of the quality of service provided to the individual jobs.  Despite the widespread use,
one often sees the concern expressed that
average flow is not the ideal quality of service measure in that an optimal average flow schedule may ``unfairly starve'' some 
longer jobs. Commonly what is desired is a quality of service measure that ``balances'' the competing priorities of optimizing average quality of service and maintaining
fairness among jobs. The mathematically most natural way to achieve this balance
would be to use the $2$-norm (or more generally the $k$-norm for some small integer $k$). 

The $k$-norms of flow time and stretch have been studied in the
scheduling theory literature in a variety of settings: on a single machine \cite{BansalP10,BansalP10goem}, multiple machines \cite{ChekuriGKK04,BussemaT06,FoxM11,ImM11,AnandGK12}, in broadcast scheduling \cite{EdmondsIM11,ChekuriIM09,GuptaIKMP10}, for parallel processors \cite{EdmondsIM11,GuptaIKMP10} and on speed scalable processors \cite{GuptaKP12}.  The choice of $k$ depends on the desired balance of
average performance with fairness.  For example, the $2$-norm is
used in the standard least-squares approach to linear regression, but the $3$-norm is used within \LaTeX to determine
the best line breaks.  Conceivably there are also situations in which one may want to choose $k<1$, say when a client wants a job to be completed quickly,
but if the job is not completed quickly, the client does not care so much about how long the job is delayed.


\paragraph{Directly Related Previous Results:} In what is essentially folklore, the following is known  for optimizing a norm of flow time offline  with release dates and preemption:
\begin{itemize}
\item the optimal $1$-norm schedule can be computed in polynomial time
by the greedy algorithm Shortest Remaining Processing Time (SRPT), and 
\item
the optimal schedule for the $\infty$-norm, of
either flow and stretch,  can be computed in polynomial time by combining a binary search over the maximum
flow or stretch and the Earliest Deadline First (EDF) scheduling algorithm, which produces a deadline feasible schedule
if one exists.
\end{itemize}
Surprisingly, despite the interest the in $k$-norms of flow time and stretch, the complexity of computing an optimal $k$-norm of flow schedule, for
$k \ne 1$ or  $\infty$, and the complexity of computing an optimal $k$-norm of stretch schedule, for any $k$, 
were all open.

\subsection{Our Results:} 
We show that for all integers $k \ge 2$, and for all $k \in (0, 1)$, the problem of finding a schedule that minimizes the $k$-norm of flow is strongly NP-hard.
Similarly, we show that for all integer $k \ge 2$, and for all $k \in (0, 1)$, the problem of finding a schedule that minimizes the $k$-norm of stretch is strongly NP-hard.
This rules out the existence of a fully polynomial time approximation scheme (FPTAS) 
for these problems unless $P= NP$. 
 See Table~\ref{table:1} for a summary.

\begin{table}
\caption{A summary of results}
\label{table:1}
 \begin{center}
  \begin{tabular}{ | c | c | c |}
    \hline
       \multicolumn{3}{|c|}{Folklore Results} \\ \hline
         & Flow & Stretch \\ \hline
   $k=1$ &  SRPT is optimal&  Open\\ \hline
    $k =\infty$ &  EDF and Binary Search&  EDF and Binary Search\\
    \hline \hline
         \multicolumn{3}{|c|}{Our Results} \\ \hline
                  & Flow & Stretch \\ \hline
        $k \in (0, 1)$ and integers  $k \in (1, \infty)$ &  NP-hard& NP-hard\\ \hline
  \end{tabular}
\end{center}
\end{table}

The starting point for our NP-hardness proofs is the NP-hardness
proof in \cite{LLLR84} for the problem of finding optimal weighted flow schedules,
$1 \mid r_j, pmtn \mid \sum_j w_j(C_j - r_j)$ in the standard 3-field scheduling notation.
In this problem, each job has a positive weight, and the quality of service measure is a weighted average
of the flow of the individual jobs. 
The NP-hardness
proof of weighted flow in \cite{LLLR84} is a reduction from 3-partition. For each element of size $x$
in the 3-partition instance, there is a job of weight $x$ and processing time $x$ released at time 0 in
the weighted flow instance. Further, in the weighted flow instance, there are intermittent streams of small jobs that partition the remaining
time into open time intervals of length equal to the partition size in the 3-partition instance.
\cite{LLLR84} shows that in this case, the best possible schedule 3-partitions the large
jobs among the open time intervals. In some sense, the reduction in \cite{LLLR84} is fragile
in that it critically relies on the equality of weights and execution times.~\footnote{We do not know
for example if weighted flow is NP-hard or in P for instances where shorter jobs have larger weights.
If this was in P, this would imply that average stretch is in P.}

In order to prove our new results, several additional ideas are needed.  We define the age of a job
to be the difference between the current time and the job's release date.  
For $k$-norms of flow, the age of a job to the $(k-1)$st power can be thought of as the job's weight
at time $t$, in that the integral over time of this quantity is the quality of service measure for the job. 
Thus the ``weight'' of a job varies over time. Our main insight is that the reduction in \cite{LLLR84}
can be extended for  $k$-norms if it is modified so that the amount of time that a job is released before 
the first open time interval is proportional to the size of the corresponding 3-partition element.  We then need 
to add a third class of jobs to make sure that the partition jobs do not run during the early part of the schedule.
 The time-varying nature of the ``weight'' of the jobs requires a more involved analysis, as we need to be able to bound
 powers of flow.
 
We note that it is easy to see that these NP-hardness easily generalize to more complicated
settings, e.g. broadcast scheduling, speed scaling and parallel processors.

The rest of the paper is structured as follows. Some moderately related results are summarized in
the next subsection. Section \ref{sec:prelim} gives some preliminary definitions. 
Section~\ref{sec:2normflow} gives the NP-hardness proof for $2$-norm of flow. 
Section~\ref{sec:summary}, briefly summarizes how to extend the $2$-norm NP-Hardness proof for the remaining problems.
Section~\ref{sec:2normstretch} gives the NP-hardness proof for $2$-norm of stretch. 
Section~\ref{sec:3normflow} gives the NP-hardness proof for the $k$-norm of flow for integer $k \ge 3$. 
Section~\ref{sec:3normstretch} gives the NP-hardness proof for the $k$-norm of stretch for integer $k \ge 3$. 
Section~\ref{sec:halfnormflow} gives the NP-hardness proof for the $k$-norm of flow for $k \in (0, 1)$. 
Section~\ref{sec:halfnormstretch} gives the NP-hardness proof for the $k$-norm of flow for $k \in (0, 1)$.

\subsection{Other Related Results}

Despite the lack of NP-completeness results, approximation algorithms and on-line algorithms have been developed for $k$-norms of flow and stretch.  
Off-line, polynomial-time approximation schemes are known for computing optimal $1$-norm of stretch schedules~\cite{BenderMuthu,ChekuriKhanna}. For $k$-norms of flow and stretch, and for weighted flow,
polynomial-time $O(\log \log Pn)$-approximation
is achievable~\cite{BansalPruhsGeometry}.
For on-line algorithms, in \cite{BansalPruhsBoat} it is shown that several  standard online scheduling algorithms,
such as SRPT, are scalable ($(1+\epsilon)$-speed $O(1)$-competitive for any fixed constant $\eps > 0$) for $k$-norms of flow and stretch.

\section{Preliminaries}
\label{sec:prelim}
In our scheduling instances, each job $i \in [n]=\{1,2,\ldots, n\}$ has a positive rational processing time $p_i$ and a rational arrival time $r_i$.  
(In the typical definition, arrival times are non-negative, but since our objective is flow time or stretch, allowing arrival times to be negative does not change 
the complexity of the problem.)
A (preemptive) schedule is a function that maps some times $t$ to a job $i$, released by $t$ time, that is run at time $t$. 
A job $i$ is completed at the first time $C_i$ when it has been scheduled for $p_i$ units of time. In the $k$-norm problem, the objective is to minimize $\sqrt[k]{\sum_{i\in [n]} (C_i - r_i)^k }$. It will be convenient to abuse terminology and use the term $k$-norm to refer to the objective $\sum_{i \in[m]} (C_i - r_i)^k$, which gives rise to the same optimal schedules as the other objective.  For the rest of this paper, we will consider this objective.
We define the increase of the $k$-norm for a job $i$ during a time period $[b, e]$ with $r_i \le e$ by $(e - r_i)^k - (b - \min( b, r_i))^k$. 
Similarly, the increase in the $k$-norm for a collection of jobs is the aggregate increase of the individual jobs.

An instance of the 3-Partition problem consists of a set  $S = \{b_1, b_2, \ldots, b_{3m} \}$ of $3m$ positive integers, and a positive integer $B$, where $B$ is
polynomially bounded in $m$. In the classical definition of 3-Partition, $b_i$ are restricted to be between $B/4$ and $B/2$. By adding some large number to each element of $S$, one can assume without loss of generality the tighter bound  that
$\frac{m}{3m+1/2}B \leq b_i \leq B/2$ for all $i$. The problem is to determine a partition of  $S$ into $m$ subsets $P_1, P_2, \ldots, P_m$ such that for any $i$ it is the case that $|P_i|=3$ and $\sum_{b_j \in P_i} b_j = B$.
The 3-Partition problem is strongly NP-complete~\cite{GJ}.

We will use the term {\em volume}  of work to refer to an amount of work.

\section{NP-hardness for 2-norm of Flow}
\label{sec:2normflow}

In this section, we show that the problem of determining if there exists a schedule
with the $2$-norm of flow less than some specified value $f$ is NP-hard by a reduction from the $3$-partition problem.
We start by describing the reduction, which uses parameters $\alpha$, $\beta$, and $\rho$, and is illustrated in Figure \ref{fig:l2}:

\begin{figure}[t]
\begin{center}
\includegraphics[scale=.6]{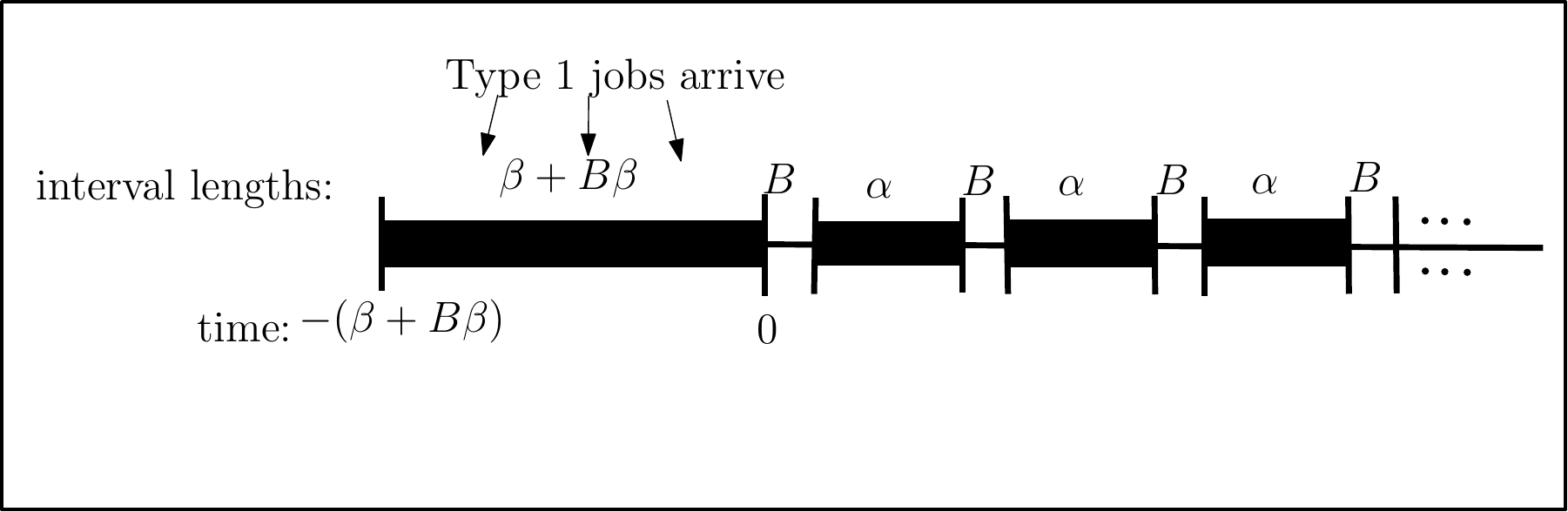}
\caption{The scheduling instance}
\label{fig:l2}
\end{center}
\end{figure}

\paragraph{The Reduction:}
\begin{itemize}
\item \textbf{Type 1 jobs:} For each integer $b_i \in S$ from the $3$-partition instance, we create a job of processing time $b_i$ and arrival time $-(\beta+\beta b_i)$. Let $T_1$ denote the set of Type $1$ jobs. 
\item \textbf{Type 2 jobs:}  During the interval $I_i$,
between time $s_i = iB+ (i-1)\alpha$ and time $s_i +  \alpha$, for $i \in [1, m-1]$, there is a job, with processing time $\rho$, released every $\rho$ time steps.
\item \textbf{Type 3 jobs:}  During the interval $I_0 = [-(\beta+\beta B),0]$ a job of processing time $\rho$ is released every $\rho$ time steps. 
\end{itemize}

Intuitively, the type 2 and type 3 jobs are so short that they must be processed essentially as they are released.
Thus we say that the times in $I_i$ are {\em closed}, while other times are {\em open}.
One can map a 3-partition to a {\em partition schedule} by scheduling the type 1 jobs corresponding  the $i$th partition in the $i$th open
time interval, and scheduling type 2 and type 3 jobs as they arrive.  
To complete the reduction, we set $f$ to be an upper bound on the 2-norm of flow for a partition schedule (this is 
proved in Lemma \ref{lem:partition upper} ):

$$f := f_{2,3} +f_o +\sum_{i=0}^n f_1(i) $$
where $f_{2,3} := \rho(\beta B+\beta+(m-1)\alpha)$,
$f_o := 6m^2B(\beta B + \beta +(m-1)B+(m-1)\alpha) + B^2$,
$f_1(i) := (3m-3i)((s_i+\beta)\alpha + \alpha^2) + 2\beta\alpha(mB-iB)$, and
$f_1(0) := \sum_{i \in T_1} (\beta+\beta b_i)^2$.
Eventually we will need that $\max(m, B) \ll \alpha \ll \beta \ll \frac{1}{\rho} \ll poly(m, B)$.
Foreshadowing slightly, more specifically we will need in the proof of Lemma \ref{lem:black}
that $\frac{1}{4m \rho} > f$, and we will need in 
Lemma \ref{lem:three} and Lemma \ref{lem:exactage} that $\alpha \beta > f_o$.
We need that the parameters are bounded by a polynomial in $m$ and $B$ so that the scheduling
instance is of polynomial size. 
We shall see that it is sufficient to set $\alpha = m^2B^3$, $\beta = m^5B^4$, and $\rho = 1/(\beta m)^3$.

\begin{lemma}
\label{lem:partition upper}
Let $A$ be an arbitrary partition schedule.
The contribution of the type 2 and type 3 jobs towards the 2-norm of flow for $A$ is, at most $f_{2,3}$.
In $A$ the increase in the 2-norm of flow of the type 1 jobs during the $I_i$, $i \in [0, m-1]$,  is at most $f_1(i)$.
In $A$ the increase in the 2-norm of the type 1 jobs during the open time intervals is at most  $f_o$.
Thus, the 2-norm of flow for $A$ is at most $f$.
\end{lemma}

\begin{proof}
We address these claims in order. 
The length of $I_0$ is $(\beta B+\beta)$, and the length of $I_i$, $i \in [1, m-1]$ is $\alpha$. 
Thus there are $(\beta B+\beta)/\rho+(m-1)\alpha/\rho$ type 2 and type 3 jobs in the instance. 
Each of these jobs contributes $\rho^2$ towards the $2$-norm of flow. 
Thus the 2-norm of flow for the type 2 and type 3 jobs in $A$ is $f_{2,3} = \rho(\beta B+\beta+(m-1)\alpha)$.

The increase in the 2-norm of flow in $A$ of the type 1 jobs during the $I_i$, $i\ge 1$,  is at most:
\begin{eqnarray*}
&&\sum_{l \in U_i} \left( (s_i+\beta+\alpha+\beta b_l)^2 - (s_i+\beta+\beta b_l)^2\right) \\
&=& \sum_{l \in U_i}\left ( 2(s_i+\beta)\alpha + \alpha^2 + 2\beta b_l\alpha \right )\\ 
&=& |U_i|((s_i+\beta)\alpha + \alpha^2) + 2\beta\alpha\sum_{l \in U_i}b_l\\
&\geq& (3m-3i)((s_i+\beta)\alpha + \alpha^2) + 2\beta\alpha\sum_{l \in U_i}b_l \;\;\;\; \mbox{[Since $|U_i| \geq 3m - 3i$]}\\
&\geq& (3m-3i)((s_i+\beta)\alpha + \alpha^2) + 2\beta\alpha(mB-iB) \;\;\;\; \mbox{[Since $\sum_{l \in U_i} b_l \geq mB-iB$]} \\
&=& f_1(i)
\end{eqnarray*}

The increase in the 2-norm of flow of the type 1 jobs during $I_0$  is $f_1(0)=\sum_{i \in T_1} (\beta+\beta b_i)^2$
since each type 1 job waits $\beta + \beta b_i$ time steps until the end of $I_0$ by construction.

The maximum increase in the 2-norm of flow for a type 1 job during an open interval is $(\beta B + \beta +mB+(m-1)\alpha)^2 - (\beta B + \beta +(m-1)B+(m-1)\alpha)^2 =2B(\beta B + \beta +(m-1)B+(m-1)\alpha) + B^2$;
this would be the increase in the last open interval if the job was released at time $-(\beta + B \beta)$.  There are $m$ open time intervals at most $3m$ jobs, so the total increase in the 2-norm for type 1 jobs during
open time intervals is upper bounded by $f_o= 6m^2B(\beta B +(m-1)B+(m-1)\alpha) + 3mB^2$.   

The last statement follows by the definition of $f$. 
\qed
\end{proof}

For the remainder of this section, let $A$ be a schedule, with $2$-norm of flow of at most $f$. To complete the proof we need to show 
that a 3-partition can be obtained by making the $i$th partition equal to the elements of $S$
corresponding to the type 1 jobs in $A$ finished between end of $I_{i-1}$ and the end of $I_i$. 
This argument is structured as follows.  Note that these lemmas are sufficient to find a valid solution to the $3$-partition instance.  This is because these lemmas show that between the end of $I_{i-1}$ and the end of $I_i$ exactly three jobs are completed and their total size is $B$.
\begin{itemize}
\item 
 Lemma \ref{lem:black} states that $A$ can only process a negligible amount of type 1 jobs during the closed time intervals $I_i$. 
 \item
 At the end of each closed time interval $I_i$,  for $i \in [0, m-1]$:
\begin{itemize}
\item 
Lemma \ref{lem:num} states that $A$ can have at most $3i$ type 1 jobs completed,
\item
Lemma \ref{lem:age} states that in $A$ the aggregate processing times of the unfinished type 1 jobs
must be at least $B(m-i)$,
\item
Lemma \ref{lem:three} states that $A$ must have at least $3i$ type 1 jobs completed,
and 
\item
Lemma \ref{lem:exactage} states that in $A$ the aggregate processing times of the unfinished type 1 jobs
can be at most $B(m-i)$.
\end{itemize}
\end{itemize}
Let $U_i$, $i \in [0, m-1]$ be the collection of type 1 jobs unfinished in $A$ by the end of $I_i$.

\begin{lemma}
\label{lem:black}
For $i \in [0, m-1]$,  the amount of time that $A$ is not processing type 2 and type 3 jobs during $I_i$
is at most $\frac{1}{2m}$.
\end{lemma}
\begin{proof}
Let $I_i$ be an interval  where a $\frac{1}{2m}$ volume of work of Type 2 or type 3 jobs that arrived during $I_i$ are not completed during $I_i$ in $A$'s schedule.  Then at least $\frac{1}{4m\rho}$ jobs wait at least $\frac{1}{4m}$ time steps to be completed. Thus the cost of the schedule is at least $\frac{1}{16m^2 \rho}$. This is strictly more than $f$. Informally this holds because $\max(m, B) \ll \alpha \ll \beta \ll \frac{1}{\rho} $.
Formally this holds by our choice of parameters and algebraic calculations. 
\qed
\end{proof}

\begin{lemma}
\label{lem:num}
For $i \in [0, m-1]$, $|U_i| \ge 3(m-i)$. 
\end{lemma}

\begin{proof}
By the end of an interval $I_i$ there are have been at exactly $iB$ time steps in the prior open time intervals.  From Lemma \ref{lem:black} at most a $1/2$ volume of work can be processed by $A$ on Type 1 jobs during closed time intervals.  Knowing that the smallest Type 1 job has size $\frac{m}{3m+1/2}B$ and $B\geq3$, the total number of jobs that can be completed before the end of $I_i$ is $\floor{(iB+1/2)\Big/(\frac{m}{3m+1/2}B)} \leq 3i$. 
\qed
\end{proof}

\begin{lemma}
\label{lem:age}
For $i \in [0, m-1]$,  $\sum_{j\in U_i} b_j \geq B(m-i)$.
\end{lemma}
\begin{proof}
By Lemma \ref{lem:black} at most a $1/2$ volume of work on Type 1 jobs can be processed by $A$ during closed time steps. 
The claim then follows by the integrality of $B$ and the elements of $S$.
\qed
\end{proof}

\begin{lemma}
\label{lem:three}
For $i \in [0, m-1]$, $|U_i| \le 3(m-i)$. 
\end{lemma}
\begin{proof}
Let $I_j$ be an interval such that $3j-1$ or less Type 1 jobs are completed by the end of $I_j$ in $A$.  The increase the 2-norm of flow
for type 1 jobs
for $A$ during $I_j$ is then:
\begin{eqnarray*}
&&\sum_{l \in U_j} \left( (s_j+\beta+\alpha+\beta b_l)^2 - (s_j+\beta+\beta b_l)^2\right) \\
&=& \sum_{l \in U_j}\left ( 2(s_j+\beta)\alpha + \alpha^2 + 2\beta b_l \alpha \right )\\ 
&=& |U_j|((s_j+\beta)\alpha + \alpha^2) + 2\beta \alpha\sum_{l \in U_j}b_l\\
&\geq& (3m-3j+1)((s_j+\beta)\alpha + \alpha^2) + 2\beta \alpha\sum_{l \in U_j}b_l \;\;\;\; \mbox{[By definition of $I_j$]}\\
&\geq& (3m-3j+1)((s_j+\beta)\alpha + \alpha^2) + 2\beta \alpha(mB-Bj) \;\;\;\; \mbox{[By Lemma \ref{lem:age}]}\\
&\geq& f_1(j) + \beta\alpha
\end{eqnarray*}
By Lemma \ref{lem:num}, and the calculations in Lemma \ref{lem:partition upper}, the increase in the 
the 2-norm of flow
for type 1 jobs
for $A$ during any $I_i$ is at least $f_1(i)$. And the 2-norm of flow for type 2 and type 3 jobs in $A$ is at least $f_{2,3}$. 
Thus to reach a contradiction, it is sufficient to show that $\beta \alpha > f_o = 6m^2B(\beta B + \beta +(m-1)B+(m-1)\alpha) + B^2$.
Informally this holds because $\max(m, B) \ll \alpha \ll \beta $. Formally this holds by our choice of parameters and algebraic calculations.
\qed
\end{proof}

\begin{lemma}
\label{lem:exactage}
For $i \in [0, m-1]$,  $\sum_{j\in U_i} b_j \geq B(m-i)$.
\end{lemma}
\begin{proof}
The claim clearly holds $i=0$. 
Assume to reach a contradiction
that there is an interval $I_j$, $j \in [1, m-1]$  such that $\sum_{l \in U_j} b_l > B(m-j)$.  Since the type 1 jobs have integral sizes, it
must be the case that $\sum_{l \in U_j} b_l \geq B(m-j)+1$.  Thus in increase in the 2-norm of flow for type 1 jobs
in $A$ during $I_j$ must be at least: 
\begin{eqnarray*}
&&\sum_{l \in U_j} \left( (s_j+\beta+\alpha+\beta b_l)^2 - (s_j+\beta+\beta b_l)^2\right) \\
&=& \sum_{l \in U_j}\left ( 2(s_j+\beta)\alpha + \alpha^2 + 2\beta b_l \alpha \right )\\ 
&=& |U_j|((s_j+\beta)\alpha + \alpha^2) + 2\beta \alpha\sum_{l \in U_j}b_l\\
&\geq& (3m-3j)((s_j+\beta)\alpha + \alpha^2) + 2\beta \alpha\sum_{l \in U_j}b_l \;\;\;\; \mbox{[By Lemma \ref{lem:num}]}\\
&\geq& (3m-3j)((s_j+\beta)\alpha + \alpha^2) + 2\beta \alpha(mB-Bj+1) \;\;\;\; \mbox{[By assumption]}\\
&\geq& f_1(j) + 2\beta\alpha
\end{eqnarray*}
By Lemma \ref{lem:num}, and the calculations in Lemma \ref{lem:partition upper}, the increase in the 
the 2-norm of flow
for type 1 jobs
for $A$ during any $I_i$ is at least $f_1(i)$. And the 2-norm of flow for type 2 and type 3 jobs in $A$ is at least $f_{2,3}$. 
Thus to reach a contradiction, it is sufficient to show that $2\beta \alpha > f_o = 6m^2B(\beta B + \beta +(m-1)B+(m-1)\alpha) + B^2$.
Informally this holds because $\max(m, B) \ll \alpha \ll \beta $. Formally this holds by our choice of parameters and algebraic calculations.
\qed
\end{proof}

\section{Summary of the Other NP-hardness Proofs}
\label{sec:summary}

In the appendices, we give the complete proofs for $k$-norm, with $k >2$ and $p \in (0,1)$, and for stretch when $k \neq 1$.  In this section, we give high level sketches of the proofs.

The proofs for all cases follow the same high level structure.  They reduce from 3-partition, and the reduction introduces 3 different types of jobs.  These classes will always play the same roles. Class 1 will contain the jobs corresponding to the 3-partition instance, released at some large negative time. Class 2 will contain jobs released during the positive time period, and will serve to partition time into intervals, each of which holds 3 type 1 jobs.  Class 3 jobs will be released during the negative time period and there will be enough of them, released frequently enough, so that the type 1 jobs cannot run during this negative time.

The proofs will differ in the particular values used and on the methods of analysis needed to obtain the proofs.  We summarize these differences for each problem below.

\subsection{Hardness Proof for the $k$-norm  of flow when $k\geq 3$.}

To emphasize the differences, we give the parameters of the reduction here.

\begin{itemize}
\item \textbf{Type 1 jobs:} For each integer $b_i \in S$ from the $3$-partition instance, we create a job of processing time $b_i$ and arrival time $-(\lambda \beta+\beta b_i)$. The value of $\beta$ is set to be $2^{10k}m^{7}B^7$.  Let $T_1$ denote the set of Type $1$ jobs. These jobs will be indexed for $i \in [3m]$. 
\item \textbf{Type 2 jobs:}  There is a job of size $\rho$ is released every $\rho$ time steps during the intervals $[iB+ (i-1)\alpha, i(B+\alpha))$ for $i$ from $1$ to $m-1$.  Here $\rho$ and $\alpha$ are set such that   $\alpha = 2^{6k}m^6B^6$ and $\rho = 1/(2m\beta\alpha)^{2k}$. 

\item \textbf{Type 3 jobs:}  During the interval $[-(\lambda \beta+\beta B),0]$ a job of size $\rho$ is released every $\rho$ time steps.  Here $\lambda = B\sqrt{\alpha} = 2^{3k}B^4 m^3$. 
\end{itemize}

Note the differences from the case when $k=2$.  Most notably, we now have that $\beta$, $\alpha$ and $\rho$ have an exponential dependency in $k$.  
 We also have a new parameter $\lambda$, which is part of the definition of the negative time period.  This period is much longer and the release date of the type $1$ jobs is significantly smaller.  We maintain the same relative  (but not absolute) values of the other parameters, adding $\lambda$, we now have  $\max(m, B) \ll \lambda \ll \alpha \ll \beta $. 
The proof uses a value of $f$ that is  
{\small
\begin{eqnarray*}
f & := & \rho^{k-1}(\beta B+\lambda \beta+(m-1)\alpha) + \sum_{l \in T_1} (b_l\beta+\beta^2)^k \\
& + &  \sum_{i =1}^{m-1} \big ( (3m - 3i)\left ( (s_i+\lambda \beta+\alpha)^{k} -(s_i+\lambda \beta)^{k}  \right ) \\
& + &  \beta k\left ((s_i+\lambda \beta+\alpha)^{k-1} -(s_i+\lambda \beta)^{k-1} \right) (mB-iB) +m2^{2k}(s_i+\lambda \beta)^{k-1}\big )  \\
& + & 3m^22^k B(\beta B + \lambda \beta + (m-1) (\alpha+B))^{k-1} \ .
\end{eqnarray*}
}
  The main technical challenge comes from the fact that the age of a job is no longer linear, but is now itself a polynomial function of degree $k-1$. In the proofs, we then evaluate the higher degree polynomial using the binomial theorem.  By the choice of parameters, the terms form a series whose values are decreasing rapidly as the exponent decreases and we are therefore able to bound the polynomial by the first two terms of the expansion.  We see the reason for the exponential dependence here, as we need the contributions to the objective function from moving the class 2 or 3 jobs by even a little bit to dominate the cost of packing the class 1 jobs.  Without the exponential dependence on $k$, we would have too large a contribution from the aging of the class 1 jobs.

\subsection{Hardness Proof for the $k$-norm when $0 < k < 1$}

We again give the parameters of the reduction.

\begin{itemize}
\item \textbf{Type 1 jobs:} For each integer $b_i \in S$ from the $3$-partition instance, we create a job of processing time $b_i$ and arrival time $-\beta+\lambda b_i$. The value of $\beta$ is set to be $(30mkB)^{5/k^2 + 2}$ and $\lambda$ is set to $\beta^{1/4}$.  Let $T_1$ denote the set of Type $1$ jobs. These jobs will be indexed for $i \in [3m]$.  Note that because $\lambda B < \beta$ and $b_i \leq B$ for all $i$, it is the case that all jobs arrive before time $0$.
\item \textbf{Type 2 jobs:}  There is a job of size $\rho$ is released every $\rho$ time steps during the intervals $[iB+ (i-1)\alpha, i(B+\alpha))$ for $i$ from $1$ to $m-1$.  Here $\rho$ and $\alpha$ are set such that $\rho = 1/(100m^4\beta)$ and $\alpha = 10\beta^{3/4}m^2B^2$. We will assume without loss of generality that $\alpha/\rho$ and $\alpha/B$ are integral. 
\item \textbf{Type 3 jobs:}  During the interval $[-\beta ,0]$ a job of size $\rho$ is released every $\rho$ time steps. We will assume that without loss of generally that $\beta/\rho$ is integral.
\end{itemize}

Again, we have a dependence on $k$, but $k$ appears in the exponent  both as $k$  and as $1/k$ in the numerator and as $1-k$ in the denominator. The proof uses a value of $f$ that is 
{\small
\begin{eqnarray*}
f  & :=  & (\beta+(m-1)\alpha)/\rho^{1-k} + \frac{3m^2kB}{( \beta -  \lambda B  )^{1-k}} + \sum_{l \in T_1} (\beta - \lambda b_l)^k  \\
& + & \sum_{i=1}^{m-1} (3m-3i) \cdot \Bigg ( \frac{k\alpha}{(s_i + \beta - \beta^{k/2} )^{1-k}} - \frac{k(1-k)(\beta^{k/2})^2}{2(s_i+\beta - \beta^{k/2})^{2-k}} \Bigg) \cdot \\
&&\qquad \qquad \qquad \qquad\Bigg(\frac{k(1-k)(\beta^{k/2})\lambda (mB-iB)}{(s_i+\beta - \beta^{k/2})^{2-k}} \Bigg)  
\end{eqnarray*}
}

In this case,  the main technical difference from the $k \geq 3$ case is that we use  a Taylor series expansion to bound polynomials that have exponents that depend on $k$.  Again, we have chosen the parameters so that the Taylor series used are rapidly decreasing and we are able to bound the series by just the first two or three terms.   This allows us to make the cost of delaying the class 2 or class 3 jobs to be prohibitively large.  

\subsection{Hardness Proof for the $2$-norm of stretch}

We now outline the approach for stretch.  Recall that stretch is flow time over processing time.  Thus, if we think about the age of a job as a weight, we now have an age that depends not only on the release date and current time, but also on the processing time.  Our first modification is to further restrict the range of values that the processing times can take on.  By adding an appropriate constant to each item in the 3-partition instance, we can assume 
without loss of generality that in the $3$-partition instance that $B/3 -\eps \leq b_i \leq B/3 + \eps$ for all $i \in [m]$ and $\eps \leq 1/(mB)^9$.  

 We then construct the identical instance as that for the $2$-norm of flow time.

Let $\Delta_s =B/3 -\eps$ and $\Delta_b = B/3 +\eps$.  Although the instance is the same as for flow, the value of $f$ will be different, with, not surprisingly terms depending on the processing time in the denominator.  More precisely, we will have terms with $\Delta_s^2$ in the denominator, since, by our restriction on the $b_i$ values, this approximates processing times well.   The value of $f$ is 
{\small 
\begin{eqnarray*}
f &  := & \sum_{i \in T_1}  \frac{1}{\Delta_s^2}(\beta + \beta  b_i)^2 + \sum_{i=1}^{m-1}( \frac{1}{\Delta_s^2}(3m-3i)((s_i+\beta)\alpha + \alpha^2) + \frac{1}{\Delta_s^2}2\beta\alpha(mB-iB)) \\
& + &     \frac{1}{\Delta_s^2}(6m^2B(\beta B +(m-1)B+(m-1)\alpha) + 3mB^2) +  (\beta B+\beta+(m-1)\alpha)/\rho \  .
\end{eqnarray*}
}

We also have terms that use the ratio of $\Delta_s$ and $\Delta_b$ which, by the choice of these parameters is close to $1$.
The proof then follows along lines similar to that for flow squared.

\subsection{Hardness Proofs for the $k$-norm of stretch, $k\geq 3$ and $k \in (0,1)$}

Given the previous proofs, the hardness proofs for other norms of stretch combine the ideas from the corresponding proofs for that norm of flow, with
the modifications made for the $2$-norm of stretch.  In particular, we restrict the range of $b_i$ and then use the exact parameters from the corresponding reduction for flow.  The analysis uses the same techniques as for flow, with the changed values to take into account the difference in objective.  Again, because of our restriction on the range of $b_i$, the objective is actually close to the corresponding flow objective, which motivates the proofs.

\section{Conclusion}

We have shown the NP-completeness the $k$-norm of flow and stretch for integers $k\geq 2$ and $k \in (0,1)$.  We believe that the techniques extend to non-integral $k > 1$, but we have not verified the details. 


Our results leaves, as the most natural open problem, the complexity of computing the optimal
$1$-norm of stretch schedule. We believe that this problem is of fundamental importance,
and is mathematically interesting. We advocate for this problem as the scheduling representative
for a rumored second-generation list of NP-hardness open problems~\cite{GJ}. 
The results of this paper can be taken as evidence that there is something uniquely interesting about
the complexity of $1$-norm of stretch schedules, and gives some explanation of the difficulties of
finding an NP-hardness proof, if in case the problem is NP-hard.  

\bibliographystyle{alpha}
\bibliography{ipco-final}

\newcommand{\etalchar}[1]{$^{#1}$}
\begin{thebibliography}{LLLRK82}

\bibitem[AGK12]{AnandGK12}
S.~Anand, Naveen Garg, and Amit Kumar.
\newblock Resource augmentation for weighted flow-time explained by dual
  fitting.
\newblock In {\em ACM-SIAM Syposium on Discrete Algorithms}, pages 1228--1241,
  2012.

\bibitem[BMR04]{BenderMuthu}
Michael Bender, S.~Muthukrishnan, and Rajmohan Rajaraman.
\newblock Approximation algorithms for average stretch scheduling.
\newblock {\em Journal of Scheduling}, 7:195--222, 2004.

\bibitem[BP10a]{BansalP10goem}
Nikhil Bansal and Kirk Pruhs.
\newblock The geometry of scheduling.
\newblock In {\em IEEE Symposium on Foundations of Computer Science}, pages
  407--414, 2010.

\bibitem[BP10b]{BansalPruhsGeometry}
Nikhil Bansal and Kirk Pruhs.
\newblock The geometry of scheduling.
\newblock In {\em Symposium on Foundations of Computer Science}, pages
  407--414, 2010.

\bibitem[BP10c]{BansalP10}
Nikhil Bansal and Kirk Pruhs.
\newblock Server scheduling to balance priorities, fairness, and average
  quality of service.
\newblock {\em SIAM J. Comput.}, 39(7):3311--3335, 2010.

\bibitem[BP10d]{BansalPruhsBoat}
Nikhil Bansal and Kirk Pruhs.
\newblock Server scheduling to balance priorities, fairness, and average
  quality of service.
\newblock {\em SIAM J. Comput.}, 39(7):3311--3335, 2010.

\bibitem[BT06]{BussemaT06}
Carl Bussema and Eric Torng.
\newblock Greedy multiprocessor server scheduling.
\newblock {\em Oper. Res. Lett.}, 34(4):451--458, 2006.

\bibitem[CGKK04]{ChekuriGKK04}
Chandra Chekuri, Ashish Goel, Sanjeev Khanna, and Amit Kumar.
\newblock Multi-processor scheduling to minimize flow time with epsilon
  resource augmentation.
\newblock In {\em ACM Symposium on Theory of Computing}, pages 363--372, 2004.

\bibitem[CIM09]{ChekuriIM09}
Chandra Chekuri, Sungjin Im, and Benjamin Moseley.
\newblock Longest wait first for broadcast scheduling [extended abstract].
\newblock In {\em WAOA}, pages 62--74, 2009.

\bibitem[CK02]{ChekuriKhanna}
Chandra Chekuri and Sanjeev Khanna.
\newblock Approximation schemes for preemptive weighted flow time.
\newblock In {\em Symposium on Theory of Computing}, pages 297--305, 2002.

\bibitem[EIM11]{EdmondsIM11}
Jeff Edmonds, Sungjin Im, and Benjamin Moseley.
\newblock Online scalable scheduling for the ;k-norms of flow time without
  conservation of work.
\newblock In {\em ACM-SIAM Syposium on Discrete Algorithms}, pages 109--119,
  2011.

\bibitem[FM11]{FoxM11}
Kyle Fox and Benjamin Moseley.
\newblock Online scheduling on identical machines using srpt.
\newblock In {\em ACM-SIAM Syposium on Discrete Algorithms}, pages 120--128,
  2011.

\bibitem[GIK{\etalchar{+}}10]{GuptaIKMP10}
Anupam Gupta, Sungjin Im, Ravishankar Krishnaswamy, Benjamin Moseley, and Kirk
  Pruhs.
\newblock Scheduling jobs with varying parallelizability to reduce variance.
\newblock In {\em Symposium on Parallelism in Algorithms and Architectures},
  pages 11--20, 2010.

\bibitem[GJ79]{GJ}
M.~R. Garey and David~S. Johnson.
\newblock {\em Computers and Intractability: A Guide to the Theory of
  NP-Completeness}.
\newblock W. H. Freeman, 1979.

\bibitem[GKP12]{GuptaKP12}
Anupam Gupta, Ravishankar Krishnaswamy, and Kirk Pruhs.
\newblock Online primal-dual for non-linear optimization with applications to
  speed scaling.
\newblock In {\em Workshop on Approximation and Online Algorithms}, 2012.

\bibitem[IM11]{ImM11}
Sungjin Im and Benjamin Moseley.
\newblock Online scalable algorithm for minimizing ;k-norms of weighted flow
  time on unrelated machines.
\newblock In {\em ACM-SIAM Syposium on Discrete Algorithms}, pages 95--108,
  2011.

\bibitem[LLLRK82]{LLLR84}
J.~Labetoulle, E.~L. Lawler, J.~K. Lenstra, and A.~H.~G. Rinnooy~Kan.
\newblock {Preemptive scheduling of uniform machines subject to release dates}.
\newblock {\em Progress in combinatorial optimization}, January 1982.

\end{thebibliography}

\appendix

\section{Hardness Proof for the $2$-norm of stretch}
\label{sec:2normstretch}
In this section , we show the hardness of the $2$-norm of stretch objective.  We reduce the $3$-partition problem to this scheduling problem.  Consider any fixed instance of the $3$-partition problem.  For this case, we will assume without loss of generality that in the $3$-partition instance that $B/3 -\eps \leq b_i \leq B/3 + \eps$ for all $i \in [m]$ and $\eps \leq 1/(mB)^9$.  We will let $\Delta_s =B/3 -\eps$ and $\Delta_b = B/3 +\eps$. We now construct the following instance of the $2$-norm problem, which is the same as that for the $2$-norm of flow time.

\begin{itemize}
\item \textbf{Type 1 jobs:} For each integer $b_i \in S$ from the $3$-partition instance, we create a job of processing time $b_i$ and arrival time $-(\beta+\beta b_i)$. The value of $\beta$ is set to be $m^5B^4$.  Let $T_1$ denote the set of Type $1$ jobs. These jobs will be indexed for $i \in [3m]$. 
\item \textbf{Type 2 jobs:}  There is a job of size $\rho$ is released every $\rho$ time steps during the intervals $[iB+ (i-1)\alpha, i(B+\alpha))$ for $i$ from $1$ to $m-1$.  Here $\rho$ and $\alpha$ are set such that $\rho = 1/(\beta m)^3$ and $\alpha = m^2B^3$. We will assume without loss of generality that $\alpha/\rho$ and $\alpha/B$ are integral. 
\item \textbf{Type 3 jobs:}  During the interval $[-(\beta+\beta B),0]$ a job of size $\rho$ is released every $\rho$ time steps. We will assume that without loss of generally that $\beta/\rho$ is integral.
\end{itemize}

This is the entire input instance.  We will let $\mathcal{I}^3$ denote the 3-Partition instance and $\mathcal{I}^\ell$ be the $2$ norm instance. Let $I_0 = [-(\beta+\beta B),0)$, $s_i = iB+ (i-1)\alpha$ and $I_i = [s_i, s_i + \alpha)$ for $i \in [m-1]$. The Type 1 jobs will be used to represent integers in the $3$-partition instance.   The Type 2 jobs will be used to ensure that an optimal schedule cannot process Type 1 during $I_i$ if there exists a valid solution to the $3$-partition instance for $i \in [m-1]$.  The Type 3 jobs are used to ensure no Type 1 job can be processed during $I_0$ in an optimal schedule. Note that during an interval $I_i$ the Type 2 or 3 jobs that arrive can be completely scheduled during this interval and require the entire interval length to be completed for all $i$.  These are the intervals which will be \emph{closed} in an optimal schedule. We call a time $t$ closed if $t \in I_i$ for $i$ from $0$ to $m-1$.  See Figure \ref{fig:l2} for a visual representation of the input instance. 

 will be to show that for a given instance of the $3$-Partition problem, there exists a schedule for the instance of the scheduling problem with an objective value of at most $f := \sum_{i \in T_1}  \frac{1}{\Delta_s^2}(\beta + \beta  b_i)^2 + \sum_{i=1}^{m-1}( \frac{1}{\Delta_s^2}(3m-3i)((s_i+\beta)\alpha + \alpha^2) + \frac{1}{\Delta_s^2}2\beta\alpha(mB-iB)) +     \frac{1}{\Delta_s^2}(6m^2B(\beta B +(m-1)B+(m-1)\alpha) + 3mB^2) +  (\beta B+\beta+(m-1)\alpha)/\rho$ if and only if there exists a valid solution to the $3$-Partition problem.  

\subsection{$2$-norm of strretch$\rightarrow$ $3$-partition:} We now show that if there is a solution to the instance $\mathcal{I}^\ell$ with an objective at most $f$ then there exists a valid solution to $\mathcal{I}^3$.  To show that this is the case, assume that there exists some fixed schedule $A$ of cost at most $f$.  Before we start with the proof, we state a few facts that will be useful throughout the proof.

\begin{fact}
\label{fact:costT1il2st}
Let $U_i$ be the set of unsatisfied Type 1 jobs at the end of interval $I_i$ in $A$.  If $|U_i| \geq 3m - 3i$ and $\sum_{l \in U_i} b_l \geq mB-iB$ then the total increase in $A$'s objective during $I_i$ due to the jobs in $U_i$ being unsatisfied during $I_i$ is at least $f_1(i) := \sum_{i \in T_1} \frac{1}{\Delta_b^2}(\beta+\beta b_i)^2(3m-3i)((s_i+\beta)\alpha + \alpha^2) + \sum_{i \in T_1}\frac{1}{\Delta_b^2}(\beta+\beta b_i)^22\beta\alpha(mB-iB)$ for $i$ from $1$ to $m-1$.  Similarly, if no Type 1 jobs are completed by time $0$ then the increase $A$'s objective for Type 1 jobs during $I_0$ is at least $f_1(0) := \sum_{i \in T_1} \frac{1}{\Delta_b^2}(\beta+\beta b_i)^2$.  
\end{fact}
\begin{proof}
The total increase in a schedule's objective due to Type 1 jobs being unsatisfied during $I_i$ is the following if $|U_i| \geq 3m - 3i$ and $\sum_{l \in U_i} b_l \geq mB-iB$.

\begin{eqnarray*}
&&\sum_{l \in U_i} \frac{1}{b_l^2}\left( (s_i+\beta+\alpha+\beta b_l)^2 - (s_i+\beta+\beta b_l)^2\right) \\
&\geq& \sum_{l \in U_i} \frac{1}{\Delta_b^2} \left ( 2(s_i+\beta)\alpha + \alpha^2 + 2\beta b_l\alpha \right )\\ 
&=& |U_i| \frac{1}{\Delta_b^2}((s_i+\beta)\alpha + \alpha^2) +  \frac{1}{\Delta_b^2}2\beta\alpha\sum_{l \in U_i}b_l\\
&\geq& \frac{1}{\Delta_b^2}(3m-3i)((s_i+\beta)\alpha + \alpha^2) + \frac{1}{\Delta_b^2}2\beta\alpha\sum_{l \in U_i}b_l \;\;\;\; \mbox{[Since $|U_i| \geq 3m - 3i$]}\\
&\geq& \frac{1}{\Delta_b^2}(3m-3i)((s_i+\beta)\alpha + \alpha^2) + \frac{1}{\Delta_b^2}2\beta\alpha(mB-iB) \;\;\;\; \mbox{[Since $\sum_{l \in U_i} b_l \geq mB-iB$]} 
\end{eqnarray*}

Now we focus on bounding the increase in a schedule's objective if no Type 1 jobs are completed by time $0$. During the interval $I_0$ a job $i \in T_1$ waits $\beta + \beta b_i$ time steps and its contribution to the objective during this interval is $\frac{1}{b_l^2}(\beta+\beta b_i)^2$.  Thus the total cost is $\sum_{i \in T_1} \frac{1}{b_l^2}(\beta+\beta b_i)^2 \leq \sum_{i \in T_1} \frac{1}{\Delta_b^2}(\beta+\beta b_i)^2$.
 
\end{proof}

\begin{fact}
\label{fact:cost23l2st}
If the algorithm completes all Type 2 and 3 jobs as soon as they arrive in first-in-first-out order then the total contribution of these jobs to the objective is $f_{2,3} := (\beta B+\beta+(m-1)\alpha)/\rho$.  This is the minimum cost any schedule can incur for Type 2 and 3 jobs.
\end{fact}
\begin{proof}
If the algorithm completes the Type 2 and 3 jobs in this way, then each of these jobs waits $\rho$ time steps to be completed. Hence, a single Type 2 or 3 job's contribution to the objective is $1$.  There are $(\beta B+\beta)/\rho+(m-1)\alpha/\rho$ Type 2 and 3 jobs total.  Thus, their total contribution is $(\beta B+\beta+(m-1)\alpha)/\rho$.
\end{proof}

Now we show that during blacked our intervals, only a small volume of Type 1 jobs can be processed.  Intuitively,  this lemma ensures that Type 1 jobs can only be processed during open times.

\begin{lemma}
\label{lem:blackl2st}
If a schedule $A$ does a $1/2$ volume of work on Type 1 jobs during $\bigcup_{i=0}^{m-1} I_i$ then the total cost of the schedule $A$ is larger than $f$. Similarly, if a schedule $A$ does not process a $1/2$ volume of work of the Type 2 or 3 jobs that arrive on an interval $I_i$ during $I_i$ then the cost of $A$'s schedule is larger than $f$ for any $i \in \{0,1,\ldots, m-1\}$.
\end{lemma}
\begin{proof}
We begin by proving the first part of the lemma.  Assume that $A$ processes a $1/2$ volume of work on Type 1 jobs during the discontinuous time interval $\bigcup_{i=0}^{m-1} I_i$.  It follows that the algorithm processes at least $1/(2m)$ volume of work of Type 1 jobs during at least one interval $I_i$.  Fix such an interval $I$.  During $I$ a job of size $\rho$ arrive every $\rho$ time steps. This implies that at least $\frac{1}{2m\rho}$ jobs have flow time at least $\frac{1}{2m}$. Thus the total cost of the schedule is at least $\frac{1}{2m\rho^3}\cdot (\frac{1}{2m})^2 \geq \beta^9/8$ which is larger than $f$ for sufficiently large $m$.

Now we prove the second part of the lemma.  Say that there is a interval $I \in \{I_0, I_1, \ldots, I_{m-1} \}$ where a $1/2$ volume of work of Type 2 or 3 jobs that arrived during $I$ are not completed during $I$ in $A$'s schedule.  Then at least $1/(4\rho)$ jobs wait at least $1/(4\rho)$ time steps to be completed. Thus the cost of the schedule is at least $1/(64\rho^5)$.  Again, this is larger than $f$ for sufficiently large $B$ and $m$.
\end{proof}

Next we observe that the total number of Type 1 that the algorithm can complete before the end of a closed interval is proportional to the number of closed intervals that have occurred so far.

\begin{lemma}
\label{lem:numl2st}
The total number of Type 1 jobs that can be completed by the end of $I_i$ is $3i$ for $i$ from $0$ to $m-1$ for any schedule $A$ with cost at most $f$.
\end{lemma}

\begin{proof}
By the end of an interval $I_i$ there are have been at exactly $iB$ time steps that are not contained in a closed interval before the end of $I_i$.  From Lemma \ref{lem:blackl2st} at most a $1/2$ volume of work can be processed by $A$ of Type 1 jobs during closed time steps.  Knowing that the smallest Type 1 job has size $\Delta_b = B/3 - \eps$, the total number of jobs that can be completed before the end of $I_i$ is $\floor{(iB+1/2)\Big/(B - \eps)} \leq 3i$ since $\eps \leq  (\frac{1}{mB})^9$. 
\end{proof}

Notice that the instantaneous increase in the objective function at a time $t$ depends on the \emph{ages} of the jobs.  Now, for a time $t\geq 0$, the age of a Type 1 job $i$ is $t+\beta+ \beta b_i$.  Our goal now is to define a bound on the total age of the Type 1 jobs during any closed interval in the schedule $A$.  Since the last lemma bounded the total number of Type 1 jobs that are unsatisfied during a closed interval, this essentially amounts to bounding the total original processing time of the unsatisfied Type 1 jobs.

\begin{lemma}
\label{lem:agel2st}
For any time $t \in I_i$, it is the case that $\sum_{j\in U_i} b_j \geq B(m-i)$ for $i$ from $0$ to $m-1$ for any schedule $A$ with cost at most $f$.
\end{lemma}
\begin{proof}
By the end of $I_i$ there are have been at exactly $iB$ time steps that are not contained in a closed interval before the end of $I_i$.  From Lemma \ref{lem:blackl2st} at most a $1/2$ volume of work on Type 1 jobs can be processed by $A$ during closed time steps. We know that the processing time of a job $i\in T_1$ is $b_i$ and therefore $\sum_{i \in T_1} b_i = mB$.  The total processing time of jobs in $T_1\setminus U_i$ can be at most $iB +1/2$ since these jobs are completed by the end of interval $I_i$.  Knowing that $i$ and $B$ are integral as well as $b_i$ for each $i \in T_1$, we know that $\sum_{j \in T_1 \setminus U_i} b_i \leq iB$.  Thus, $ \sum_{j\in U_i} b_j = mB - \sum_{j \in T_i \setminus U_i} b_i \geq mB - iB$.
\end{proof}

Next we show that by the end of every closed time interval $I_i$ for $i$ from $1$ to $m-1$ there must be at least $3i$ jobs completed.

\begin{lemma}
\label{lem:threel2st}
By the end of interval $I_i$ there must be at least $3i$ Type 1 jobs complete in any schedule $A$ that has cost at most $f$.
\end{lemma}
\begin{proof}
For the sake of contradiction, say that there is an interval $I_j$ such that $3j-1$ or less Type 1 jobs are completed by the end of $I_j$ in $A$.  We begin by bounding the increase in $A$'s objective for $T_1$ jobs during closed time intervals. Recall that $s_i$ is the start time of interval $I_i$. For any interval $I_i$ where $i \neq j$, the total increase in $A$'s objective due to Type 1 jobs being unsatisfied during $I_i$ is at least $f_1(i) = (3m-3i)((s_i+\beta)\alpha + \alpha^2) + 2\beta \alpha(mB-Bi)$ by Lemma \ref{lem:numl2st}, Lemma \ref{lem:agel2st}, and Fact \ref{fact:costT1il2st}. Similarly, the increase in $A$'s objective during $I_j$ due to jobs in $U_j$ being unsatisfied is,

\begin{eqnarray*}
&&\sum_{l \in U_j} \frac{1}{b_l^2} \left( (s_j+\beta+\alpha+\beta b_l)^2 - (s_j+\beta+\beta b_l)^2\right)\\
 &=& \sum_{l \in U_j}\left ( 2(s_j+\beta)\alpha + \alpha^2 + 2\beta b_l \alpha \right )\\ 
&\geq& |U_j|\frac{1}{\Delta_b^2}((s_j+\beta)\alpha + \alpha^2) + \frac{1}{\Delta_b^2}2\beta \alpha\sum_{l \in U_j}b_l\\
&\geq& \frac{1}{\Delta_b^2}(3m-3j+1)((s_j+\beta)\alpha + \alpha^2) + \frac{1}{\Delta_b^2}2\beta \alpha\sum_{l \in U_j}b_l \;\;\;\; \mbox{[By definition of $I_j$]}\\
&\geq& \frac{1}{\Delta_b^2}(3m-3j+1)((s_j+\beta)\alpha + \alpha^2) + \frac{1}{\Delta_b^2}2\beta \alpha(mB-Bj) \;\;\;\; \mbox{[Lemma \ref{lem:agel2st}]}\\
&\geq& f_1(j) + \frac{1}{\Delta_b^2}\beta\alpha
\end{eqnarray*}

Finally, the minimum cost any algorithm can incur due to Type 2 and 3 jobs is $f_{2,3}$ as stated in Fact \ref{fact:cost23l2st}.  Thus, the algorithm $A$ would have total cost at least $\sum_{i = 0}^{m-1}f_1(i) + f_{2,3} + \frac{1}{\Delta_b^2}\beta\alpha  > f$  since $\Delta_s = B/3-\eps$, $\Delta_b = B/3 + \eps$, $\eps <  (\frac{1}{mB})^9$ and $B$ and $m$ are sufficiently large.
\end{proof}

\begin{lemma}
\label{lem:exactagel2st}
It must be the case that $\sum_{j \in U_i} b_j \leq B(m-i)$ for $i$ from $0$ to $m-1$ for any schedule $A$ with cost at most $f$.  
\end{lemma}
\begin{proof}
Clearly the lemma holds true to the interval $I_0$, since the lemma implies that in this case all jobs in $T_1$ could possibly be incomplete at the end of $I_0$.  For the sake of contradiction say that there is an interval $I_j$ such that for a schedule $A$ of cost at most $f$, $\sum_{l \in U_j} b_l > B(m-j)$.  Since $b_l$ is integral for all jobs $l \in T_1$, it is the case that $\sum_{l \in U_j} b_l \geq B(m-j)+1$.  Consider the increase in $A$'s objective during $I_j$ for Type 1 jobs.  This is the following, 

\begin{eqnarray*}
&&\sum_{l \in U_j} \left( (s_j+\beta+\alpha+\beta b_l)^2 - (s_j+\beta+\beta b_l)^2\right) \\
&=& \sum_{l \in U_j}\left ( 2(s_j+\beta)\alpha + \alpha^2 + 2\beta b_l \alpha \right )\\ 
&=& |U_j|((s_j+\beta)\alpha + \alpha^2) + 2\beta \alpha\sum_{l \in U_j}b_l\\
&\geq& (3m-3j)((s_j+\beta)\alpha + \alpha^2) + 2\beta \alpha\sum_{l \in U_j}b_l \;\;\;\; \mbox{[Lemma \ref{lem:numl2st}]}\\
&\geq& (3m-3j)((s_j+\beta)\alpha + \alpha^2) + 2\beta \alpha(mB-Bj+1) \;\;\;\; \mbox{[By definition of $I_j$]}\\
&\geq& f_1(j) + 2\beta\alpha
\end{eqnarray*}

Now we know that the increase in the algorithm's objective due to Type 1 jobs being unsatisfied during $I_i$ where $i\neq j$ is at least $f_1(i)$ by Lemma\ref{lem:numl2st}, Lemma \ref{lem:agel2st}, and Fact \ref{fact:costT1il2st}.  Further, the minimum cost any schedule can incur for Type 2 and 3 jobs is $f_{2,3}$.  Thus, the cost of $A$'s schedule is at least $\sum_{i = 0}^{m-1}f_1(i) + f_{2,3} + 2\beta\alpha$. Knowing that $f = \sum_{i = 0}^{m-1}f_1(i) + f_{2,3} + 6m^2B(\beta B + \beta +(m-1)B+(m-1)\alpha) + 3mB^2$ and that $\beta\alpha > 6m^2B(\beta B+\beta +(m-1)B+(m-1)\alpha) + 3mB^2$, we have a contradiction to the the fact that $A$ has cost at most $f$.

\end{proof}

Now we are finally ready to prove the theorem.

\begin{theorem}
If the schedule $A$ has cost at most $f$ then there exists a valid solution to the 3-partition instance.
\end{theorem}
\begin{proof}
By Lemma \ref{lem:threel2st} and Lemma \ref{lem:numl2st} we know that $|T_i \setminus U_i|= 3i$.  This implies that $3$ jobs are completed between the end of $I_i$ and the end of $I_{i+1}$ for $i$ from $0$ to $m-1$ and $3$ jobs are completed after $I_{m-1}$. Let $P_i$ be the three jobs completed between the end of $I_i$ and the end of $I_{i+1}$ and $P_m$ be the remaining $3$ jobs. Lemma \ref{lem:agel2st} and Lemma \ref{lem:exactagel2st} state that $\sum_{l \in T_i \setminus U_i} b_l = Bm-Bi$ for $i$ from $1$ to $m-1$.  This implies that $\sum_{j \in P_i} b_j = B$.  Thus, $P_i$ contains exactly $3$ jobs such that their total processing time is exactly $B$ for $i$ from $1$ to $m$.  Hence, $P_1, \ldots, P_m$ corresponds to a solution to the 3-partition problem.
\end{proof}

\subsection{$3$-partition of stretch $\rightarrow$ $2$-norm:}  Now we show that there if there is a valid solution to the $3$-partition then there is a solution to the $2$-norm with an objective at most $f$.   Consider a valid solution to the $3$-partition problem $P_1, P_2, \ldots, P_m$.  Consider the following solution to $2$-norm problem instance.  Each job of Type 2 and 3 is scheduled as soon as it arrives.  Using Fact \ref{fact:cost23l2st} the total cost for these jobs is $f_{2,3}$. Now consider Type 1 jobs.  The three Type 1 jobs corresponding to integers in $P_i$ are scheduled during $[(i-1)(B+\alpha), iB+(i-1)\alpha)$.   Note that these jobs can be exactly scheduled during this interval.  

Now we bound the cost of Type 1 jobs by separating the intervals where the jobs could possibly be waiting. First we bound the cost Type 1 jobs accumulate during closed time intervals.

\begin{lemma}
\label{lem:cacostl2st}
For the schedule $\cA$ the total cost Type 1 jobs accumulate during closed time intervals is at most $\sum_{i \in T_1}  \frac{1}{\Delta_s^2}(\beta + \beta  b_i)^2 + \sum_{i=1}^{m-1}( \frac{1}{\Delta_s^2}(3m-3i)((s_i+\beta)\alpha + \alpha^2) + \frac{1}{\Delta_s^2}2\beta\alpha(mB-iB)) $.
\end{lemma}

\begin{proof}

Consider an interval $I_i$ where $1 \leq i \leq m-1$.  Let $U^\cA_i$ be the set of Type 1 jobs that are unsatisfied win the schedule $\cA$ during $I_i$.  The total accumulated cost during $I_i$ for these jobs is at most the following knowing that $|U_i^\cA|= 3m - 3i$ and $\sum_{l \in U^\cA_i} b_l = mB-iB$.

\begin{eqnarray*}
&&\sum_{l \in U_i} \frac{1}{b_l^2}\left( (s_i+\beta+\alpha+\beta b_l)^2 - (s_i+\beta+\beta b_l)^2\right) \\
&\geq& \sum_{l \in U_i} \frac{1}{\Delta_s^2} \left ( 2(s_i+\beta)\alpha + \alpha^2 + 2\beta b_l\alpha \right )\\ 
&=& |U_i| \frac{1}{\Delta_s^2}((s_i+\beta)\alpha + \alpha^2) +  \frac{1}{\Delta_b^2}2\beta\alpha\sum_{l \in U_i}b_l\\
&\geq& \frac{1}{\Delta_s^2}(3m-3i)((s_i+\beta)\alpha + \alpha^2) + \frac{1}{\Delta_s^2}2\beta\alpha\sum_{l \in U_i}b_l \;\;\;\; \mbox{[Since $|U_i| \geq 3m - 3i$]}\\
&\geq& \frac{1}{\Delta_s^2}(3m-3i)((s_i+\beta)\alpha + \alpha^2) + \frac{1}{\Delta_s^2}2\beta\alpha(mB-iB) \;\;\;\; \mbox{[Since $\sum_{l \in U_i} b_l \geq mB-iB$]} 
\end{eqnarray*}

Now consider the interval $I_0$.  All Type 1 jobs are unsatisfied by time $0$, thus the total cost $\cA$ accumulates for these jobs during $I_0$ is $\sum_{i \in T_1} \frac{1}{b_i^2}(\beta + \beta  b_i)^2 \leq \sum_{i \in T_1} \frac{1}{\Delta_s^2}(\beta + \beta  b_i)^2$.  This completes the proof of the lemma.

\end{proof}

The only cost that has not been accounted for is the increase due to Type 1 jobs waiting in open out time intervals.  By convexity of objective function and the fact that the earliest a Type 1 job arrives is later than $-(\beta B+\beta)$,  the most a job can contribute to the objective during a maximal contiguous open time interval is $\frac{1}{\Delta_s^2}(\beta B + \beta +mB+(m-1)\alpha)^2 - \frac{1}{\Delta_s^2}(\beta B + \beta +(m-1)B+(m-1)\alpha)^2 =\frac{1}{\Delta_s^2}(2B(\beta B + \beta +(m-1)B+(m-1)\alpha) + B^2)$.  There are $m$ such time intervals at most $3m$ jobs, so this cost can be upper bounded by $\frac{1}{\Delta_s^2}(6m^2B(\beta B +(m-1)B+(m-1)\alpha) + 3mB^2)$.  This and Lemma~\ref{lem:cacostl2st} imply that, the cost of schedule $\cA$ is at most $f$ and we have the following theorem.

\begin{theorem}
If there exists a valid solution to the 3-Partition instance then there is a schedule for the $2$-norm of stretch problem instance with an objective value at most $f$.
\end{theorem}

\section{Hardness Proof for the $k$-norm  of flow when $k\geq 3$.}
\label{sec:3normflow}
In this section, we show the hardness of the $k$-norm objective for fixed integral $k \geq 3$.  We reduce the $3$-partition problem to this scheduling problem.  Consider any fixed instance of the $3$-partition problem. We now construct the following instance of the $k$-norm problem.

\begin{itemize}
\item \textbf{Type 1 jobs:} For each integer $b_i \in S$ from the $3$-partition instance, we create a job of processing time $b_i$ and arrival time $-(\lambda \beta+\beta b_i)$. The value of $\beta$ is set to be $2^{10k}m^{7}B^7$.  Let $T_1$ denote the set of Type $1$ jobs. These jobs will be indexed for $i \in [3m]$. 
\item \textbf{Type 2 jobs:}  There is a job of size $\rho$ is released every $\rho$ time steps during the intervals $[iB+ (i-1)\alpha, i(B+\alpha))$ for $i$ from $1$ to $m-1$.  Here $\rho$ and $\alpha$ are set such that   $\alpha = 2^{6k}m^6B^6$ and $\rho = 1/(2m\beta\alpha)^{2k}$. We will assume without loss of generality that $\alpha/\rho$ and $\alpha/B$ are integral. 
\item \textbf{Type 3 jobs:}  During the interval $[-(\lambda \beta+\beta B),0]$ a job of size $\rho$ is released every $\rho$ time steps.  Here $\lambda = B\sqrt{\alpha} = 2^{3k}B^4 m^3$. We will assume that without loss of generally that $\beta/\rho$ is integral.
\end{itemize}

This is the entire input instance.  We will let $\mathcal{I}^3$ denote the 3-Partition instance and $\mathcal{I}^\ell$ be the $k$ norm instance. Let $I_0 = [-(\lambda \beta+\beta B),0)$, $s_i = iB+ (i-1)\alpha$ and $I_i = [s_i, s_i + \alpha)$ for $i \in [m-1]$. The intuition behind the instance $\mathcal{I}^\ell$ is essentially the same as the $2$-norm instance of the previous section.  The Type 1 jobs will be used to represent integers in the $3$-partition instance.   The Type 2 jobs will be used to ensure that an optimal schedule cannot process Type 1 during $I_i$ if there exists a valid solution to the $3$-partition instance for $i \in [m-1]$.  The Type 3 jobs are used to ensure no Type 1 job can be processed during $I_0$ in an optimal schedule. Note that during an interval $I_i$ the Type 2 or 3 jobs that arrive can be completely scheduled during this interval and require the entire interval length to be completed for all $i$.  These are the intervals which will be \emph{closed} in an optimal schedule. We call a time $t$ closed if $t \in I_i$ for $i$ from $0$ to $m-1$.  

Our goal will be to show that for a given instance of the $3$-Partition problem, there exists a schedule for the instance of the scheduling problem with an objective value of at most $f :=\rho^{k-1}(\beta B+\lambda \beta+(m-1)\alpha) + \sum_{l \in T_1} (b_l\beta+\beta^2)^k + \sum_{i =1}^{m-1} \big ( (3m - 3i)\left ( (s_i+\lambda \beta+\alpha)^{k} -(s_i+\lambda \beta)^{k}  \right ) + \beta k\left ((s_i+\lambda \beta+\alpha)^{k-1} -(s_i+\lambda \beta)^{k-1} \right) (mB-iB) +m2^{2k}(s_i+\lambda \beta)^{k-1}\big )  + 3m^22^k B(\beta B + \lambda \beta + (m-1) (\alpha+B))^{k-1}$ if and only if there exists a valid solution to the $3$-Partition problem.  

\subsection{$k$-norm $\rightarrow$ $3$-partition:} We now show that if there is a solution to the instance $\mathcal{I}^\ell$ with an objective at most $f$ then there exists a valid solution to $\mathcal{I}^3$.  To show that this is the case, assume that there exists some fixed schedule $A$ of cost at most $f$.  Before we start with the proof, we state a few facts that will be useful throughout the proof.

\begin{fact}
\label{fact:costT1ilk}
Let $U_i$ be the set of unsatisfied Type 1 jobs at the end of interval $I_i$ in $A$.  If $|U_i| \geq 3m - 3i$ and $\sum_{l \in U_i} b_l \geq mB-iB$ then the total increase in $A$'s objective during $I_i$ due to the jobs in $U_i$ being unsatisfied during $I_i$ is at least $f_1(i) := (3m - 3i)\left ( (s_i+\lambda \beta+\alpha)^{k} -(s_i+\lambda \beta)^{k}  \right ) + \beta k\left ((s_i+\lambda \beta+\alpha)^{k-1} -(s_i+\lambda \beta)^{k-1} \right) (mB-iB)$ for $i$ from $1$ to $m-1$.  Similarly, if no Type 1 jobs are completed by time $0$ then the increase $A$'s objective for Type 1 jobs during $I_0$ is at least $f_1(0) := \sum_{i \in T_1} (\lambda \beta+\beta b_i)^k$.  
\end{fact}
\begin{proof}
The total increase in a schedule's objective due to Type 1 jobs being unsatisfied during $I_i$ is the following if $|U_i| \geq 3m - 3i$, $\sum_{l \in U_i} b_l \geq mB-iB$ and $1 \leq i \leq m-1$.

\begin{eqnarray*}
&&\sum_{l \in U_i} \left( (s_i+\lambda \beta+\alpha+\beta b_l)^k - (s_i+\lambda \beta+\beta b_l)^k\right) \\
&=& \sum_{l \in U_i}\left ( \sum_{x=0}^k \binom{k}{x} (\beta b_l)^{x} \left ( (s_i+\lambda \beta+\alpha)^{k-x} -(s_i+\lambda \beta)^{k-x} \right) \right )  \;\;\;\; \mbox{[Binomial theorem]}\\ 
&\geq&  \sum_{l \in U_i}\left ( (s_i+\lambda \beta+\alpha)^{k} -(s_i+\lambda \beta)^{k} + \beta b_l k\left ((s_i+\lambda \beta+\alpha)^{k-1} -(s_i+\lambda \beta)^{k-1} \right)  \right )\\
&=&|U_i|\left ( (s_i+\lambda \beta+\alpha)^{k} -(s_i+\lambda \beta)^{k}  \right ) +   \beta  k \left ((s_i+\lambda \beta+\alpha)^{k-1} -(s_i+\lambda \beta)^{k-1} \right)\sum_{l \in U_i}b_l \\
&\geq& (3m - 3i)\left ( (s_i+\lambda \beta+\alpha)^{k} -(s_i+\lambda \beta)^{k}  \right ) +   \beta k\left ((s_i+\lambda \beta+\alpha)^{k-1} -(s_i+\lambda \beta)^{k-1} \right)\sum_{l \in U_i}b_l\\ && \;\;\;\; \mbox{[Since $|U_i| \geq 3m - 3i$]}\\
&\geq& (3m - 3i)\left ( (s_i+\lambda \beta+\alpha)^{k} -(s_i+\lambda \beta)^{k}  \right ) + \beta k\left ((s_i+\lambda \beta+\alpha)^{k-1} -(s_i+\lambda \beta)^{k-1} \right) (mB-iB)  \\&&\;\;\;\; \mbox{[Since $\sum_{l \in U_i} b_l \geq mB-iB$]} 
\end{eqnarray*}

This completes the proof for an interval $I_i$ for for $1 \leq i \leq m-1$.  Now we focus on bounding the increase in a schedule's objective if no Type 1 jobs are completed by time $0$. During the interval $I_0$ a job $i \in T_1$ waits $\beta + \beta b_i$ time steps and its contribution to the objective during this interval is $(\lambda \beta+\beta b_i)^k$.  Thus the total cost is $\sum_{i \in T_1} (\lambda \beta+\beta b_i)^k$.
 
\end{proof}

\begin{fact}
\label{fact:cost23lk}
If the algorithm completes all Type 2 and 3 jobs as soon as they arrive in first-in-first-out order then the total contribution of these jobs to the objective is $f_{2,3} := \rho^{k-1}(\beta B+\lambda \beta+(m-1)\alpha)$.  This is the minimum cost any schedule can incur for Type 2 and 3 jobs.
\end{fact}
\begin{proof}
If the algorithm completes the Type 2 and 3 jobs in this way, then each of these jobs waits $\rho$ time steps to be completed. Hence, a single Type 2 or 3 job's contribution to the objective is $\rho^k$.  There are $(\beta B+\lambda \beta)/\rho+(m-1)\alpha/\rho$ Type 2 and 3 jobs total.  Thus, their total contribution is $\rho^{k-1}(\beta B+\lambda \beta+(m-1)\alpha)$.
\end{proof}

Now we show that during closed intervals, only a small volume of Type 1 jobs can be processed.  Intuitively,  this lemma ensures that Type 1 jobs can only be processed during open times.

\begin{lemma}
\label{lem:blacklk}
If a schedule $A$ does a $1/2$ volume of work on Type 1 jobs during $\bigcup_{i=0}^{m-1} I_i$ then the total cost of the schedule $A$ is larger than $f$. Similarly, if a schedule $A$ does not process a $1/2$ volume of work of the Type 2 or 3 jobs that arrive on an interval $I_i$ during $I_i$ then the cost of $A$'s schedule is larger than $f$ for any $i \in \{0,1,\ldots, m-1\}$.
\end{lemma}
\begin{proof}
We begin by proving the first part of the lemma.  Assume that $A$ processes a $1/2$ volume of work on Type 1 jobs during the discontinuous time interval $\bigcup_{i=0}^{m-1} I_i$.  It follows that the algorithm processes at least $1/(2m)$ volume of work of Type 1 jobs during at least one interval $I_i$ where $0 \leq i \leq m-1$.  Fix such an interval $I$.  During $I$ a job of size $\rho$ arrive every $\rho$ time steps. This implies that at least $\frac{1}{2m\rho}$ jobs have flow time at least $\frac{1}{2m}$. Thus the total cost of the schedule is at least $\frac{1}{2m\rho}\cdot (\frac{1}{2m})^k \geq (\beta\alpha)^{2k}$ which is larger than $f$ for sufficiently large $m$.

Now we prove the second part of the lemma.  Say that there is a interval $I \in \{I_0, I_1, \ldots, I_{m-1} \}$ where a $1/2$ volume of work of Type 2 or 3 jobs that arrived during $I$ are not completed during $I$ in $A$'s schedule.  Then at least $1/(4\rho)$ jobs wait at least $1/(4\rho)$ time steps to be completed. Thus the cost of the schedule is at least $1/(4\rho)^{k+1}$.  Again, this is larger than $f$ for sufficiently large $B$ and $m$.
\end{proof}

Next we observe that the total number of Type 1 that the algorithm can complete before the end of a closed interval is proportional to the number of closed intervals that have occurred so far.

\begin{lemma}
\label{lem:numlk}
The total number of Type 1 jobs that can be completed by the end of $I_i$ is $3i$ for $i$ from $0$ to $m-1$ for any schedule $A$ with cost at most $f$.
\end{lemma}

\begin{proof}
By the end of an interval $I_i$ there are have been at exactly $iB$ time steps that are not contained in a closed interval before the end of $I_i$.  From Lemma \ref{lem:blacklk} at most a $1/2$ volume of work can be processed by $A$ of Type 1 jobs during closed time steps.  Knowing that the smallest Type 1 job has size $\frac{m}{3m+1/2}B$ and $B\geq3$, the total number of jobs that can be completed before the end of $I_i$ is $\floor{(iB+1/2)\Big/(\frac{m}{3m+1/2}B)} \leq 3i$. 
\end{proof}

Notice that the instantaneous increase in the objective function at a time $t$ depends on the \emph{ages} of the jobs.  Now, for a time $t\geq 0$, the age of a Type 1 job $i$ is $t+\lambda \beta+ \beta b_i$.  Our goal now is to define a bound on the total age of the Type 1 jobs during any closed interval in the schedule $A$.  Since the last lemma bounded the total number of Type 1 jobs that are unsatisfied during a closed interval, this essentially amounts to bounding the total original processing time of the unsatisfied Type 1 jobs.

\begin{lemma}
\label{lem:agelk}
For any time $t \in I_i$, it is the case that $\sum_{j\in U_i} b_j \geq B(m-i)$ for $i$ from $0$ to $m-1$ for any schedule $A$ with cost at most $f$.
\end{lemma}
\begin{proof}
By the end of $I_i$ there are have been at exactly $iB$ time steps that are not contained in a closed interval before the end of $I_i$.  From Lemma \ref{lem:blacklk} at most a $1/2$ volume of work on Type 1 jobs can be processed by $A$ during closed time steps. We know that the processing time of a job $i\in T_1$ is $b_i$ and therefore $\sum_{i \in T_1} b_i = mB$.  The total processing time of jobs in $T_1\setminus U_i$ can be at most $iB +1/2$ since these jobs are completed by the end of interval $I_i$.  Knowing that $i$ and $B$ are integral as well as $b_i$ for each $i \in T_1$, we know that $\sum_{j \in T_1 \setminus U_j} b_i \leq iB$.  Thus, $ \sum_{j\in U_i} b_j = mB - \sum_{j \in T_i \setminus U_i} b_j \geq mB - iB$.
\end{proof}

Next we show that by the end of every closed time interval $I_i$ for $i$ from $1$ to $m-1$ there must be at least $3i$ jobs completed.

\begin{lemma}
\label{lem:threelk}
By the end of interval $I_i$ there must be at least $3i$ Type 1 jobs complete in any schedule $A$ that has cost at most $f$.
\end{lemma}
\begin{proof}
For the sake of contradiction, say that there is an interval $I_j$ such that $3j-1$ or less Type 1 jobs are completed by the end of $I_j$ in $A$.  We begin by bounding the increase in $A$'s objective for $T_1$ jobs during closed time intervals. Recall that $s_i$ is the start time of interval $I_i$. For any interval $I_i$ where $i \neq j$, the total increase in $A$'s objective due to Type 1 jobs being unsatisfied during $I_i$ is at least $f_1(i) $ by Lemma \ref{lem:numlk}, Lemma \ref{lem:agelk}, and Fact \ref{fact:costT1ilk}. Similarly, the increase in $A$'s objective during $I_j$ due to jobs in $U_j$ being unsatisfied is,

\begin{eqnarray}
&&\sum_{l \in U_j} \left( (s_j+\lambda \beta+\alpha+\beta b_l)^k - (s_j+\lambda \beta+\beta b_l)^k\right) \nonumber\\
&=& \sum_{l \in U_j}\left ( \sum_{x=0}^k \binom{k}{x} (\beta b_l)^{x} \left ( (s_j+\lambda \beta+\alpha)^{k-x} -(s_j+\lambda \beta)^{k-x} \right) \right )  \;\;\;\; \mbox{[Binomial theorem]}\nonumber\\ 
&\geq&  \sum_{l \in U_j}\left ( (s_j+\lambda \beta+\alpha)^{k} -(s_j+\lambda \beta)^{k} + \beta b_l \left ((s_j+\lambda \beta+\alpha)^{k-1} -(s_j+\lambda \beta)^{k-1} \right)  \right )\nonumber\\
&=&|U_j|\left ( (s_j+\lambda \beta+\alpha)^{k} -(s_j+\lambda \beta)^{k}  \right ) +  \beta k   \left ((s_j+\lambda \beta+\alpha)^{k-1} -(s_j+\lambda \beta)^{k-1} \right)\sum_{l \in U_j}b_l  \nonumber\\
&\geq& (3m - 3i+1)\left ( (s_j+\lambda \beta+\alpha)^{k} -(s_j+\lambda \beta)^{k}  \right ) +   \beta k\left ((s_j+\lambda \beta+\alpha)^{k-1} -(s_j+\lambda \beta)^{k-1} \right)\sum_{l \in U_j}b_l \nonumber\\ && \;\;\;\; \mbox{[Since $|U_j| \geq 3m - 3i+1$]}\nonumber\\
&\geq& (3m - 3i+1)\left ( (s_j+\lambda \beta+\alpha)^{k} -(s_j+\lambda \beta)^{k}  \right ) + \beta k(mB-iB) \left ((s_j+\lambda \beta+\alpha)^{k-1} -(s_j+\lambda \beta)^{k-1} \right) \nonumber \\ \nonumber&&\;\;\;\; \mbox{[Lemma \ref{lem:agelk}]} \\
&\geq& f_1(j)+  (s_j+\lambda \beta+\alpha)^{k} -(s_j+\lambda \beta)^{k} \nonumber
\end{eqnarray}

Finally, the minimum cost any algorithm can incur due to Type 2 and 3 jobs is $f_{2,3}$ as stated in Fact \ref{fact:cost23lk}.  Thus, the algorithm $A$ would have total cost at least $\sum_{i = 0}^{m-1}f_1(i) + f_{2,3} + (s_j+\lambda \beta+\alpha)^{k} -(s_j+\lambda \beta)^{k} $.  Notice that $f = f_{2,3} + \sum_{i=0}^{m-1}f_1(i) + \sum_{i=1}^{m-1} m2^{2k}(s_i+\lambda \beta)^{k-1} + 3m^22^k B(\beta B + \lambda \beta + (m-1) (\alpha+B))^{k-1}$.   Basic algebra shows that $(s_j+\lambda \beta+\alpha)^{k} -(s_j+\lambda \beta)^{k} >  \sum_{i=1}^{m-1} m2^{2k}(s_i+\lambda \beta)^{k-1}  + 3m^22^k B(\beta B + \lambda \beta + (m-1) (\alpha+B))^{k-1}$  for any $j$ and sufficiently large $m$ and $B$. Therefore this contradicts the assumption on the objective value of $A$'s schedule.

\end{proof}

\begin{lemma}
\label{lem:exactagelk}
It must be the case that $\sum_{j \in U_i} b_j \leq B(m-i)$ for $i$ from $0$ to $m-1$ for any schedule $A$ with cost at most $f$.  
\end{lemma}
\begin{proof}
Clearly the lemma holds true to the interval $I_0$, since the lemma implies that in this case all jobs in $T_1$ could possibly be incomplete at the end of $I_0$.  For the sake of contradiction say that there is an interval $I_j$ such that for a schedule $A$ of cost at most $f$, $\sum_{l \in U_j} b_l > B(m-j)$.  Since $b_l$ is integral for all jobs $l \in T_1$, it is the case that $\sum_{l \in U_j} b_l \geq B(m-j)+1$.  Consider the increase in $A$'s objective during $I_j$ for Type 1 jobs.  This is the following, 

\begin{eqnarray}
&&\sum_{l \in U_j} \left( (s_j+\lambda \beta+\alpha+\beta b_l)^k - (s_j+\lambda \beta+\beta b_l)^k\right) \nonumber\\
&=& \sum_{l \in U_j}\left ( \sum_{x=0}^k \binom{k}{x} (\beta b_l)^{x} \left ( (s_j+\lambda \beta+\alpha)^{k-x} -(s_j+\lambda \beta)^{k-x} \right) \right )  \;\;\;\; \mbox{[Binomial theorem]}\nonumber\\ 
&\geq&  \sum_{l \in U_j}\left ( (s_j+\lambda \beta+\alpha)^{k} -(s_j+\lambda \beta)^{k} + \beta b_l \left ((s_j+\lambda \beta+\alpha)^{k-1} -(s_j+\lambda \beta)^{k-1} \right)  \right )\nonumber\\
&=&|U_j|\left ( (s_j+\lambda \beta+\alpha)^{k} -(s_j+\lambda \beta)^{k}  \right ) +  \beta k\left ((s_j+\lambda \beta+\alpha)^{k-1} -(s_j+\lambda \beta)^{k-1} \right)  \sum_{l \in U_j}  b_l\nonumber\\
&\geq& (3m - 3i)\left ( (s_j+\lambda \beta+\alpha)^{k} -(s_j+\lambda \beta)^{k}  \right ) +   \beta k \left ((s_j+\lambda \beta+\alpha)^{k-1} -(s_j+\lambda \beta)^{k-1} \right) \sum_{l \in U_j}  b_l \nonumber\\ && \;\;\;\; \mbox{[Lemma \ref{lem:numlk}]}\nonumber\\
&\geq& (3m - 3i)\left ( (s_j+\lambda \beta+\alpha)^{k} -(s_j+\lambda \beta)^{k}  \right ) + \beta k(mB-jB+1) \left ((s_j+\lambda \beta+\alpha)^{k-1} -(s_j+\lambda \beta)^{k-1} \right) \nonumber \\ \nonumber&&\;\;\;\; \mbox{[$\sum_{l \in U_j} b_l \geq B(m-j)+1$]} \\
&\geq& f_1(j)+  \beta k \left ((s_j+\lambda \beta+\alpha)^{k-1} -(s_j+\lambda \beta)^{k-1} \right) \nonumber
\end{eqnarray}
Now we know that the increase in the algorithm's objective due to Type 1 jobs being unsatisfied during $I_i$ where $i\neq j$ is at least $f_1(i)$ by Lemma \ref{lem:numlk}, Lemma \ref{lem:agelk}, and Fact \ref{fact:costT1ilk}.  Further, the minimum cost any schedule can incur for Type 2 and 3 jobs is $f_{2,3}$.  Thus, the cost of $A$'s schedule is at least $\sum_{i = 0}^{m-1}f_1(i) + f_{2,3} +   \beta k \left ((s_j+\lambda \beta+\alpha)^{k-1} -(s_j+\lambda \beta)^{k-1} \right)$. Knowing that $f = f_{2,3} + \sum_{i=0}^{m-1}f_1(i) + \sum_{i=1}^{m-1} m2^{2k}(s_i+\lambda \beta)^{k-1}  + 3m^22^k B(\beta B + \lambda \beta + (m-1) (\alpha+B))^{k-1}$ and that $  \beta k \left ((s_j+\lambda \beta+\alpha)^{k-1} -(s_j+\lambda \beta)^{k-1} \right) >  \sum_{i=1}^{m-1}m2^{2k}(s_i+\lambda \beta)^{k-1} + 3m^22^k B(\beta B + \lambda \beta + (m-1) (\alpha+B))^{k-1}$, we have a contradiction to the  assumption that $A$ has cost at most $f$.

\end{proof}

Now we are finally ready to prove the theorem.

\begin{theorem}
If the schedule $A$ has cost at most $f$ then there exists a valid solution to the 3-partition instance.
\end{theorem}
\begin{proof}
By Lemma \ref{lem:threelk} and Lemma \ref{lem:numlk} we know that $|T_i \setminus U_i|= 3i$.  This implies that $3$ jobs are completed between the end of $I_i$ and the end of $I_{i+1}$ for $i$ from $0$ to $m-1$ and $3$ jobs are completed after $I_{m-1}$. Let $P_i$ be the three jobs completed between the end of $I_i$ and the end of $I_{i+1}$ and $P_m$ be the remaining $3$ jobs. Lemma \ref{lem:agelk} and Lemma \ref{lem:exactagelk} state that $\sum_{l \in T_i \setminus U_i} b_l = Bm-Bi$ for $i$ from $1$ to $m-1$.  This implies that $\sum_{j \in P_i} b_j = B$.  Thus, $P_i$ contains exactly $3$ jobs such that their total processing time is exactly $B$ for $i$ from $1$ to $m$.  Hence, $P_1, \ldots, P_m$ corresponds to a solution to the 3-partition problem.
\end{proof}

\subsection{$3$-partition $\rightarrow$ $k$-norm:}  Now we show that there if there is a valid solution to the $3$-partition then there is a solution to the $k$-norm with an objective at most $f$.   Consider a valid solution to the $3$-partition problem $P_1, P_2, \ldots, P_m$.  Consider the following solution to $k$-norm problem instance.  Each job of Type 2 and 3 is scheduled as soon as it arrives.  Now consider Type 1 jobs.  The three Type 1 jobs corresponding to integers in $P_i$ are scheduled during $[(i-1)(B+\alpha), iB+(i-1)\alpha)$.   Note that these jobs can be exactly scheduled during this interval.  Let this schedule be denoted as $\mathcal{A}$.

Now we bound the cost of Type 1 jobs by separating the intervals where the jobs could possibly be waiting. First we bound the cost due to Type 1 jobs waiting during closed time intervals.

\begin{lemma}
\label{lem:type1blacklk}
For the schedule $\mathcal{A}$ the total cost Type 1 jobs accumulate during closed time intervals is $\sum_{l \in T_1} (b_l\beta+\lambda \beta)^k + \sum_{i =1}^{m-1} f_1(i) + m2^{2k}(s_i+\lambda \beta)^{k-1}$.
\end{lemma}
\begin{proof}
First consider an interval $I_i$ where $1 \leq i \leq m-1$.  Let $U^{\cA}_i$ be the set of Type 1 jobs that are unsatisfied in the schedule $\cA$ during $I_i$.  The total accumulated cost during $I_i$ for these jobs can be bounded as follows. Note that by definition of $\cA$ we know that $|U^{\cA}_i| = 3m-3i$ and $\sum_{l \in U_i^{\cA}} b_l =  mB-iB$.

\begin{eqnarray*}
&&\sum_{l \in U^{\cA}_i} \left( (s_i+\lambda \beta+\alpha+\beta b_l)^k - (s_i+\lambda \beta+\beta b_l)^k\right) \\
&=& \sum_{l \in U^{\cA}_i}\left ( \sum_{x=0}^k \binom{k}{x} (\beta b_l)^{x} \left ( (s_i+\lambda \beta+\alpha)^{k-x} -(s_i+\lambda \beta)^{k-x} \right) \right )  \;\;\;\; \mbox{[Binomial theorem]}\\ 
&=&  \sum_{l \in U^{\cA}_i}\left ( (s_i+\lambda \beta+\alpha)^{k} -(s_i+\lambda \beta)^{k} + k\beta b_l \left ((s_i+\lambda \beta+\alpha)^{k-1} -(s_i+\lambda \beta)^{k-1} \right)  \right )\\
&& \;\;\;\; + \sum_{l \in U^{\cA}_i}\left ( \sum_{x=2}^k \binom{k}{x} (\beta b_l)^{x} \left ( (s_i+\lambda \beta+\alpha)^{k-x} -(s_i+\lambda \beta)^{k-x} \right) \right )\\
&=& (3m - 3i)\left ( (s_i+\lambda \beta+\alpha)^{k} -(s_i+\lambda \beta)^{k}  \right ) +   \beta k  \left ((s_i+\lambda \beta+\alpha)^{k-1} -(s_i+\lambda \beta)^{k-1} \right)\sum_{l \in U^{\cA}_i} b_l\\ 
&& \;\;\;\; + \sum_{l \in U^{\cA}_i}\left ( \sum_{x=2}^k \binom{k}{x} (\beta b_l)^{x} \left ( (s_i+\lambda \beta+\alpha)^{k-x} -(s_i+\lambda \beta)^{k-x} \right) \right )\;\;\;\; \mbox{[Since $|U^{\cA}_i| = 3m - 3i$]}\\
&=& (3m - 3i)\left ( (s_i+\lambda \beta+\alpha)^{k} -(s_i+\lambda \beta)^{k}  \right ) + \beta k(mB-iB) \left ((s_i+\lambda \beta+\alpha)^{k-1} -(s_i+\lambda \beta)^{k-1} \right) \\
&& \;\;\;\; + \sum_{l \in U^{\cA}_i}\left ( \sum_{x=2}^k \binom{k}{x} (\beta b_l)^{x} \left ( (s_i+\lambda \beta+\alpha)^{k-x} -(s_i+\lambda \beta)^{k-x} \right) \right )\;\;\;\; \mbox{[Since $\sum_{l \in U^{\cA}_i} b_l = mB-iB$]} \\
&=& f_1(i) + \sum_{l \in U^{\cA}_i}\left ( \sum_{x=2}^k \binom{k}{x} (\beta b_l)^{x} \left ( (s_i+\lambda \beta+\alpha)^{k-x} -(s_i+\lambda \beta)^{k-x} \right) \right )\\
&=& f_1(i) + \sum_{l \in U^{\cA}_i}\left ( \sum_{x=2}^k \binom{k}{x} (\beta b_l)^{x} \sum_{y=1}^{k-x} \binom{k-x}{y}\alpha^y(s_i+\lambda \beta)^{k-x-y} \right )  \;\;\;\; \mbox{[Binomial thoerem]}\\
&\leq& f_1(i) + \sum_{l \in U^{\cA}_i}\left ( \sum_{x=2}^k \binom{k}{x} (\beta b_l)^{x} 2^{k-x}\alpha(s_i+\lambda \beta)^{k-x-1} \right )  \;\;\;\; \mbox{[$\alpha < \lambda \beta$ and  $\sum_{x=2}^k \binom{k}{x} < 2^{k-x}$]}\\
&\leq& f_1(i) + \sum_{l \in U^{\cA}_i}\left ( 2^k (\beta b_l)^2 2^{k-2}\alpha(s_i+\lambda \beta)^{k-3} \right )  \;\;\;\; \mbox{[$\beta b_l < \lambda \beta$  for all $l$ and $\sum_{x=2}^k \binom{k}{x}  < 2^k$ ]}\\
&\leq& f_1(i) + \sum_{l \in U^{\cA}_i} 2^{2k-2}(s_i+\lambda \beta)^{k-1}  \;\;\;\; \mbox{[$\beta^2 B^2 \alpha< (\lambda \beta)^2$ and $b_l < B$ for all $l$]}\\
&\leq& f_1(i) + m2^{2k}(s_i+\lambda \beta)^{k-1} \;\;\;\; \mbox{[$|U^{\cA}_i| \leq 3m$]}
\end{eqnarray*}

Now consider the interval $I_0$.  No Type 1 job is worked on during $I_0$ and all Type 1 jobs arrive on $I_0$.  Knowing this, the total increase in the objective due to Type 1 jobs waiting during $I_i$ is $\sum_{l \in T_1} (b_l\beta+\lambda \beta)^k$.  This completes the proof of the lemma.
\end{proof}

Next we bound the bound the cost Type 1 jobs accumulate during open time intervals.

\begin{lemma}
\label{lem:type1nonblacklk}
For the schedule $\cA$ the total cost Type 1 jobs accumulate during open times is $3m^22^k B(\beta B + \lambda \beta + (m-1) (\alpha+B))^{k-1}$.
\end{lemma}
\begin{proof}
Fix any Type 1 jobs $l$. The most a Type 1 job will accumulate during a maximal contiguous open time interval is during the $B$ time steps after all closed time intervals end.  This because after this time all Type 1 jobs are completed, the objective function is convex and every maximal contiguous open time interval has length $B$.  Knowing that the arrival time of any Type 1 job $l$ is after time $(\beta B +\lambda \beta)$, the most job $l$ can accumulate in a single maximal contiguous open time interval is the following.

\begin{eqnarray*}
& & (\beta B + \lambda \beta + \alpha(m-1) +mB)^k - (\beta B + \lambda \beta + (m-1)(\alpha+B))^k\\
&=& \sum_{x=1}^{k} \binom{k}{x} B^x(\beta B + \lambda \beta + (m-1) (\alpha+B))^{k-x} \;\;\;\; \mbox{[Binomial theorem]}\\
&\leq& 2^k B(\beta B + \lambda \beta + (m-1) (\alpha+B))^{k-1} \;\;\;\; \mbox{[$B < \beta$ and $ \sum_{x=1}^{k} \binom{k}{x}  < 2^k$]}
\end{eqnarray*}

Knowing that there are $3m$ jobs and $m$ maximal non-contiguous open time intervals, the total increase in $\cA$'s objective due to the cost Type 1 jobs accumulate during these time intervals is at most $3m^22^k B(\beta B + \lambda \beta + (m-1) (\alpha+B))^{k-1}$.
\end{proof}

Now we are ready to bound the cost of the schedule $\cA$.  Using Fact \ref{fact:cost23lk} the total cost for Type 2 and 3 jobs in $\cA$ is $f_{2,3}$. This and lemmas \ref{lem:type1blacklk} and \ref{lem:type1nonblacklk} implies that the total cost of $\cA$ is less than $f_{2,3} + \sum_{l \in T_1} (b_l\beta+\lambda \beta)^k + \sum_{i =1}^{m-1} f_1(i) + m2^{2k}(s_i+\lambda \beta)^{k-1}   + 3m^2k^k B(\beta B + \lambda \beta + (m-1) (\alpha+B))^{k-1} =f$.

\begin{theorem}
If there exists a valid solution to the 3-Partition instance then there is a schedule $\mathcal{A}$ for the $k$-norm problem instance with an objective value at most $f$.
\end{theorem}

\section{Hardness Proof for the $k$-norm when $0 < k < 1$}
\label{sec:halfnormflow}
In this section, we show the hardness of the $k$-norm objective for positive $k < 1$.  We reduce the $3$-partition problem to this scheduling problem.  Consider any fixed instance of the $3$-partition problem. We now construct the following instance of the $k$-norm problem.

\begin{itemize}
\item \textbf{Type 1 jobs:} For each integer $b_i \in S$ from the $3$-partition instance, we create a job of processing time $b_i$ and arrival time $-\beta+\lambda b_i$. The value of $\beta$ is set to be $(30mkB)^{5/k^2 + 2}$ and $\lambda$ is set to $\beta^{1/4}$.  Let $T_1$ denote the set of Type $1$ jobs. These jobs will be indexed for $i \in [3m]$.  Note that because $\lambda B < \beta$ and $b_i \leq B$ for all $i$, it is the case that all jobs arrive before time $0$.
\item \textbf{Type 2 jobs:}  There is a job of size $\rho$ is released every $\rho$ time steps during the intervals $[iB+ (i-1)\alpha, i(B+\alpha))$ for $i$ from $1$ to $m-1$.  Here $\rho$ and $\alpha$ are set such that $\rho = 1/(100m^4\beta)$ and $\alpha = 10\beta^{3/4}m^2B^2$. We will assume without loss of generality that $\alpha/\rho$ and $\alpha/B$ are integral. 
\item \textbf{Type 3 jobs:}  During the interval $[-\beta ,0]$ a job of size $\rho$ is released every $\rho$ time steps. We will assume that without loss of generally that $\beta/\rho$ is integral.
\end{itemize}

This is the entire input instance.  We will let $\mathcal{I}^3$ denote the 3-Partition instance and $\mathcal{I}^\ell$ be the $k$ norm instance. Let $I_0 = [\beta,0)$, $s_i = iB+ (i-1)\alpha$ and $I_i = [s_i, s_i + \alpha)$ for $i \in [m-1]$. The Type 1 jobs will be used to represent integers in the $3$-partition instance.   The Type 2 jobs will be used to ensure that an optimal schedule cannot process Type 1 during $I_i$ if there exists a valid solution to the $3$-partition instance for $i \in [m-1]$.  The Type 3 jobs are used to ensure no Type 1 job can be processed during $I_0$ in an optimal schedule. Note that during an interval $I_i$ the Type 2 or 3 jobs that arrive can be completely scheduled during this interval and require the entire interval length to be completed for all $i$.  These are the intervals which will be \emph{closed} in an optimal schedule. We call a time $t$ closed if $t \in I_i$ for $i$ from $0$ to $m-1$.  

Our goal will be to show that for a given instance of the $3$-Partition problem, there exists a schedule for the instance of the scheduling problem with an objective value of at most $f :=  (\beta+(m-1)\alpha)/\rho^{1-k} + \frac{3m^2kB}{( \beta -  \lambda B  )^{1-k}} + \sum_{l \in T_1} (\beta - \lambda b_l)^k + \sum_{i=1}^{m-1} \left ( (3m-3i)\Bigg ( \frac{k\alpha}{(s_i + \beta - \beta^{k/2} )^{1-k}} - \frac{k(1-k)(\beta^{k/2})^2}{2(s_i+\beta - \beta^{k/2})^{2-k}} \Bigg) + \frac{k(1-k)(\beta^{k/2})\lambda (mB-iB)}{(s_i+\beta - \beta^{k/2})^{2-k}} \right ) $ if and only if there exists a valid solution to the $3$-Partition problem.   

\subsection{$k$-norm $\rightarrow$ $3$-partition:} We now show that if there is a solution to the instance $\mathcal{I}^\ell$ with an objective at most $f$ then there exists a valid solution to $\mathcal{I}^3$.  To show that this is the case, assume that there exists some fixed schedule $A$ of cost at most $f$.  Before we start with the proof, we state a few facts that will be useful throughout the proof.

\begin{fact}
\label{fact:taylor}
For any $x,y>0$ and $0 < k < 1$ where $2x < y$ it is the case that:
\begin{itemize}
\item $(x+y)^k \leq y^k + k\frac{x}{y^{1-k}}$
\item $(x+y)^k \geq y^k + k\frac{x}{y^{1-k}} - k(1-k)\frac{x^2}{2y^{2-k}}$
\end{itemize}
\end{fact}
\begin{proof}
Let $g(x) = (x+y)^k$ and $g^i(x)$ denote the $i$th derivative of $g(x)$.   It can be seen that when $i \geq 1$ and $0 < k < 1$,
$$g^i(x) = (-1)^{i-1}  \frac{k\prod_{j=1}^{i-1} (j-k)}{(x+y)^{i-k}}$$
The Taylor series expansion of $g(x)$ at $x= 0$  is $\sum_{i=0}^\infty \frac{g^i(0) x^i}{i!} = y^k + \sum_{i=1}^\infty (-1)^{i-1} (\frac{x^i}{i!} ) \frac{k\prod_{j=1}^{i-1} (j-k)}{y^{i-k}}$.  We now show that the absolute value of the $l$th term in the Taylor series expansion is at least as large as the sum of the absolute values of all the later terms when $l \geq 1$.  For the sake of contradiction, assume that this is false.  Then we have the following

\begin{eqnarray*}
&& (\frac{x^l}{l!} ) \frac{k\prod_{j=1}^{l-1} (j-k)}{y^{l-k}} <   \sum_{i=l+1}^\infty (\frac{x^i}{i!} ) \frac{k\prod_{j=1}^{i-1} (j-k)}{y^{i-k}} \\
&\Rightarrow& (\frac{1}{l!} )  <   \sum_{i=l+1}^\infty (\frac{x^{i-l}}{i!} ) \frac{\prod_{j=l}^{i-1} (j-k)}{y^{i-l}} \\
&\Rightarrow& (\frac{1}{l!} )  <   \sum_{i=l+1}^\infty (\frac{x^{i-l}}{i!} ) \frac{\prod_{j=l}^{i-1} j}{y^{i-l}} \;\;\;\; \mbox{[$0 < k < 1$]}\\
&\Rightarrow&1  <   \sum_{i=l+1}^\infty  \frac{lx^{i-l}}{iy^{i-l}} \\
&\Rightarrow&1  <   \sum_{i=l+1}^\infty  \frac{x^{i-l}}{y^{i-l}} \;\;\;\; \mbox{[$i > l$]}\\
&\Rightarrow&1  <   \sum_{i=l+1}^\infty  \frac{x^{i-l}}{(2x)^{i-l}} \;\;\;\; \mbox{[$y > 2x$]} \\
&\Rightarrow&1  <   \sum_{i=l+1}^\infty  \frac{1}{2^{i-l}} \\
\end{eqnarray*}

Now we know that $\sum_{i=l+1}^\infty  \frac{1}{2^{i-l}} $ can be at most $1$ and therefore we have a contradiction.  Thus, the absolute value of the $l$th term in the Taylor series expansion is at least as large as the sum of the absolute values of all the later terms.  Using this and the Taylor expansion implies the desired inequalities.
\end{proof}

\begin{fact}
\label{fact:costT1ism}
Let $U_i$ be the set of unsatisfied Type 1 jobs at the end of interval $I_i$ in $A$.  If $|U_i| \geq 3m - 3i$ and $\sum_{l \in U_i} b_l \geq mB-iB$ then the total increase in $A$'s objective during $I_i$ due to the jobs in $U_i$ being unsatisfied during $I_i$ is at least $f_1(i) := (3m-3i)\bigg ( \frac{k\alpha}{(s_i + \beta - \beta^{k/2} )^{1-k}} - \frac{k(1-k)(\beta^{k/2}+\alpha)^2}{2(s_i+\beta - \beta^{k/2})^{2-k}} \bigg) + \frac{k(1-k)(\beta^{k/2}+\alpha)\lambda (mB-iB)}{(s_i+\beta- \beta^{k/2})^{2-k}}  -  m\frac{k(1-k)(\lambda B)^2}{2(s_i+ \beta - \beta^{k/2})^{2-k}} $ for $i$ from $1$ to $m-1$.  Similarly, if no Type 1 jobs are completed by time $0$ then the increase $A$'s objective for Type 1 jobs during $I_0$ is $f_1(0) := \sum_{i \in T_1} (\beta-\lambda b_i)^k$.  
\end{fact}
\begin{proof}
The total increase in a schedule's objective due to Type 1 jobs being unsatisfied during $I_i$ is the following if $|U_i| \geq 3m - 3i$ and $\sum_{l \in U_i} b_l \geq mB-iB$.

\begin{eqnarray*}
&&\sum_{l \in U_i} \left( (s_i+\beta+\alpha-\lambda b_l)^k - (s_i+\beta-\lambda b_l)^k\right) \\
&\geq& \sum_{l \in U_i}\Bigg( \Bigg ( (s_i+ \beta - \beta^{k/2})^k+ \frac{k(\beta^{k/2}+\alpha-\lambda b_l)}{(s_i + \beta - \beta^{k/2} )^{1-k}} - \frac{k(1-k)(\beta^{k/2}+\alpha-\lambda b_l)^2}{2(s_i+\beta - \beta^{k/2})^{2-k}} \Bigg)\\
&&\;\;\;\; - \Bigg ( (s_i+ \beta - \beta^{k/2})^k+ \frac{k(\beta^{k/2}-\lambda b_l)}{(s_i + \beta - \beta^{k/2} )^{1-k}} \Bigg)\Bigg )   \;\;\;\; \mbox{[Fact \ref{fact:taylor}, $b_l \leq B$ for all $l$ and $2(\beta^{k/2}+\alpha) \leq \beta - \beta^{k/2}$]}\\
&=& \sum_{l \in U_i} \Bigg ( \frac{k\alpha}{(s_i + \beta - \beta^{k/2} )^{1-k}} - \frac{k(1-k)(\beta^{k/2}+\alpha-\lambda b_l)^2}{2(s_i+\beta - \beta^{k/2})^{2-k}} \Bigg)\\
&=& |U_i| \Bigg ( \frac{k\alpha}{(s_i + \beta - \beta^{k/2} )^{1-k}} - \frac{k(1-k)(\beta^{k/2}+\alpha)^2}{2(s_i+\beta - \beta^{k/2})^{2-k}} \Bigg) + \sum_{l \in U_i} \frac{k(1-k)(2((\beta^{k/2}+\alpha)\lambda b_l)-(\lambda b_l)^2)}{2(s_i+\beta - \beta^{k/2})^{2-k}} \\
&\geq& (3m-3i)\Bigg ( \frac{k\alpha}{(s_i + \beta - \beta^{k/2} )^{1-k}} - \frac{k(1-k)(\beta^{k/2}+\alpha)^2}{2(s_i+\beta - \beta^{k/2})^{2-k}} \Bigg) + \frac{k(1-k)(\beta^{k/2}+\alpha)\lambda (mB-iB)}{(s_i+\beta - \beta^{k/2})^{2-k}} \\
&& -\sum_{l \in U_i} \frac{k(1-k)(\lambda B)^2}{2(s_i+\beta - \beta^{k/2})^{2-k}} \;\;\;\; \mbox{[$|U_i| = mB-iB$, $\sum_{l \in U_i} b_l = mB-iB$ and $b_l < B$ for all $l$]}\\
&\geq& (3m-3i)\Bigg ( \frac{k\alpha}{(s_i + \beta - \beta^{k/2} )^{1-k}} - \frac{k(1-k)(\beta^{k/2}+\alpha)^2}{2(s_i+\beta - \beta^{k/2})^{2-k}} \Bigg) + \frac{k(1-k)(\beta^{k/2}+\alpha)\lambda (mB-iB)}{(s_i+\beta - \beta^{k/2})^{2-k}} \\
&& - m\frac{k(1-k)(\lambda B)^2}{2(s_i+\beta - \beta^{k/2})^{2-k}} \;\;\;\; \mbox{[$|U_i| \leq m$]}
\end{eqnarray*}

Now we focus on bounding the increase in a schedule's objective if no Type 1 jobs are completed by time $0$. During the interval $I_0$ a job $i \in T_1$ waits $\beta - \lambda b_i$ time steps and its contribution to the objective during this interval is $(\beta-\lambda b_i)^k$.  Thus the total cost is $\sum_{i \in T_1} (\beta-\lambda b_i)^k$.
 
\end{proof}

\begin{fact}
\label{fact:cost23sm}
If the algorithm completes all Type 2 and 3 jobs as soon as they arrive in first-in-first-out order then the total contribution of these jobs to the objective is $f_{2,3} := (\beta+(m-1)\alpha)/\rho^{1-k}$ when $0< k <1$.  This is the minimum cost any schedule can incur for Type 2 and 3 jobs.
\end{fact}
\begin{proof}
If the algorithm completes the Type 2 and 3 jobs in this way, then each of these jobs waits $\rho$ time steps to be completed. Hence, a single Type 2 or 3 job's contribution to the objective is $\rho^k$.  There are $\beta/\rho+(m-1)\alpha/\rho$ Type 2 and 3 jobs total.  Thus, their total contribution is $(\beta+(m-1)\alpha)/\rho^{1-k}$ when $0 < k < 1$.
\end{proof}

Now we show that during blacked our intervals, only a small volume of Type 1 jobs can be processed.  Intuitively,  this lemma ensures that Type 1 jobs can only be processed during open times.

\begin{lemma}
\label{lem:blacksm}
If a schedule $A$ does a $1/2$ volume of work on Type 1 jobs during $\bigcup_{i=0}^{m-1} I_i$ then the total cost of the schedule $A$ is larger than $f$. Similarly, if a schedule $A$ does not process a $1/2$ volume of work of the Type 2 or 3 jobs that arrive on an interval $I_i$ during $I_i$ then the cost of $A$'s schedule is larger than $f$ for any $i \in \{0,1,\ldots, m-1\}$.
\end{lemma}
\begin{proof}
We begin by proving the first part of the lemma.  Assume that $A$ processes a $1/2$ volume of work on Type 1 jobs during the discontinuous time interval $\bigcup_{i=0}^{m-1} I_i$.  It follows that the algorithm processes at least $1/(2m)$ volume of work of Type 1 jobs during at least one interval $I_i$.  Fix such an interval $I$.  During $I$ a job of size $\rho$ arrive every $\rho$ time steps. This implies that at least $\frac{1}{2m\rho}$ jobs have flow time at least $\frac{1}{2m}$. Thus the total cost of the schedule is at least $\frac{1}{2m\rho}\cdot (\frac{1}{2m})^k \geq 25m^2\beta$ which is larger than $f$ for sufficiently large $m$.

Now we prove the second part of the lemma.  Say that there is a interval $I \in \{I_0, I_1, \ldots, I_{m-1} \}$ where a $1/2$ volume of work of Type 2 or 3 jobs that arrived during $I$ are not completed during $I$ in $A$'s schedule.  Then at least $1/(4\rho)$ jobs wait at least $1/(4\rho)$ time steps to be completed. Thus the cost of the schedule is at least $1/(4\rho)^{1+k}$.  Again, this is larger than $f$ for sufficiently large $B$ and $m$.
\end{proof}

Next we observe that the total number of Type 1 that the algorithm can complete before the end of a closed interval is proportional to the number of closed intervals that have occurred so far.

\begin{lemma}
\label{lem:numsm}
The total number of Type 1 jobs that can be completed by the end of $I_i$ is $3i$ for $i$ from $0$ to $m-1$ for any schedule $A$ with cost at most $f$.
\end{lemma}

\begin{proof}
By the end of an interval $I_i$ there are have been at exactly $iB$ time steps that are not contained in a closed interval before the end of $I_i$.  From Lemma \ref{lem:blacksm} at most a $1/2$ volume of work can be processed by $A$ of Type 1 jobs during closed time steps.  Knowing that the smallest Type 1 job has size $\frac{m}{3m+1/2}B$ and $B\geq3$, the total number of jobs that can be completed before the end of $I_i$ is $\floor{(iB+1/2)\Big/(\frac{m}{3m+1/2}B)} \leq 3i$. 
\end{proof}

Notice that the instantaneous increase in the objective function at a time $t$ depends on the \emph{ages} of the jobs.  Now, for a time $t\geq 0$, the age of a Type 1 job $i$ is $t+\beta- \lambda b_i$.  Our goal now is to define a bound on the total age of the Type 1 jobs during any closed interval in the schedule $A$.  Since the last lemma bounded the total number of Type 1 jobs that are unsatisfied during a closed interval, this essentially amounts to bounding the total original processing time of the unsatisfied Type 1 jobs.

\begin{lemma}
\label{lem:agesm}
For any time $t \in I_i$, it is the case that $\sum_{j\in U_i} b_j \geq B(m-i)$ for $i$ from $0$ to $m-1$ for any schedule $A$ with cost at most $f$.
\end{lemma}
\begin{proof}
By the end of $I_i$ there are have been at exactly $iB$ time steps that are not contained in a closed interval before the end of $I_i$.  From Lemma \ref{lem:blacksm} at most a $1/2$ volume of work on Type 1 jobs can be processed by $A$ during closed time steps. We know that the processing time of a job $i\in T_1$ is $b_i$ and therefore $\sum_{i \in T_1} b_i = mB$.  The total processing time of jobs in $T_1\setminus U_i$ can be at most $iB +1/2$ since these jobs are completed by the end of interval $I_i$.  Knowing that $i$ and $B$ are integral as well as $b_i$ for each $i \in T_1$, we know that $\sum_{j \in T_1 \setminus U_i} b_j \leq iB$.  Thus, $ \sum_{j\in U_i} b_j = mB - \sum_{j \in T_i \setminus U_i} b_j \geq mB - iB$.
\end{proof}

Next we show that by the end of every closed time interval $I_i$ for $i$ from $1$ to $m-1$ there must be at least $3i$ jobs completed.

\begin{lemma}
\label{lem:threesm}
By the end of interval $I_i$ there must be at least $3i$ Type 1 jobs complete in any schedule $A$ that has cost at most $f$.
\end{lemma}
\begin{proof}
For the sake of contradiction, say that there is an interval $I_j$ such that $3j-1$ or less Type 1 jobs are completed by the end of $I_j$ in $A$.  We begin by bounding the increase in $A$'s objective for $T_1$ jobs during closed time intervals. Recall that $s_i$ is the start time of interval $I_i$. For any interval $I_i$ where $i \neq j$, the total increase in $A$'s objective due to Type 1 jobs being unsatisfied during $I_i$ is at least $f_1(i)  $ by Lemma \ref{lem:numsm}, Lemma \ref{lem:agesm}, and Fact \ref{fact:costT1ism}. Similarly, the increase in $A$'s objective during $I_j$ due to jobs in $U_j$ being unsatisfied is,

\begin{eqnarray*}
&&\sum_{l \in U_j} \left( (s_j+\beta+\alpha-\lambda b_l)^k - (s_j+\beta-\lambda b_l)^k\right) \\
&\geq& \sum_{l \in U_j}\Bigg( \Bigg ( (s_j+ \beta - \beta^{k/2})^k+ \frac{k(\beta^{k/2}+\alpha-\lambda b_l)}{(s_j + \beta - \beta^{k/2} )^{1-k}} - \frac{k(1-k)(\beta^{k/2}+\alpha-\lambda b_l)^2}{2(s_j+\beta - \beta^{k/2})^{2-k}} \Bigg)\\
&&\;\;\;\; - \Bigg ( (s_j+ \beta - \beta^{k/2})^k+ \frac{k(\beta^{k/2}-\lambda b_l)}{(s_j + \beta - \beta^{k/2} )^{1-k}} \Bigg)\Bigg )   \;\;\;\; \mbox{[Fact \ref{fact:taylor}, $b_l \leq B$ for all $l$ and $2(\beta^{k/2}+ \alpha) \leq \beta - \beta^{k/2}$]}\\
&=& \sum_{l \in U_j} \Bigg ( \frac{k\alpha}{(s_j + \beta - \beta^{k/2} )^{1-k}} - \frac{k(1-k)(\beta^{k/2}+\alpha-\lambda b_l)^2}{2(s_j+\beta - \beta^{k/2})^{2-k}} \Bigg)\\
&=& |U_j| \Bigg ( \frac{k\alpha}{(s_j + \beta - \beta^{k/2} )^{1-k}} - \frac{k(1-k)(\beta^{k/2}+\alpha)^2}{2(s_j+\beta - \beta^{k/2})^{2-k}} \Bigg) + \sum_{l \in U_j} \frac{k(1-k)(2((\beta^{k/2}+\alpha)\lambda b_l)-(\lambda b_l)^2)}{2(s_j+\beta - \beta^{k/2})^{2-k}} \\
&\geq& (3m-3j+1)\Bigg ( \frac{k\alpha}{(s_j + \beta - \beta^{k/2} )^{1-k}} - \frac{k(1-k)(\beta^{k/2}+\alpha)^2}{2(s_j+\beta - \beta^{k/2})^{2-k}} \Bigg) + \frac{k(1-k)(\beta^{k/2}+\alpha)\lambda (mB-iB)}{(s_j+\beta - \beta^{k/2})^{2-k}} \\
&& -\sum_{l \in U_j} \frac{k(1-k)(\lambda B)^2}{2(s_j+\beta - \beta^{k/2})^{2-k}} \;\;\;\; \mbox{[Lemma \ref{lem:agesm}, the definition of $I_j$ and $b_l < B$ for all $l$ ]}\\
&\geq& (3m-3j+1)\Bigg ( \frac{k\alpha}{(s_j + \beta - \beta^{k/2} )^{1-k}} - \frac{k(1-k)(\beta^{k/2}+\alpha)^2}{2(s_j+\beta - \beta^{k/2})^{2-k}} \Bigg) + \frac{k(1-k)(\beta^{k/2}+\alpha)\lambda (mB-iB)}{(s_j+\beta - \beta^{k/2})^{2-k}} \\
&& -m\frac{k(1-k)(\lambda B)^2}{2(s_j+\beta - \beta^{k/2})^{2-k}} \;\;\;\; \mbox{[$|U_j| \leq m$]}
\end{eqnarray*}

Finally, the minimum cost any algorithm can incur due to Type 2 and 3 jobs is $f_{2,3}$ as stated in Fact \ref{fact:cost23sm}.  Thus, the algorithm $A$ would have total cost at least $\sum_{i = 0}^{m-1}f_1(i) + f_{2,3} + ( \frac{k\alpha}{(s_j + \beta - \beta^{k/2} )^{1-k}} - \frac{k(1-k)(\beta^{k/2}+\alpha)^2}{2(s_j+\beta - \beta^{k/2})^{2-k}} ) $.  Notice that $f = \sum_{i = 0}^{m-1}f_1(i) + f_{2,3} + \frac{3m^2kB}{( \beta - \beta \lambda B  )^{1-k}} - \sum_{i  =1}^{m-1} m\frac{k(1-k)(\lambda B)^2}{2(s_i+\beta - \beta^{k/2})^{2-k}}$.  It can be seen that $( \frac{k\alpha}{(s_j + \beta - \beta^{k/2} )^{1-k}} - \frac{k(1-k)(\beta^{k/2}+\alpha)^2}{2(s_j+\beta - \beta^{k/2})^{2-k}} )    > \frac{3m^2kB}{( \beta -  \lambda B  )^{1-k}}+ \sum_{i  =1}^{m-1}m\frac{k(1-k)(\lambda B)^2}{2(s_i+\beta - \beta^{k/2})^{2-k}}$ for sufficiently large $m$ and $B$ therefore this contradicts the assumption on the objective value of $A$'s schedule.

\end{proof}

\begin{lemma}
\label{lem:exactagesm}
It must be the case that $\sum_{j \in U_i} b_j \leq B(m-i)$ for $i$ from $0$ to $m-1$ for any schedule $A$ with cost at most $f$.  
\end{lemma}
\begin{proof}
Clearly the lemma holds true to the interval $I_0$, since the lemma implies that in this case all jobs in $T_1$ could possibly be incomplete at the end of $I_0$.  For the sake of contradiction say that there is an interval $I_j$ such that for a schedule $A$ of cost at most $f$, $\sum_{l \in U_j} b_l > B(m-j)$.  Since $b_l$ is integral for all jobs $l \in T_1$, it is the case that $\sum_{l \in U_j} b_l \geq B(m-j)+1$.  Consider the increase in $A$'s objective during $I_j$ for Type 1 jobs.  This is the following, 

\begin{eqnarray*}
&&\sum_{l \in U_j} \left( (s_j+\beta+\alpha-\lambda b_l)^k - (s_j+\beta-\lambda b_l)^k\right) \\
&\geq& \sum_{l \in U_j}\Bigg( \Bigg ( (s_j+ \beta - \beta^{k/2})^k+ \frac{k(\beta^{k/2}+\alpha-\lambda b_l)}{(s_j + \beta - \beta^{k/2} )^{1-k}} - \frac{k(1-k)(\beta^{k/2}+\alpha-\lambda b_l)^2}{2(s_j+\beta - \beta^{k/2})^{2-k}} \Bigg)\\
&&\;\;\;\; - \Bigg ( (s_j+ \beta - \beta^{k/2})^k+ \frac{k(\beta^{k/2}-\lambda b_l)}{(s_j + \beta - \beta^{k/2} )^{1-k}} \Bigg)\Bigg )   \;\;\;\; \mbox{[Fact \ref{fact:taylor}, $b_l \leq B$ for all $l$ and $2(\beta^{k/2} + \alpha) \leq \beta - \beta^{k/2}$]}\\
&=& \sum_{l \in U_j} \Bigg ( \frac{k\alpha}{(s_j + \beta - \beta^{k/2} )^{1-k}} - \frac{k(1-k)(\beta^{k/2}+\alpha-\lambda b_l)^2}{2(s_j+\beta - \beta^{k/2})^{2-k}} \Bigg)\\
&=& |U_j| \Bigg ( \frac{k\alpha}{(s_j + \beta - \beta^{k/2} )^{1-k}} - \frac{k(1-k)(\beta^{k/2}+\alpha)^2}{2(s_j+\beta - \beta^{k/2})^{2-k}} \Bigg) + \sum_{l \in U_j} \frac{k(1-k)(2((\beta^{k/2}+\alpha)\lambda b_l)-(\lambda b_l)^2)}{2(s_j+\beta - \beta^{k/2})^{2-k}} \\
&\geq& (3m-3j)\Bigg ( \frac{k\alpha}{(s_j + \beta - \beta^{k/2} )^{1-k}} - \frac{k(1-k)(\beta^{k/2}+\alpha)^2}{2(s_j+\beta - \beta^{k/2})^{2-k}} \Bigg) + \frac{k(1-k)(\beta^{k/2}+\alpha)\lambda (mB-iB+1)}{(s_j+\beta - \beta^{k/2})^{2-k}} \\
&& -\sum_{l \in U_j} \frac{k(1-k)(\lambda B)^2}{2(s_j+\beta - \beta^{k/2})^{2-k}} \;\;\;\; \mbox{[Lemma \ref{lem:numsm}, the definition of $I_j$ and $b_l < B$ for all $l$ ]}\\
&\geq& (3m-3j)\Bigg ( \frac{k\alpha}{(s_j + \beta - \beta^{k/2} )^{1-k}} - \frac{k(1-k)(\beta^{k/2}+\alpha)^2}{2(s_j+\beta - \beta^{k/2})^{2-k}} \Bigg) + \frac{k(1-k)(\beta^{k/2}+\alpha)\lambda (mB-iB+1)}{(s_j+\beta - \beta^{k/2})^{2-k}} \\
&& - m \frac{k(1-k)(\lambda B)^2}{2(s_j+\beta - \beta^{k/2})^{2-k}} \;\;\;\; \mbox{[$|U_j| \leq m$]}
\end{eqnarray*}

Now we know that the increase in the algorithm's objective due to Type 1 jobs being unsatisfied during $I_i$ where $i\neq j$ is at least $f_1(i)$ by Lemma \ref{lem:numsm}, Lemma \ref{lem:agesm}, and Fact \ref{fact:costT1ism}.  Further, the minimum cost any schedule can incur for Type 2 and 3 jobs is $f_{2,3}$.  Thus, the cost of $A$'s schedule is at least $\sum_{i = 0}^{m-1}f_1(i) + f_{2,3} + \frac{k(1-k)(\beta^{k/2}+\alpha)\lambda}{(s_j+\beta - \beta^{k/2})^{2-k}}$. Knowing that $f = \sum_{i = 0}^{m-1}f_1(i) + f_{2,3} + \frac{3m^2kB}{( \beta -  \lambda B  )^{1-k}} + \sum_{i  =1}^{m-1}m\frac{k(1-k)(\lambda B)^2}{2(s_i+\beta - \beta^{k/2})^{2-k}}$ and that $\frac{k(1-k)(\beta^{k/2}+\alpha)\lambda}{(s_j+\beta - \beta^{k/2})^{2-k}} >\frac{3m^2kB}{( \beta -  \lambda B  )^{1-k}} + \sum_{i  =1}^{m-1}m\frac{k(1-k)(\lambda B)^2}{2(s_i+\beta - \beta^{k/2})^{2-k}}$, we have a contradiction to the the fact that $A$ has cost at most $f$.
\end{proof}

Now we are finally ready to prove the theorem.

\begin{theorem}
If the schedule $A$ has cost at most $f$ then there exists a valid solution to the 3-partition instance.
\end{theorem}
\begin{proof}
By Lemma \ref{lem:threesm} and Lemma \ref{lem:numsm} we know that $|T_i \setminus U_i|= 3i$.  This implies that $3$ jobs are completed between the end of $I_i$ and the end of $I_{i+1}$ for $i$ from $0$ to $m-1$ and $3$ jobs are completed after $I_{m-1}$. Let $P_i$ be the three jobs completed between the end of $I_i$ and the end of $I_{i+1}$ and $P_m$ be the remaining $3$ jobs. Lemma \ref{lem:agesm} and Lemma \ref{lem:exactagesm} state that $\sum_{l \in T_i \setminus U_i} b_l = Bm-Bi$ for $i$ from $1$ to $m-1$.  This implies that $\sum_{j \in P_i} b_j = B$.  Thus, $P_i$ contains exactly $3$ jobs such that their total processing time is exactly $B$ for $i$ from $1$ to $m$.  Hence, $P_1, \ldots, P_m$ corresponds to a solution to the 3-partition problem.
\end{proof}

\subsection{$3$-partition $\rightarrow$ $k$-norm:}  Now we show that there if there is a valid solution to the $3$-partition then there is a solution to the $k$-norm with an objective at most $f$.   Consider a valid solution to the $3$-partition problem $P_1, P_2, \ldots, P_m$.  Consider the following solution to $k$-norm problem instance.  Each job of Type 2 and 3 is scheduled as soon as it arrives. Now consider Type 1 jobs.  The three Type 1 jobs corresponding to integers in $P_i$ are scheduled during $[(i-1)(B+\alpha), iB+(i-1)\alpha)$.   Note that these jobs can be exactly scheduled during this interval.  Let $\mathcal{A}$ denote this schedule.

Now we bound the cost of Type 1 jobs in the schedule $\mathcal{A}$ by separating the intervals where the jobs could possibly be waiting.

\begin{lemma}
\label{lem:type1blacksm}
For the schedule $\mathcal{A}$ the total cost Type 1 jobs accumulate during closed times is less than $\sum_{l \in T_1} (\beta - \lambda b_l)^k + (3m-3i)\Bigg ( \frac{k\alpha}{(s_i + \beta - \beta^{k/2} )^{1-k}} - \frac{k(1-k)(\beta^{k/2})^2}{2(s_i+\beta - \beta^{k/2})^{2-k}} \Bigg) + \frac{k(1-k)(\beta^{k/2})\lambda (mB-iB)}{(s_i+\beta - \beta^{k/2})^{2-k}}$.
\end{lemma}
\begin{proof}
Consider an interval $I_i$ where $1 \leq i \leq m-1$.  Let $U^{\mathcal{A}}_i$ be the set of Type 1 jobs that are unsatisfied in the schedule $\mathcal{A}$ during $I_i$.  The total accumulated cost during $I_i$ for these jobs can be bounded as follows knowing that $|U^{\mathcal{A}}_i| = 3m-3i$ and $\sum_{l \in U^{\mathcal{A}}_i } b_l = mB - iB $.

 \begin{eqnarray*}
&&\sum_{l \in U_i} \left( (s_i+\beta+\alpha-\lambda b_l)^k - (s_i+\beta-\lambda b_l)^k\right) \\
&\leq& \sum_{l \in U_i}\Bigg( \Bigg ( (s_i+ \beta - \beta^{k/2})^k+ \frac{k(\beta^{k/2}+\alpha-\lambda b_l)}{(s_i + \beta - \beta^{k/2} )^{1-k}} \Bigg) - \Bigg ( (s_i+ \beta - \beta^{k/2})^k+ \frac{k(\beta^{k/2}-\lambda b_l)}{(s_i + \beta - \beta^{k/2} )^{1-k}}\\
&& \;\;\;\;  - \frac{k(1-k)(\beta^{k/2}-\lambda b_l)^2}{2(s_i+\beta - \beta^{k/2})^{2-k}} \Bigg)\Bigg )   \;\;\;\; \mbox{[Fact \ref{fact:taylor}, $b_l \leq B$ for all $l$ and $2(\beta^{k/2}+ \alpha) \leq \beta - \beta^{k/2}$]}\\
&=& \sum_{l \in U_i} \Bigg ( \frac{k\alpha}{(s_i + \beta - \beta^{k/2} )^{1-k}} - \frac{k(1-k)(\beta^{k/2}-\lambda b_l)^2}{2(s_i+\beta - \beta^{k/2})^{2-k}} \Bigg)\\
&=& |U_i| \Bigg ( \frac{k\alpha}{(s_i + \beta - \beta^{k/2} )^{1-k}} - \frac{k(1-k)(\beta^{k/2})^2}{2(s_i+\beta - \beta^{k/2})^{2-k}} \Bigg) + \sum_{l \in U_i} \frac{k(1-k)(2((\beta^{k/2})\lambda b_l)-(\lambda b_l)^2)}{2(s_i+\beta - \beta^{k/2})^{2-k}} \\
&\geq& (3m-3i)\Bigg ( \frac{k\alpha}{(s_i + \beta - \beta^{k/2} )^{1-k}} - \frac{k(1-k)(\beta^{k/2})^2}{2(s_i+\beta - \beta^{k/2})^{2-k}} \Bigg) + \frac{k(1-k)(\beta^{k/2})\lambda (mB-iB)}{(s_i+\beta - \beta^{k/2})^{2-k}} \\
&& \;\;\;\; \mbox{[$|U^{\mathcal{A}}_i| = mB-iB$ and $\sum_{l \in U^{\mathcal{A}}_i} b_l = mB-iB$ ]}
\end{eqnarray*}

Now consider the interval $I_0$.  No Type 1 jobs is completed by the end of $I_0$, so the total cost accumulated during this interval is $\sum_{l \in T_1} (\beta - \lambda b_l)^k$.
\end{proof}

\begin{lemma}
\label{lem:type1nonblacksm}
For the schedule $\mathcal{A}$ the total cost Type 1 jobs accumulate during open times is less than $\frac{3m^2kB}{( \beta -  \lambda B  )^{1-k}}$.
\end{lemma}
\begin{proof}
Fix any Type 1 job $l$. The most a Type 1 job will accumulate during a maximal contiguous open time interval is during the $B$ time steps time $0$.  This because the objective function is concave and every maximal contiguous open interval has length $B$.  Knowing that the arrival time of any Type 1 job $l$ is before time $(\beta -\lambda \beta B)$, the most job $l$ can accumulate in a single maximal contiguous open time interval is the following.

 \begin{eqnarray*}
&& (\beta -  \lambda B +B)^k - (\beta -  \lambda B )^k \\
&\leq&  \frac{kB}{( \beta -  \lambda B  )^{1-k}} \;\;\;\; \mbox{[Fact \ref{fact:taylor} and $2B \leq \beta$]}\\
\end{eqnarray*}

There are $3m$ jobs total and $m$ open maximal contiguous time intervals.  Thus, the total cost the schedule $\mathcal{A}$ can accumulate during open times is less than $\frac{3m^2kB}{( \beta -  \lambda B  )^{1-k}}$.
\end{proof}

Now we can bound the total cost of $\mathcal{A}$.  Using Fact \ref{fact:cost23sm} the total cost for Type 2 and 3 jobs in $\mathcal{A}$ is $f_{2,3}$.  This and lemmas \ref{lem:type1blacksm} and \ref{lem:type1nonblacksm} imply that the total cost of $\mathcal{A}$ is less than $f_{2,3} + \frac{3m^2kB}{( \beta -  \lambda B  )^{1-k}} + \sum_{l \in T_1} (\beta - \lambda b_l)^k + (3m-3i)\Bigg ( \frac{k\alpha}{(s_i + \beta - \beta^{k/2} )^{1-k}} - \frac{k(1-k)(\beta^{k/2})^2}{2(s_i+\beta - \beta^{k/2})^{2-k}} \Bigg) + \frac{k(1-k)(\beta^{k/2})\lambda (mB-iB)}{(s_i+\beta - \beta^{k/2})^{2-k}} = f$.

\begin{theorem}
If there exists a valid solution to the 3-Partition instance then there is a schedule for the $k$-norm problem instance with an objective value at most $f$.
\end{theorem}

\section{Hardness Proof for the $k$-norm of stretch when $k\geq 3$.}
\label{sec:3normstretch}

In this section, we show the hardness of the $k$-norm of stretch objective for fixed integral $k \geq 3$.  We reduce the $3$-partition problem to this scheduling problem.  For this case, we will assume without loss of generality that in the $3$-partition instance that $B/3 -\eps \leq b_i \leq B/3 + \eps$ for all $i \in [m]$ and $\eps \leq 1/(2mB)^{20k}$.  We will let $\Delta_s =B/3 -\eps$ and $\Delta_b = B/3 +\eps$.  Consider any fixed instance of the $3$-partition problem. We now construct the following instance of the $k$-norm problem.

\begin{itemize}
\item \textbf{Type 1 jobs:} For each integer $b_i \in S$ from the $3$-partition instance, we create a job of processing time $b_i$ and arrival time $-(\lambda \beta+\beta b_i)$. The value of $\beta$ is set to be $2^{10k}m^{7}B^7$.  Let $T_1$ denote the set of Type $1$ jobs. These jobs will be indexed for $i \in [3m]$. 
\item \textbf{Type 2 jobs:}  There is a job of size $\rho$ is released every $\rho$ time steps during the intervals $[iB+ (i-1)\alpha, i(B+\alpha))$ for $i$ from $1$ to $m-1$.  Here $\rho$ and $\alpha$ are set such that   $\alpha = 2^{6k}m^6B^6$ and $\rho = 1/(2m\beta\alpha)^{2k}$. We will assume without loss of generality that $\alpha/\rho$ and $\alpha/B$ are integral. 
\item \textbf{Type 3 jobs:}  During the interval $[-(\lambda \beta+\beta B),0]$ a job of size $\rho$ is released every $\rho$ time steps.  Here $\lambda = B\sqrt{\alpha} = 2^{3k}B^4 m^3$. We will assume that without loss of generally that $\beta/\rho$ is integral.
\end{itemize}

This is the entire input instance.  We will let $\mathcal{I}^3$ denote the 3-Partition instance and $\mathcal{I}^\ell$ be the $k$ norm instance. Let $I_0 = [-(\lambda \beta+\beta B),0)$, $s_i = iB+ (i-1)\alpha$ and $I_i = [s_i, s_i + \alpha)$ for $i \in [m-1]$. The intuition behind the instance $\mathcal{I}^\ell$ is essentially the same as the $\ell_2$-norm instance of the previous section.  The Type 1 jobs will be used to represent integers in the $3$-partition instance.   The Type 2 jobs will be used to ensure that an optimal schedule cannot process Type 1 during $I_i$ if there exists a valid solution to the $3$-partition instance for $i \in [m-1]$.  The Type 3 jobs are used to ensure no Type 1 job can be processed during $I_0$ in an optimal schedule. Note that during an interval $I_i$ the Type 2 or 3 jobs that arrive can be completely scheduled during this interval and require the entire interval length to be completed for all $i$.  These are the intervals which will be \emph{closed} in an optimal schedule. We call a time $t$ closed if $t \in I_i$ for $i$ from $0$ to $m-1$.  

Our goal will be to show that for a given instance of the $3$-Partition problem, there exists a schedule for the instance of the scheduling problem with an objective value of at most $f :=  (\beta B+\lambda \beta+(m-1)\alpha)/\rho+ \sum_{l \in T_1} (b_l\beta+\beta^2)^k + \sum_{i =1}^{m-1} \frac{1}{\Delta_s^k}\big ( (3m - 3i)\left ( (s_i+\lambda \beta+\alpha)^{k} -(s_i+\lambda \beta)^{k}  \right ) + \beta k\left ((s_i+\lambda \beta+\alpha)^{k-1} -(s_i+\lambda \beta)^{k-1} \right) (mB-iB) +m2^{2k}(s_i+\lambda \beta)^{k-1}\big )  +  \frac{1}{\Delta_s^k} 3m^22^k B(\beta B + \lambda \beta + (m-1) (\alpha+B))^{k-1}$ if and only if there exists a valid solution to the $3$-Partition problem.

\subsection{$k$-norm of stretch $\rightarrow$ $3$-partition:} We now show that if there is a solution to the instance $\mathcal{I}^\ell$ with an objective at most $f$ then there exists a valid solution to $\mathcal{I}^3$.  To show that this is the case, assume that there exists some fixed schedule $A$ of cost at most $f$.  Before we start with the proof, we state a few facts that will be useful throughout the proof.

\begin{fact}
\label{fact:costT1ilkst}
Let $U_i$ be the set of unsatisfied Type 1 jobs at the end of interval $I_i$ in $A$.  If $|U_i| \geq 3m - 3i$ and $\sum_{l \in U_i} b_l \geq mB-iB$ then the total increase in $A$'s objective during $I_i$ due to the jobs in $U_i$ being unsatisfied during $I_i$ is at least $f_1(i) := \frac{1}{\Delta_b^k}(3m - 3i)\left ( (s_i+\lambda \beta+\alpha)^{k} -(s_i+\lambda \beta)^{k}  \right ) + \frac{1}{\Delta_b^k}\beta k\left ((s_i+\lambda \beta+\alpha)^{k-1} -(s_i+\lambda \beta)^{k-1} \right) (mB-iB)$ for $i$ from $1$ to $m-1$.  Similarly, if no Type 1 jobs are completed by time $0$ then the increase $A$'s objective for Type 1 jobs during $I_0$ is at least $f_1(0) := \sum_{i \in T_1} \frac{1}{\Delta_b^k}(\lambda \beta+\beta b_i)^k$.  
\end{fact}
\begin{proof}
The total increase in a schedule's objective due to Type 1 jobs being unsatisfied during $I_i$ is the following if $|U_i| \geq 3m - 3i$, $\sum_{l \in U_i} b_l \geq mB-iB$ and $1 \leq i \leq m-1$.

\begin{eqnarray*}
&&\sum_{l \in U_i} \frac{1}{b_l^k}\left( (s_i+\lambda \beta+\alpha+\beta b_l)^k - (s_i+\lambda \beta+\beta b_l)^k\right) \\
&=& \sum_{l \in U_i}\frac{1}{b_l^k} \left ( \sum_{x=0}^k \binom{k}{x} (\beta b_l)^{x} \left ( (s_i+\lambda \beta+\alpha)^{k-x} -(s_i+\lambda \beta)^{k-x} \right) \right )  \;\;\;\; \mbox{[Binomial theorem]}\\ 
&\geq& \sum_{l \in U_i}\frac{1}{\Delta_b^k} \left ( \sum_{x=0}^k \binom{k}{x} (\beta b_l)^{x} \left ( (s_i+\lambda \beta+\alpha)^{k-x} -(s_i+\lambda \beta)^{k-x} \right) \right )  \;\;\;\; \mbox{[$b_l \leq \Delta^b$ fror all $l$]}\\ 
&\geq&  \sum_{l \in U_i} \frac{1}{\Delta_b^k}\left ( (s_i+\lambda \beta+\alpha)^{k} -(s_i+\lambda \beta)^{k} + \beta b_l k\left ((s_i+\lambda \beta+\alpha)^{k-1} -(s_i+\lambda \beta)^{k-1} \right)  \right )\\
&=&|U_i| \frac{1}{\Delta_b^k}\left ( (s_i+\lambda \beta+\alpha)^{k} -(s_i+\lambda \beta)^{k}  \right ) +  \frac{1}{\Delta_b^k} \beta  k \left ((s_i+\lambda \beta+\alpha)^{k-1} -(s_i+\lambda \beta)^{k-1} \right)\sum_{l \in U_i}b_l \\
&\geq& \frac{1}{\Delta_b^k}(3m - 3i)\left ( (s_i+\lambda \beta+\alpha)^{k} -(s_i+\lambda \beta)^{k}  \right ) +   \frac{1}{\Delta_b^k}\beta k\left ((s_i+\lambda \beta+\alpha)^{k-1} -(s_i+\lambda \beta)^{k-1} \right)\sum_{l \in U_i}b_l\\ && \;\;\;\; \mbox{[Since $|U_i| \geq 3m - 3i$]}\\
&\geq& \frac{1}{\Delta_b^k}(3m - 3i)\left ( (s_i+\lambda \beta+\alpha)^{k} -(s_i+\lambda \beta)^{k}  \right ) + \frac{1}{\Delta_b^k}\beta k\left ((s_i+\lambda \beta+\alpha)^{k-1} -(s_i+\lambda \beta)^{k-1} \right) (mB-iB)  \\&&\;\;\;\; \mbox{[Since $\sum_{l \in U_i} b_l \geq mB-iB$]} 
\end{eqnarray*}

This completes the proof for an interval $I_i$ for for $1 \leq i \leq m-1$.  Now we focus on bounding the increase in a schedule's objective if no Type 1 jobs are completed by time $0$. During the interval $I_0$ a job $i \in T_1$ waits $\beta + \beta b_i$ time steps and its contribution to the objective during this interval is $(\lambda \beta+\beta b_i)^k$.  Thus the total cost is $\sum_{i \in T_1} \frac{1}{b_l^k} (\lambda \beta+\beta b_i)^k \geq \sum_{i \in T_1} \frac{1}{\Delta_b^k} (\lambda \beta+\beta b_i)^k$.
 
\end{proof}

\begin{fact}
\label{fact:cost23lkst}
If the algorithm completes all Type 2 and 3 jobs as soon as they arrive in first-in-first-out order then the total contribution of these jobs to the objective is $f_{2,3} :=(\beta B+\lambda \beta+(m-1)\alpha)/\rho$.  This is the minimum cost any schedule can incur for Type 2 and 3 jobs.
\end{fact}
\begin{proof}
If the algorithm completes the Type 2 and 3 jobs in this way, then each of these jobs waits $\rho$ time steps to be completed. Hence, a single Type 2 or 3 job's contribution to the objective is $1$.  There are $(\beta B+\lambda \beta)/\rho+(m-1)\alpha/\rho$ Type 2 and 3 jobs total.  Thus, their total contribution is $(\beta B+\lambda \beta+(m-1)\alpha)/\rho$.
\end{proof}

Now we show that during closed intervals, only a small volume of Type 1 jobs can be processed.  Intuitively,  this lemma ensures that Type 1 jobs can only be processed during open times.

\begin{lemma}
\label{lem:blacklkst}
If a schedule $A$ does a $1/2$ volume of work on Type 1 jobs during $\bigcup_{i=0}^{m-1} I_i$ then the total cost of the schedule $A$ is larger than $f$. Similarly, if a schedule $A$ does not process a $1/2$ volume of work of the Type 2 or 3 jobs that arrive on an interval $I_i$ during $I_i$ then the cost of $A$'s schedule is larger than $f$ for any $i \in \{0,1,\ldots, m-1\}$.
\end{lemma}
\begin{proof}
We begin by proving the first part of the lemma.  Assume that $A$ processes a $1/2$ volume of work on Type 1 jobs during the discontinuous time interval $\bigcup_{i=0}^{m-1} I_i$.  It follows that the algorithm processes at least $1/(2m)$ volume of work of Type 1 jobs during at least one interval $I_i$ where $0 \leq i \leq m-1$.  Fix such an interval $I$.  During $I$ a job of size $\rho$ arrive every $\rho$ time steps. This implies that at least $\frac{1}{2m\rho}$ jobs have flow time at least $\frac{1}{2m}$. Thus the total cost of the schedule is at least $\frac{1}{2m\rho^{k+1}}\cdot (\frac{1}{2m})^k \geq (\beta\alpha)^{k^2}$ which is larger than $f$ for sufficiently large $m$.

Now we prove the second part of the lemma.  Say that there is a interval $I \in \{I_0, I_1, \ldots, I_{m-1} \}$ where a $1/2$ volume of work of Type 2 or 3 jobs that arrived during $I$ are not completed during $I$ in $A$'s schedule.  Then at least $1/(4\rho)$ jobs wait at least $1/(4\rho)$ time steps to be completed. Thus the cost of the schedule is at least $1/(4^{k+1}\rho^{2k+1})$.  Again, this is larger than $f$ for sufficiently large $B$ and $m$.
\end{proof}

Next we observe that the total number of Type 1 that the algorithm can complete before the end of a closed interval is proportional to the number of closed intervals that have occurred so far.

\begin{lemma}
\label{lem:numlkst}
The total number of Type 1 jobs that can be completed by the end of $I_i$ is $3i$ for $i$ from $0$ to $m-1$ for any schedule $A$ with cost at most $f$.
\end{lemma}

\begin{proof}
By the end of an interval $I_i$ there are have been at exactly $iB$ time steps that are not contained in a closed interval before the end of $I_i$.  From Lemma \ref{lem:blacklkst} at most a $1/2$ volume of work can be processed by $A$ of Type 1 jobs during closed time steps.  Knowing that the smallest Type 1 job has size $\Delta_s = B/3 -\eps$, the total number of jobs that can be completed before the end of $I_i$ is $\floor{(iB+1/2)\Big/(B/3 -\eps)} \leq 3i$ since $\eps \leq 1/(2mB)^{20k}$. 
\end{proof}

Notice that the instantaneous increase in the objective function at a time $t$ depends on the \emph{ages} of the jobs.  Now, for a time $t\geq 0$, the age of a Type 1 job $i$ is $t+\lambda \beta+ \beta b_i$.  Our goal now is to define a bound on the total age of the Type 1 jobs during any closed interval in the schedule $A$.  Since the last lemma bounded the total number of Type 1 jobs that are unsatisfied during a closed interval, this essentially amounts to bounding the total original processing time of the unsatisfied Type 1 jobs.

\begin{lemma}
\label{lem:agelkst}
For any time $t \in I_i$, it is the case that $\sum_{j\in U_i} b_j \geq B(m-i)$ for $i$ from $0$ to $m-1$ for any schedule $A$ with cost at most $f$.
\end{lemma}
\begin{proof}
By the end of $I_i$ there are have been at exactly $iB$ time steps that are not contained in a closed interval before the end of $I_i$.  From Lemma \ref{lem:blacklkst} at most a $1/2$ volume of work on Type 1 jobs can be processed by $A$ during closed time steps. We know that the processing time of a job $i\in T_1$ is $b_i$ and therefore $\sum_{i \in T_1} b_i = mB$.  The total processing time of jobs in $T_1\setminus U_i$ can be at most $iB +1/2$ since these jobs are completed by the end of interval $I_i$.  Knowing that $i$ and $B$ are integral as well as $b_i$ for each $i \in T_1$, we know that $\sum_{j \in T_1 \setminus U_j} b_i \leq iB$.  Thus, $ \sum_{j\in U_i} b_j = mB - \sum_{j \in T_i \setminus U_i} b_j \geq mB - iB$.
\end{proof}

Next we show that by the end of every closed time interval $I_i$ for $i$ from $1$ to $m-1$ there must be at least $3i$ jobs completed.

\begin{lemma}
\label{lem:threelkst}
By the end of interval $I_i$ there must be at least $3i$ Type 1 jobs complete in any schedule $A$ that has cost at most $f$.
\end{lemma}
\begin{proof}
For the sake of contradiction, say that there is an interval $I_j$ such that $3j-1$ or less Type 1 jobs are completed by the end of $I_j$ in $A$.  We begin by bounding the increase in $A$'s objective for $T_1$ jobs during closed time intervals. Recall that $s_i$ is the start time of interval $I_i$. For any interval $I_i$ where $i \neq j$, the total increase in $A$'s objective due to Type 1 jobs being unsatisfied during $I_i$ is at least $f_1(i) $ by Lemma \ref{lem:numlkst}, Lemma \ref{lem:agelkst}, and Fact \ref{fact:costT1ilkst}. Similarly, the increase in $A$'s objective during $I_j$ due to jobs in $U_j$ being unsatisfied is,

\begin{eqnarray}
&&\sum_{l \in U_j} \frac{1}{b_l^k}\left( (s_j+\lambda \beta+\alpha+\beta b_l)^k - (s_j+\lambda \beta+\beta b_l)^k\right) \nonumber\\
&=& \sum_{l \in U_j}\frac{1}{b_l^k}\left ( \sum_{x=0}^k \binom{k}{x} (\beta b_l)^{x} \left ( (s_j+\lambda \beta+\alpha)^{k-x} -(s_j+\lambda \beta)^{k-x} \right) \right )  \;\;\;\;   \mbox{[Binomial theorem]}   \nonumber\\ 
&=& \sum_{l \in U_j}\frac{1}{\Delta_b^k} \left ( \sum_{x=0}^k \binom{k}{x} (\beta b_l)^{x} \left ( (s_j+\lambda \beta+\alpha)^{k-x} -(s_j+\lambda \beta)^{k-x} \right) \right )  \;\;\;\; \mbox{[$b_l \leq \Delta_b$ for all $l$]} \nonumber\\ 
&\geq&  \sum_{l \in U_j} \frac{1}{\Delta_b^k} \left ( (s_j+\lambda \beta+\alpha)^{k} -(s_j+\lambda \beta)^{k} + \beta b_l \left ((s_j+\lambda \beta+\alpha)^{k-1} -(s_j+\lambda \beta)^{k-1} \right)  \right )\nonumber\\
&=&|U_j| \frac{1}{\Delta_b^k} \left ( (s_j+\lambda \beta+\alpha)^{k} -(s_j+\lambda \beta)^{k}  \right ) + \frac{1}{\Delta_b^k} \beta k   \left ((s_j+\lambda \beta+\alpha)^{k-1} -(s_j+\lambda \beta)^{k-1} \right)\sum_{l \in U_j}b_l  \nonumber\\
&\geq& \frac{1}{\Delta_b^k} (3m - 3i+1)\left ( (s_j+\lambda \beta+\alpha)^{k} -(s_j+\lambda \beta)^{k}  \right ) +  \frac{1}{\Delta_b^k} \beta k\left ((s_j+\lambda \beta+\alpha)^{k-1} -(s_j+\lambda \beta)^{k-1} \right)\sum_{l \in U_j}b_l \nonumber\\ && \;\;\;\; \mbox{[Since $|U_j| \geq 3m - 3i+1$]}\nonumber\\
&\geq& \frac{1}{\Delta_b^k}(3m - 3i+1)\left ( (s_j+\lambda \beta+\alpha)^{k} -(s_j+\lambda \beta)^{k}  \right ) + \frac{1}{\Delta_b^k}\beta k(mB-iB) \left ((s_j+\lambda \beta+\alpha)^{k-1} -(s_j+\lambda \beta)^{k-1} \right) \nonumber \\ \nonumber&&\;\;\;\; \mbox{[Lemma \ref{lem:agelkst}]} \\
&\geq& f_1(j)+ \frac{1}{\Delta_b^k} (s_j+\lambda \beta+\alpha)^{k} -(s_j+\lambda \beta)^{k} \nonumber
\end{eqnarray}

Finally, the minimum cost any algorithm can incur due to Type 2 and 3 jobs is $f_{2,3}$ as stated in Fact \ref{fact:cost23lkst}.  Thus, the algorithm $A$ would have total cost at least $\sum_{i = 0}^{m-1}f_1(i) + f_{2,3} + \frac{1}{\Delta_b^k}(s_j+\lambda \beta+\alpha)^{k} -(s_j+\lambda \beta)^{k}> f $. This contradicts the assumption on the objective value of $A$'s schedule.

\end{proof}

\begin{lemma}
\label{lem:exactagelkst}
It must be the case that $\sum_{j \in U_i} b_j \leq B(m-i)$ for $i$ from $0$ to $m-1$ for any schedule $A$ with cost at most $f$.  
\end{lemma}
\begin{proof}
Clearly the lemma holds true to the interval $I_0$, since the lemma implies that in this case all jobs in $T_1$ could possibly be incomplete at the end of $I_0$.  For the sake of contradiction say that there is an interval $I_j$ such that for a schedule $A$ of cost at most $f$, $\sum_{l \in U_j} b_l > B(m-j)$.  Since $b_l$ is integral for all jobs $l \in T_1$, it is the case that $\sum_{l \in U_j} b_l \geq B(m-j)+1$.  Consider the increase in $A$'s objective during $I_j$ for Type 1 jobs.  This is the following, 

\begin{eqnarray}
&&\sum_{l \in U_j}\frac{1}{b_l^k} \left( (s_j+\lambda \beta+\alpha+\beta b_l)^k - (s_j+\lambda \beta+\beta b_l)^k\right) \nonumber\\
&=& \sum_{l \in U_j} \frac{1}{b_l^k} \left ( \sum_{x=0}^k \binom{k}{x} (\beta b_l)^{x} \left ( (s_j+\lambda \beta+\alpha)^{k-x} -(s_j+\lambda \beta)^{k-x} \right) \right )  \;\;\;\; \mbox{[Binomial theorem]}\nonumber\\ 
&\geq& \sum_{l \in U_j} \frac{1}{\Delta_b^k} \left ( \sum_{x=0}^k \binom{k}{x} (\beta b_l)^{x} \left ( (s_j+\lambda \beta+\alpha)^{k-x} -(s_j+\lambda \beta)^{k-x} \right) \right )  \;\;\;\;\mbox{[$b_l \leq \Delta_b$ for all $l$]}\nonumber\\ 
&\geq&  \sum_{l \in U_j} \frac{1}{\Delta_b^k}\left ( (s_j+\lambda \beta+\alpha)^{k} -(s_j+\lambda \beta)^{k} + \beta b_l \left ((s_j+\lambda \beta+\alpha)^{k-1} -(s_j+\lambda \beta)^{k-1} \right)  \right )\nonumber\\
&=&|U_j| \frac{1}{\Delta_b^k}\left ( (s_j+\lambda \beta+\alpha)^{k} -(s_j+\lambda \beta)^{k}  \right ) +  \frac{1}{\Delta_b^k}\beta k\left ((s_j+\lambda \beta+\alpha)^{k-1} -(s_j+\lambda \beta)^{k-1} \right)  \sum_{l \in U_j}  b_l\nonumber\\
&\geq& \frac{1}{\Delta_b^k}(3m - 3i)\left ( (s_j+\lambda \beta+\alpha)^{k} -(s_j+\lambda \beta)^{k}  \right ) +   \frac{1}{\Delta_b^k}\beta k \left ((s_j+\lambda \beta+\alpha)^{k-1} -(s_j+\lambda \beta)^{k-1} \right) \sum_{l \in U_j}  b_l \nonumber\\ && \;\;\;\; \mbox{[Lemma \ref{lem:numlkst}]}\nonumber\\
&\geq& \frac{1}{\Delta_b^k}(3m - 3i)\left ( (s_j+\lambda \beta+\alpha)^{k} -(s_j+\lambda \beta)^{k}  \right ) + \frac{1}{\Delta_b^k}\beta k(mB-jB+1) \left ((s_j+\lambda \beta+\alpha)^{k-1} -(s_j+\lambda \beta)^{k-1} \right) \nonumber \\ \nonumber&&\;\;\;\; \mbox{[$\sum_{l \in U_j} b_l \geq B(m-j)+1$]} \\
&\geq& f_1(j)+  \frac{1}{\Delta_b^k}\beta k \left ((s_j+\lambda \beta+\alpha)^{k-1} -(s_j+\lambda \beta)^{k-1} \right) \nonumber
\end{eqnarray}
Now we know that the increase in the algorithm's objective due to Type 1 jobs being unsatisfied during $I_i$ where $i\neq j$ is at least $f_1(i)$ by Lemma \ref{lem:numlkst}, Lemma \ref{lem:agelkst}, and Fact \ref{fact:costT1ilkst}.  Further, the minimum cost any schedule can incur for Type 2 and 3 jobs is $f_{2,3}$.  Thus, the cost of $A$'s schedule is at least $\sum_{i = 0}^{m-1}f_1(i) + f_{2,3} +  \frac{1}{\Delta_b^k}  \beta k \left ((s_j+\lambda \beta+\alpha)^{k-1} -(s_j+\lambda \beta)^{k-1} \right) > f$.  This  contradicts the  assumption that $A$ has cost at most $f$.

\end{proof}

Now we are finally ready to prove the theorem.

\begin{theorem}
If the schedule $A$ has cost at most $f$ then there exists a valid solution to the 3-partition instance.
\end{theorem}
\begin{proof}
By Lemma \ref{lem:threelkst} and Lemma \ref{lem:numlkst} we know that $|T_i \setminus U_i|= 3i$.  This implies that $3$ jobs are completed between the end of $I_i$ and the end of $I_{i+1}$ for $i$ from $0$ to $m-1$ and $3$ jobs are completed after $I_{m-1}$. Let $P_i$ be the three jobs completed between the end of $I_i$ and the end of $I_{i+1}$ and $P_m$ be the remaining $3$ jobs. Lemma \ref{lem:agelkst} and Lemma \ref{lem:exactagelkst} state that $\sum_{l \in T_i \setminus U_i} b_l = Bm-Bi$ for $i$ from $1$ to $m-1$.  This implies that $\sum_{j \in P_i} b_j = B$.  Thus, $P_i$ contains exactly $3$ jobs such that their total processing time is exactly $B$ for $i$ from $1$ to $m$.  Hence, $P_1, \ldots, P_m$ corresponds to a solution to the 3-partition problem.
\end{proof}

\subsection{$3$-partition of stretch $\rightarrow$ $k$-norm:}  Now we show that there if there is a valid solution to the $3$-partition then there is a solution to the $k$-norm with an objective at most $f$.   Consider a valid solution to the $3$-partition problem $P_1, P_2, \ldots, P_m$.  Consider the following solution to $k$-norm problem instance.  Each job of Type 2 and 3 is scheduled as soon as it arrives.  Now consider Type 1 jobs.  The three Type 1 jobs corresponding to integers in $P_i$ are scheduled during $[(i-1)(B+\alpha), iB+(i-1)\alpha)$.   Note that these jobs can be exactly scheduled during this interval.  Let this schedule be denoted as $\mathcal{A}$.

Now we bound the cost of Type 1 jobs by separating the intervals where the jobs could possibly be waiting. First we bound the cost due to Type 1 jobs waiting during closed time intervals.

\begin{lemma}
\label{lem:type1blacklkst}
For the schedule $\mathcal{A}$ the total cost Type 1 jobs accumulate during closed time intervals is $\sum_{l \in T_1} \frac{1}{\Delta_s^k}(b_l\beta+\lambda \beta)^k + \sum_{i =1}^{m-1}\frac{\Delta^k_b}{\Delta_s^k} f_1(i) + \frac{1}{\Delta^k_s}m2^{2k}(s_i+\lambda \beta)^{k-1}$.
\end{lemma}
\begin{proof}
First consider an interval $I_i$ where $1 \leq i \leq m-1$.  Let $U^{\cA}_i$ be the set of Type 1 jobs that are unsatisfied in the schedule $\cA$ during $I_i$.  The total accumulated cost during $I_i$ for these jobs can be bounded as follows. Note that by definition of $\cA$ we know that $|U^{\cA}_i| = 3m-3i$ and $\sum_{l \in U_i^{\cA}} b_l =  mB-iB$.

\begin{eqnarray*}
&&\sum_{l \in U^{\cA}_i} \frac{1}{b_l^k} \left( (s_i+\lambda \beta+\alpha+\beta b_l)^k - (s_i+\lambda \beta+\beta b_l)^k\right) \\
&=& \sum_{l \in U^{\cA}_i}  \frac{1}{b_l^k} \left ( \sum_{x=0}^k \binom{k}{x} (\beta b_l)^{x} \left ( (s_i+\lambda \beta+\alpha)^{k-x} -(s_i+\lambda \beta)^{k-x} \right) \right )  \;\;\;\; \mbox{[Binomial theorem]}\\ 
&=& \sum_{l \in U^{\cA}_i}  \frac{1}{\Delta^k_s} \left ( \sum_{x=0}^k \binom{k}{x} (\beta b_l)^{x} \left ( (s_i+\lambda \beta+\alpha)^{k-x} -(s_i+\lambda \beta)^{k-x} \right) \right )  \;\;\;\; \mbox{[$\Delta_s \leq b_l$ for all $l$]}\\ 
&=&  \sum_{l \in U^{\cA}_i}\frac{1}{\Delta^k_s} \left ( (s_i+\lambda \beta+\alpha)^{k} -(s_i+\lambda \beta)^{k} + k\beta b_l \left ((s_i+\lambda \beta+\alpha)^{k-1} -(s_i+\lambda \beta)^{k-1} \right)  \right )\\
&& \;\;\;\; + \sum_{l \in U^{\cA}_i}\frac{1}{\Delta^k_s}  \left ( \sum_{x=2}^k \binom{k}{x} (\beta b_l)^{x} \left ( (s_i+\lambda \beta+\alpha)^{k-x} -(s_i+\lambda \beta)^{k-x} \right) \right )\\
&=&\frac{1}{\Delta^k_s}  (3m - 3i)\left ( (s_i+\lambda \beta+\alpha)^{k} -(s_i+\lambda \beta)^{k}  \right ) +   \beta k  \left ((s_i+\lambda \beta+\alpha)^{k-1} -(s_i+\lambda \beta)^{k-1} \right)\sum_{l \in U^{\cA}_i} b_l\\ 
&& \;\;\;\; +\frac{1}{\Delta^k_s}  \sum_{l \in U^{\cA}_i}\left ( \sum_{x=2}^k \binom{k}{x} (\beta b_l)^{x} \left ( (s_i+\lambda \beta+\alpha)^{k-x} -(s_i+\lambda \beta)^{k-x} \right) \right )\;\;\;\; \mbox{[Since $|U^{\cA}_i| = 3m - 3i$]}\\
&=&\frac{1}{\Delta^k_s}  (3m - 3i)\left ( (s_i+\lambda \beta+\alpha)^{k} -(s_i+\lambda \beta)^{k}  \right ) + \beta k(mB-iB) \left ((s_i+\lambda \beta+\alpha)^{k-1} -(s_i+\lambda \beta)^{k-1} \right) \\
&& \;\;\;\; +\frac{1}{\Delta^k_s}  \sum_{l \in U^{\cA}_i}\left ( \sum_{x=2}^k \binom{k}{x} (\beta b_l)^{x} \left ( (s_i+\lambda \beta+\alpha)^{k-x} -(s_i+\lambda \beta)^{k-x} \right) \right )\;\;\;\; \mbox{[Since $\sum_{l \in U^{\cA}_i} b_l = mB-iB$]} \\
&=&\frac{\Delta^k_b}{\Delta^k_s}  f_1(i) +\frac{1}{\Delta^k_s}  \sum_{l \in U^{\cA}_i}\left ( \sum_{x=2}^k \binom{k}{x} (\beta b_l)^{x} \left ( (s_i+\lambda \beta+\alpha)^{k-x} -(s_i+\lambda \beta)^{k-x} \right) \right )\\
&=&\frac{\Delta^k_b}{\Delta^k_s}  f_1(i) + \frac{1}{\Delta^k_s} \sum_{l \in U^{\cA}_i}\left ( \sum_{x=2}^k \binom{k}{x} (\beta b_l)^{x} \sum_{y=1}^{k-x} \binom{k-x}{y}\alpha^y(s_i+\lambda \beta)^{k-x-y} \right )  \;\;\;\; \mbox{[Binomial thoerem]}\\
&\leq& \frac{\Delta^k_b}{\Delta^k_s} f_1(i) + \frac{1}{\Delta^k_s} \sum_{l \in U^{\cA}_i}\left ( \sum_{x=2}^k \binom{k}{x} (\beta b_l)^{x} 2^{k-x}\alpha(s_i+\lambda \beta)^{k-x-1} \right )  \;\;\;\; \mbox{[$\alpha < \lambda \beta$ and  $\sum_{x=2}^k \binom{k}{x} < 2^{k-x}$]}\\
&\leq& \frac{\Delta^k_b}{\Delta^k_s} f_1(i) +\frac{1}{\Delta^k_s}  \sum_{l \in U^{\cA}_i}\left ( 2^k (\beta b_l)^2 2^{k-2}\alpha(s_i+\lambda \beta)^{k-3} \right )  \;\;\;\; \mbox{[$\beta b_l < \lambda \beta$  for all $l$ and $\sum_{x=2}^k \binom{k}{x}  < 2^k$ ]}\\
&\leq& \frac{\Delta^k_b}{\Delta^k_s} f_1(i) + \frac{1}{\Delta^k_s} \sum_{l \in U^{\cA}_i} 2^{2k-2}(s_i+\lambda \beta)^{k-1}  \;\;\;\; \mbox{[$\beta^2 B^2 \alpha< (\lambda \beta)^2$ and $b_l < B$ for all $l$]}\\
&\leq& \frac{\Delta^k_b}{\Delta^k_s} f_1(i) + \frac{1}{\Delta^k_s} m2^{2k}(s_i+\lambda \beta)^{k-1} \;\;\;\; \mbox{[$|U^{\cA}_i| \leq 3m$]}
\end{eqnarray*}

Now consider the interval $I_0$.  No Type 1 job is worked on during $I_0$ and all Type 1 jobs arrive on $I_0$.  Knowing this, the total increase in the objective due to Type 1 jobs waiting during $I_i$ is $\sum_{l \in T_1} \frac{1}{b_l^k} (b_l\beta+\lambda \beta)^k \leq \sum_{l \in T_1} \frac{1}{\Delta_s^k} (b_l\beta+\lambda \beta)^k$.  This completes the proof of the lemma.
\end{proof}

Next we bound the bound the cost Type 1 jobs accumulate during open time intervals.

\begin{lemma}
\label{lem:type1nonblacklkst}
For the schedule $\cA$ the total cost Type 1 jobs accumulate during open times is $\frac{1}{\Delta_s^k}3m^22^k B(\beta B + \lambda \beta + (m-1) (\alpha+B))^{k-1}$.
\end{lemma}
\begin{proof}
Fix any Type 1 jobs $l$. The most a Type 1 job will accumulate during a maximal contiguous open time interval is during the $B$ time steps after all closed time intervals end.  This because after this time all Type 1 jobs are completed, the objective function is convex and every maximal contiguous open time interval has length $B$.  Knowing that the arrival time of any Type 1 job $l$ is after time $(\beta B +\lambda \beta)$ and $b_l \geq \Delta_s$, the most job $l$ can accumulate in a single maximal contiguous open time interval is the following.

\begin{eqnarray*}
& &\frac{1}{\Delta_s^k} (\beta B + \lambda \beta + \alpha(m-1) +mB)^k - \frac{1}{\Delta_s^k}(\beta B + \lambda \beta + (m-1)(\alpha+B))^k\\
&=&\frac{1}{\Delta_s^k} \sum_{x=1}^{k} \binom{k}{x} B^x(\beta B + \lambda \beta + (m-1) (\alpha+B))^{k-x} \;\;\;\; \mbox{[Binomial theorem]}\\
&\leq&\frac{1}{\Delta_s^k} 2^k B(\beta B + \lambda \beta + (m-1) (\alpha+B))^{k-1} \;\;\;\; \mbox{[$B < \beta$ and $ \sum_{x=1}^{k} \binom{k}{x}  < 2^k$]}
\end{eqnarray*}

Knowing that there are $3m$ jobs and $m$ maximal non-contiguous open time intervals, the total increase in $\cA$'s objective due to the cost Type 1 jobs accumulate during these time intervals is at most $\frac{1}{\Delta_s^k}3m^22^k B(\beta B + \lambda \beta + (m-1) (\alpha+B))^{k-1}$.
\end{proof}

Now we are ready to bound the cost of the schedule $\cA$.  Using Fact \ref{fact:cost23lkst} the total cost for Type 2 and 3 jobs in $\cA$ is $f_{2,3}$. This and lemmas \ref{lem:type1blacklkst} and \ref{lem:type1nonblacklkst} implies that the total cost of $\cA$ is less than $f_{2,3} + \sum_{l \in T_1} \frac{1}{\Delta_s^k} (b_l\beta+\lambda \beta)^k + \sum_{i =1}^{m-1} (\frac{\Delta_b^k}{\Delta_s^k}f_1(i) + \frac{1}{\Delta_s^k}m2^{2k}(s_i+\lambda \beta)^{k-1} )  +\frac{1}{\Delta_s^k}3m^2k^k B(\beta B + \lambda \beta + (m-1) (\alpha+B))^{k-1} =f$.

\begin{theorem}
If there exists a valid solution to the 3-Partition instance then there is a schedule $\mathcal{A}$ for the $k$-norm problem of stretch instance with an objective value at most $f$.
\end{theorem}

\section{Hardness Proof for the $k$-norm of stretch when $0 < k < 1$}
\label{sec:halfnormstretch}
In this section, we show the hardness of the $k$-norm objective for positive $k < 1$.  We reduce the $3$-partition problem to this scheduling problem.   We will assume without loss of generality that in the $3$-partition instance $B/3 - \eps \leq b_i \leq B/3 + \eps$ for all $i \in [m]$ and $\eps \leq 1/(2mB)^{20/k^3}$.  We let $\Delta_s = B/3 -\eps$ and $\Delta_n = B/3 + \eps$.  Consider any fixed instance of the $3$-partition problem. We now construct the following instance of the $k$-norm problem.

\begin{itemize}
\item \textbf{Type 1 jobs:} For each integer $b_i \in S$ from the $3$-partition instance, we create a job of processing time $b_i$ and arrival time $-\beta+\lambda b_i$. The value of $\beta$ is set to be $(30mkB)^{5/k^2 + 2}$ and $\lambda$ is set to $\beta^{1/4}$.  Let $T_1$ denote the set of Type $1$ jobs. These jobs will be indexed for $i \in [3m]$.  Note that because $\lambda B < \beta$ and $b_i \leq B$ for all $i$, it is the case that all jobs arrive before time $0$.
\item \textbf{Type 2 jobs:}  There is a job of size $\rho$ is released every $\rho$ time steps during the intervals $[iB+ (i-1)\alpha, i(B+\alpha))$ for $i$ from $1$ to $m-1$.  Here $\rho$ and $\alpha$ are set such that $\rho = 1/(100m^4\beta)$ and $\alpha = 10\beta^{3/4}m^2B^2$. We will assume without loss of generality that $\alpha/\rho$ and $\alpha/B$ are integral. 
\item \textbf{Type 3 jobs:}  During the interval $[-\beta ,0]$ a job of size $\rho$ is released every $\rho$ time steps. We will assume that without loss of generally that $\beta/\rho$ is integral.
\end{itemize}

This is the entire input instance.  We will let $\mathcal{I}^3$ denote the 3-Partition instance and $\mathcal{I}^\ell$ be the $k$ norm instance. Let $I_0 = [\beta,0)$, $s_i = iB+ (i-1)\alpha$ and $I_i = [s_i, s_i + \alpha)$ for $i \in [m-1]$. The Type 1 jobs will be used to represent integers in the $3$-partition instance.   The Type 2 jobs will be used to ensure that an optimal schedule cannot process Type 1 during $I_i$ if there exists a valid solution to the $3$-partition instance for $i \in [m-1]$.  The Type 3 jobs are used to ensure no Type 1 job can be processed during $I_0$ in an optimal schedule. Note that during an interval $I_i$ the Type 2 or 3 jobs that arrive can be completely scheduled during this interval and require the entire interval length to be completed for all $i$.  These are the intervals which will be \emph{closed} in an optimal schedule. We call a time $t$ closed if $t \in I_i$ for $i$ from $0$ to $m-1$.  

Our goal will be to show that for a given instance of the $3$-Partition problem, there exists a schedule for the instance of the scheduling problem with an objective value of at most $f :=  (\beta+(m-1)\alpha)/\rho + \frac{3m^2kB}{\Delta^k_s( \beta -  \lambda B  )^{1-k}} + \sum_{l \in T_1}\frac{1}{\Delta^k_s} (\beta - \lambda b_l)^k + (3m-3i)\frac{1}{\Delta^k_s}\Bigg ( \frac{k\alpha}{(s_i + \beta - \beta^{k/2} )^{1-k}} - \frac{k(1-k)(\beta^{k/2})^2}{2(s_i+\beta - \beta^{k/2})^{2-k}} \Bigg) + \frac{1}{\Delta^k_s}\frac{k(1-k)(\beta^{k/2})\lambda (mB-iB)}{(s_i+\beta - \beta^{k/2})^{2-k}}  $ if and only if there exists a valid solution to the $3$-Partition problem.   

\subsection{$k$-norm stretch $\rightarrow$ $3$-partition:} We now show that if there is a solution to the instance $\mathcal{I}^\ell$ with an objective at most $f$ then there exists a valid solution to $\mathcal{I}^3$.  To show that this is the case, assume that there exists some fixed schedule $A$ of cost at most $f$.  Before we start with the proof, we state a few facts that will be useful throughout the proof.

\begin{fact}
\label{fact:costT1ismstlk}
Let $U_i$ be the set of unsatisfied Type 1 jobs at the end of interval $I_i$ in $A$.  If $|U_i| \geq 3m - 3i$ and $\sum_{l \in U_i} b_l \geq mB-iB$ then the total increase in $A$'s objective during $I_i$ due to the jobs in $U_i$ being unsatisfied during $I_i$ is at least $f_1(i) := (3m-3i)\frac{1}{\Delta_b^k} \bigg ( \frac{k\alpha}{(s_i + \beta - \beta^{k/2} )^{1-k}} - \frac{k(1-k)(\beta^{k/2}+\alpha)^2}{2(s_i+\beta - \beta^{k/2})^{2-k}} \bigg) + \frac{1}{\Delta_b^k}\frac{k(1-k)(\beta^{k/2}+\alpha)\lambda (mB-iB)}{(s_i+\beta- \beta^{k/2})^{2-k}}  -  \frac{1}{\Delta_b^k}m\frac{k(1-k)(\lambda B)^2}{2(s_i+ \beta - \beta^{k/2})^{2-k}} $ for $i$ from $1$ to $m-1$.  Similarly, if no Type 1 jobs are completed by time $0$ then the increase $A$'s objective for Type 1 jobs during $I_0$ is at least $f_1(0) := \sum_{i \in T_1} \frac{1}{\Delta_b^k} (\beta-\lambda b_i)^k$.  
\end{fact}
\begin{proof}
The total increase in a schedule's objective due to Type 1 jobs being unsatisfied during $I_i$ is the following if $|U_i| \geq 3m - 3i$ and $\sum_{l \in U_i} b_l \geq mB-iB$.

\begin{eqnarray*}
&&\sum_{l \in U_i} \frac{1}{b_l^k} \left( (s_i+\beta+\alpha-\lambda b_l)^k - (s_i+\beta-\lambda b_l)^k\right) \\
&\geq& \sum_{l \in U_i} \frac{1}{b_l^k}\Bigg( \Bigg ( (s_i+ \beta - \beta^{k/2})^k+ \frac{k(\beta^{k/2}+\alpha-\lambda b_l)}{(s_i + \beta - \beta^{k/2} )^{1-k}} - \frac{k(1-k)(\beta^{k/2}+\alpha-\lambda b_l)^2}{2(s_i+\beta - \beta^{k/2})^{2-k}} \Bigg)\\
&&\;\;\;\; - \Bigg ( (s_i+ \beta - \beta^{k/2})^k+ \frac{k(\beta^{k/2}-\lambda b_l)}{(s_i + \beta - \beta^{k/2} )^{1-k}} \Bigg)\Bigg )   \;\;\;\; \mbox{[Fact \ref{fact:taylor}, $b_l \leq B$ for all $l$ and $2(\beta^{k/2}+\alpha) \leq \beta - \beta^{k/2}$]}\\
&=& \sum_{l \in U_i} \frac{1}{b_l^k} \Bigg ( \frac{k\alpha}{(s_i + \beta - \beta^{k/2} )^{1-k}} - \frac{k(1-k)(\beta^{k/2}+\alpha-\lambda b_l)^2}{2(s_i+\beta - \beta^{k/2})^{2-k}} \Bigg)\\
&=& \sum_{l \in U_i} \frac{1}{\Delta_b^k} \Bigg ( \frac{k\alpha}{(s_i + \beta - \beta^{k/2} )^{1-k}} - \frac{k(1-k)(\beta^{k/2}+\alpha-\lambda b_l)^2}{2(s_i+\beta - \beta^{k/2})^{2-k}} \Bigg) \;\;\;\; \mbox{[$b_l \leq \frac{1}{\Delta_b^k}$ for all $l$]}\\
&=& |U_i|\frac{1}{\Delta_b^k} \Bigg ( \frac{k\alpha}{(s_i + \beta - \beta^{k/2} )^{1-k}} - \frac{k(1-k)(\beta^{k/2}+\alpha)^2}{2(s_i+\beta - \beta^{k/2})^{2-k}} \Bigg) + \sum_{l \in U_i} \frac{1}{\Delta_b^k}\frac{k(1-k)(2((\beta^{k/2}+\alpha)\lambda b_l)-(\lambda b_l)^2)}{2(s_i+\beta - \beta^{k/2})^{2-k}} \\
&\geq& (3m-3i)\frac{1}{\Delta_b^k}\Bigg ( \frac{k\alpha}{(s_i + \beta - \beta^{k/2} )^{1-k}} - \frac{k(1-k)(\beta^{k/2}+\alpha)^2}{2(s_i+\beta - \beta^{k/2})^{2-k}} \Bigg) +\frac{1}{\Delta_b^k} \frac{k(1-k)(\beta^{k/2}+\alpha)\lambda (mB-iB)}{(s_i+\beta - \beta^{k/2})^{2-k}} \\
&& -\sum_{l \in U_i} \frac{1}{\Delta_b^k}\frac{k(1-k)(\lambda B)^2}{2(s_i+\beta - \beta^{k/2})^{2-k}} \;\;\;\; \mbox{[$|U_i| = mB-iB$, $\sum_{l \in U_i} b_l = mB-iB$ and $b_l < B$ for all $l$]}\\
&\geq& (3m-3i)\frac{1}{\Delta_b^k}\Bigg ( \frac{k\alpha}{(s_i + \beta - \beta^{k/2} )^{1-k}} - \frac{k(1-k)(\beta^{k/2}+\alpha)^2}{2(s_i+\beta - \beta^{k/2})^{2-k}} \Bigg) + \frac{1}{\Delta_b^k}\frac{k(1-k)(\beta^{k/2}+\alpha)\lambda (mB-iB)}{(s_i+\beta - \beta^{k/2})^{2-k}} \\
&& - \frac{1}{\Delta_b^k}m\frac{k(1-k)(\lambda B)^2}{2(s_i+\beta - \beta^{k/2})^{2-k}} \;\;\;\; \mbox{[$|U_i| \leq m$]}
\end{eqnarray*}

Now we focus on bounding the increase in a schedule's objective if no Type 1 jobs are completed by time $0$. During the interval $I_0$ a job $i \in T_1$ waits $\beta - \lambda b_i$ time steps and its contribution to the objective during this interval is $(\beta-\lambda b_i)^k$.  Thus the total cost is $\sum_{i \in T_1} \frac{1}{b_i^k} (\beta-\lambda b_i)^k \geq\sum_{i \in T_1} \frac{1}{\Delta_b^k} (\beta-\lambda b_i)^k  $.
\end{proof}

\begin{fact}
\label{fact:cost23smstlk}
If the algorithm completes all Type 2 and 3 jobs as soon as they arrive in first-in-first-out order then the total contribution of these jobs to the objective is $f_{2,3} := (\beta+(m-1)\alpha)/\rho$ when $0< k <1$.  This is the minimum cost any schedule can incur for Type 2 and 3 jobs.
\end{fact}
\begin{proof}
If the algorithm completes the Type 2 and 3 jobs in this way, then each of these jobs waits $\rho$ time steps to be completed. Hence, a single Type 2 or 3 job's contribution to the objective is $1$.  There are $\beta/\rho+(m-1)\alpha/\rho$ Type 2 and 3 jobs total.  Thus, their total contribution is $(\beta+(m-1)\alpha)/\rho$ when $0 < k < 1$.
\end{proof}

Now we show that during blacked our intervals, only a small volume of Type 1 jobs can be processed.  Intuitively,  this lemma ensures that Type 1 jobs can only be processed during open times.

\begin{lemma}
\label{lem:blacksmstlk}
If a schedule $A$ does a $1/2$ volume of work on Type 1 jobs during $\bigcup_{i=0}^{m-1} I_i$ then the total cost of the schedule $A$ is larger than $f$. Similarly, if a schedule $A$ does not process a $1/2$ volume of work of the Type 2 or 3 jobs that arrive on an interval $I_i$ during $I_i$ then the cost of $A$'s schedule is larger than $f$ for any $i \in \{0,1,\ldots, m-1\}$.
\end{lemma}
\begin{proof}
We begin by proving the first part of the lemma.  Assume that $A$ processes a $1/2$ volume of work on Type 1 jobs during the discontinuous time interval $\bigcup_{i=0}^{m-1} I_i$.  It follows that the algorithm processes at least $1/(2m)$ volume of work of Type 1 jobs during at least one interval $I_i$.  Fix such an interval $I$.  During $I$ a job of size $\rho$ arrive every $\rho$ time steps. This implies that at least $\frac{1}{2m\rho}$ jobs have flow time at least $\frac{1}{2m}$. Thus the total cost of the schedule is at least $\frac{1}{2m\rho^{1+k}}\cdot (\frac{1}{2m})^k \geq 25m^2\beta$ which is larger than $f$ for sufficiently large $m$.

Now we prove the second part of the lemma.  Say that there is a interval $I \in \{I_0, I_1, \ldots, I_{m-1} \}$ where a $1/2$ volume of work of Type 2 or 3 jobs that arrived during $I$ are not completed during $I$ in $A$'s schedule.  Then at least $1/(4\rho)$ jobs wait at least $1/(4\rho)$ time steps to be completed. Thus the cost of the schedule is at least $1/(4^{1+k}\rho^{1+2k})$.  Again, this is larger than $f$ for sufficiently large $B$ and $m$.
\end{proof}

Next we observe that the total number of Type 1 that the algorithm can complete before the end of a closed interval is proportional to the number of closed intervals that have occurred so far.

\begin{lemma}
\label{lem:numsmstlk}
The total number of Type 1 jobs that can be completed by the end of $I_i$ is $3i$ for $i$ from $0$ to $m-1$ for any schedule $A$ with cost at most $f$.
\end{lemma}

\begin{proof}
By the end of an interval $I_i$ there are have been at exactly $iB$ time steps that are not contained in a closed interval before the end of $I_i$.  From Lemma~\ref{lem:blacksmstlk}  at most a $1/2$ volume of work can be processed by $A$ of Type 1 jobs during closed time steps.  Knowing that the smallest Type 1 job has size $\frac{m}{3m+1/2}B$ and $B\geq3$, the total number of jobs that can be completed before the end of $I_i$ is $\floor{(iB+1/2)\Big/(\frac{m}{3m+1/2}B)} \leq 3i$. 
\end{proof}

Notice that the instantaneous increase in the objective function at a time $t$ depends on the \emph{ages} of the jobs.  Now, for a time $t\geq 0$, the age of a Type 1 job $i$ is $t+\beta- \lambda b_i$.  Our goal now is to define a bound on the total age of the Type 1 jobs during any closed interval in the schedule $A$.  Since the last lemma bounded the total number of Type 1 jobs that are unsatisfied during a closed interval, this essentially amounts to bounding the total original processing time of the unsatisfied Type 1 jobs.

\begin{lemma}
\label{lem:agesmstlk}
For any time $t \in I_i$, it is the case that $\sum_{j\in U_i} b_j \geq B(m-i)$ for $i$ from $0$ to $m-1$ for any schedule $A$ with cost at most $f$.
\end{lemma}
\begin{proof}
By the end of $I_i$ there are have been at exactly $iB$ time steps that are not contained in a closed interval before the end of $I_i$.  From Lemma~\ref{lem:blacksmstlk}  at most a $1/2$ volume of work on Type 1 jobs can be processed by $A$ during closed time steps. We know that the processing time of a job $i\in T_1$ is $b_i$ and therefore $\sum_{i \in T_1} b_i = mB$.  The total processing time of jobs in $T_1\setminus U_i$ can be at most $iB +1/2$ since these jobs are completed by the end of interval $I_i$.  Knowing that $i$ and $B$ are integral as well as $b_i$ for each $i \in T_1$, we know that $\sum_{j \in T_1 \setminus U_i} b_j \leq iB$.  Thus, $ \sum_{j\in U_i} b_j = mB - \sum_{j \in T_i \setminus U_i} b_j \geq mB - iB$.
\end{proof}

Next we show that by the end of every closed time interval $I_i$ for $i$ from $1$ to $m-1$ there must be at least $3i$ jobs completed.

\begin{lemma}
\label{lem:threesmstlk}
By the end of interval $I_i$ there must be at least $3i$ Type 1 jobs complete in any schedule $A$ that has cost at most $f$.
\end{lemma}
\begin{proof}
For the sake of contradiction, say that there is an interval $I_j$ such that $3j-1$ or less Type 1 jobs are completed by the end of $I_j$ in $A$.  We begin by bounding the increase in $A$'s objective for $T_1$ jobs during closed time intervals. Recall that $s_i$ is the start time of interval $I_i$. For any interval $I_i$ where $i \neq j$, the total increase in $A$'s objective due to Type 1 jobs being unsatisfied during $I_i$ is at least $f_1(i)  $ by Lemma \ref{lem:numsmstlk}, Lemma \ref{lem:agesmstlk}, and Fact \ref{fact:costT1ismstlk}. Similarly, the increase in $A$'s objective during $I_j$ due to jobs in $U_j$ being unsatisfied is,

\begin{eqnarray*}
&&\sum_{l \in U_j} \frac{1}{b_l^k} \left( (s_j+\beta+\alpha-\lambda b_l)^k - (s_j+\beta-\lambda b_l)^k\right) \\
&\geq& \sum_{l \in U_j} \frac{1}{b_l^k}\Bigg( \Bigg ( (s_j+ \beta - \beta^{k/2})^k+ \frac{k(\beta^{k/2}+\alpha-\lambda b_l)}{(s_j + \beta - \beta^{k/2} )^{1-k}} - \frac{k(1-k)(\beta^{k/2}+\alpha-\lambda b_l)^2}{2(s_j+\beta - \beta^{k/2})^{2-k}} \Bigg)\\
&&\;\;\;\; - \Bigg ( (s_j+ \beta - \beta^{k/2})^k+ \frac{k(\beta^{k/2}-\lambda b_l)}{(s_j + \beta - \beta^{k/2} )^{1-k}} \Bigg)\Bigg )   \;\;\;\; \mbox{[Fact \ref{fact:taylor}, $b_l \leq B$ for all $l$ and $2(\beta^{k/2}+ \alpha) \leq \beta - \beta^{k/2}$]}\\
&=& \sum_{l \in U_j}  \frac{1}{b_l^k}\Bigg ( \frac{k\alpha}{(s_j + \beta - \beta^{k/2} )^{1-k}} - \frac{k(1-k)(\beta^{k/2}+\alpha-\lambda b_l)^2}{2(s_j+\beta - \beta^{k/2})^{2-k}} \Bigg)\\
&=& \sum_{l \in U_j}  \frac{1}{\Delta_b^k}\Bigg ( \frac{k\alpha}{(s_j + \beta - \beta^{k/2} )^{1-k}} - \frac{k(1-k)(\beta^{k/2}+\alpha-\lambda b_l)^2}{2(s_j+\beta - \beta^{k/2})^{2-k}} \Bigg) \;\;\;\; \mbox{[Since $\Delta_b \geq b_l$ for all $l$]}\\
&=& |U_j| \frac{1}{\Delta_b^k} \Bigg ( \frac{k\alpha}{(s_j + \beta - \beta^{k/2} )^{1-k}} - \frac{k(1-k)(\beta^{k/2}+\alpha)^2}{2(s_j+\beta - \beta^{k/2})^{2-k}} \Bigg) + \sum_{l \in U_j}  \frac{1}{\Delta_b^k}\frac{k(1-k)(2((\beta^{k/2}+\alpha)\lambda b_l)-(\lambda b_l)^2)}{2(s_j+\beta - \beta^{k/2})^{2-k}} \\
&\geq& (3m-3j+1) \frac{1}{\Delta_b^k}\Bigg ( \frac{k\alpha}{(s_j + \beta - \beta^{k/2} )^{1-k}} - \frac{k(1-k)(\beta^{k/2}+\alpha)^2}{2(s_j+\beta - \beta^{k/2})^{2-k}} \Bigg) +  \frac{1}{\Delta_b^k}\frac{k(1-k)(\beta^{k/2}+\alpha)\lambda (mB-iB)}{(s_j+\beta - \beta^{k/2})^{2-k}} \\
&& -\sum_{l \in U_j}  \frac{1}{\Delta_b^k}\frac{k(1-k)(\lambda B)^2}{2(s_j+\beta - \beta^{k/2})^{2-k}} \;\;\;\; \mbox{[Lemma \ref{lem:agesmstlk}, the definition of $I_j$ and $b_l < B$ for all $l$ ]}\\
&\geq& (3m-3j+1) \frac{1}{\Delta_b^k}\Bigg ( \frac{k\alpha}{(s_j + \beta - \beta^{k/2} )^{1-k}} - \frac{k(1-k)(\beta^{k/2}+\alpha)^2}{2(s_j+\beta - \beta^{k/2})^{2-k}} \Bigg) +  \frac{1}{\Delta_b^k}\frac{k(1-k)(\beta^{k/2}+\alpha)\lambda (mB-iB)}{(s_j+\beta - \beta^{k/2})^{2-k}} \\
&& -m \frac{1}{\Delta_b^k}\frac{k(1-k)(\lambda B)^2}{2(s_j+\beta - \beta^{k/2})^{2-k}} \;\;\;\; \mbox{[$|U_j| \leq m$]}
\end{eqnarray*}

Finally, the minimum cost any algorithm can incur due to Type 2 and 3 jobs is $f_{2,3}$ as stated in Fact \ref{fact:cost23smstlk}.  Thus, the algorithm $A$ would have total cost at least $\sum_{i = 0}^{m-1}f_1(i) + f_{2,3} + ( \frac{k\alpha}{(s_j + \beta - \beta^{k/2} )^{1-k}} -  \frac{1}{\Delta_b^k}\frac{k(1-k)(\beta^{k/2}+\alpha)^2}{2(s_j+\beta - \beta^{k/2})^{2-k}} ) $.  Notice that $f = \sum_{i = 0}^{m-1}f_1(i) + f_{2,3} + \frac{3m^2kB}{( \beta - \beta \lambda B  )^{1-k}} - \sum_{i  =1}^{m-1} m\frac{k(1-k)(\lambda B)^2}{2(s_i+\beta - \beta^{k/2})^{2-k}}$.  For sufficiently large $m$ and $B$ this is larger than $f$ contradicts the assumption on the objective value of $A$'s schedule.

\end{proof}

\begin{lemma}
\label{lem:exactagesmstlk}
It must be the case that $\sum_{j \in U_i} b_j \leq B(m-i)$ for $i$ from $0$ to $m-1$ for any schedule $A$ with cost at most $f$.  
\end{lemma}
\begin{proof}
Clearly the lemma holds true to the interval $I_0$, since the lemma implies that in this case all jobs in $T_1$ could possibly be incomplete at the end of $I_0$.  For the sake of contradiction say that there is an interval $I_j$ such that for a schedule $A$ of cost at most $f$, $\sum_{l \in U_j} b_l > B(m-j)$.  Since $b_l$ is integral for all jobs $l \in T_1$, it is the case that $\sum_{l \in U_j} b_l \geq B(m-j)+1$.  Consider the increase in $A$'s objective during $I_j$ for Type 1 jobs.  This is the following, 

\begin{eqnarray*}
&&\sum_{l \in U_j} \frac{1}{b_l^k}\left( (s_j+\beta+\alpha-\lambda b_l)^k - (s_j+\beta-\lambda b_l)^k\right) \\
&\geq& \sum_{l \in U_j} \frac{1}{b_l^k} \Bigg( \Bigg ( (s_j+ \beta - \beta^{k/2})^k+ \frac{k(\beta^{k/2}+\alpha-\lambda b_l)}{(s_j + \beta - \beta^{k/2} )^{1-k}} - \frac{k(1-k)(\beta^{k/2}+\alpha-\lambda b_l)^2}{2(s_j+\beta - \beta^{k/2})^{2-k}} \Bigg)\\
&&\;\;\;\; - \Bigg ( (s_j+ \beta - \beta^{k/2})^k+ \frac{k(\beta^{k/2}-\lambda b_l)}{(s_j + \beta - \beta^{k/2} )^{1-k}} \Bigg)\Bigg )   \;\;\;\; \mbox{[Fact \ref{fact:taylor}, $b_l \leq B$ for all $l$ and $2(\beta^{k/2} + \alpha) \leq \beta - \beta^{k/2}$]}\\
&=& \sum_{l \in U_j} \frac{1}{b_l^k}\Bigg ( \frac{k\alpha}{(s_j + \beta - \beta^{k/2} )^{1-k}} - \frac{k(1-k)(\beta^{k/2}+\alpha-\lambda b_l)^2}{2(s_j+\beta - \beta^{k/2})^{2-k}} \Bigg)\\
&=& \sum_{l \in U_j} \frac{1}{\Delta_b^k}\Bigg ( \frac{k\alpha}{(s_j + \beta - \beta^{k/2} )^{1-k}} - \frac{k(1-k)(\beta^{k/2}+\alpha-\lambda b_l)^2}{2(s_j+\beta - \beta^{k/2})^{2-k}} \Bigg) \;\;\;\; \mbox{[Since $b_l \leq \Delta^b$ for all $l$]}\\
&=& |U_j| \frac{1}{\Delta_b^k} \Bigg ( \frac{k\alpha}{(s_j + \beta - \beta^{k/2} )^{1-k}} - \frac{k(1-k)(\beta^{k/2}+\alpha)^2}{2(s_j+\beta - \beta^{k/2})^{2-k}} \Bigg) + \sum_{l \in U_j} \frac{1}{\Delta_b^k} \frac{k(1-k)(2((\beta^{k/2}+\alpha)\lambda b_l)-(\lambda b_l)^2)}{2(s_j+\beta - \beta^{k/2})^{2-k}} \\
&\geq& (3m-3j) \frac{1}{\Delta_b^k} \Bigg ( \frac{k\alpha}{(s_j + \beta - \beta^{k/2} )^{1-k}} - \frac{k(1-k)(\beta^{k/2}+\alpha)^2}{2(s_j+\beta - \beta^{k/2})^{2-k}} \Bigg) +  \frac{1}{\Delta_b^k}\frac{k(1-k)(\beta^{k/2}+\alpha)\lambda (mB-iB+1)}{(s_j+\beta - \beta^{k/2})^{2-k}} \\
&& -\sum_{l \in U_j} \frac{1}{\Delta_b^k} \frac{k(1-k)(\lambda B)^2}{2(s_j+\beta - \beta^{k/2})^{2-k}} \;\;\;\; \mbox{[Lemma \ref{lem:numsmstlk}, the definition of $I_j$ and $b_l < B$ for all $l$ ]}\\
&\geq& (3m-3j) \frac{1}{\Delta_b^k}\Bigg ( \frac{k\alpha}{(s_j + \beta - \beta^{k/2} )^{1-k}} - \frac{k(1-k)(\beta^{k/2}+\alpha)^2}{2(s_j+\beta - \beta^{k/2})^{2-k}} \Bigg) + \frac{1}{\Delta_b^k} \frac{k(1-k)(\beta^{k/2}+\alpha)\lambda (mB-iB+1)}{(s_j+\beta - \beta^{k/2})^{2-k}} \\
&& - m  \frac{1}{\Delta_b^k}\frac{k(1-k)(\lambda B)^2}{2(s_j+\beta - \beta^{k/2})^{2-k}} \;\;\;\; \mbox{[$|U_j| \leq m$]}
\end{eqnarray*}

Now we know that the increase in the algorithm's objective due to Type 1 jobs being unsatisfied during $I_i$ where $i\neq j$ is at least $f_1(i)$ by Lemma \ref{lem:numsmstlk}, Lemma \ref{lem:agesmstlk}, and Fact \ref{fact:costT1ismstlk}.  Further, the minimum cost any schedule can incur for Type 2 and 3 jobs is $f_{2,3}$.  Thus, the cost of $A$'s schedule is at least $\sum_{i = 0}^{m-1}f_1(i) + f_{2,3} +  \frac{1}{\Delta_b^k}\frac{k(1-k)(\beta^{k/2}+\alpha)\lambda}{(s_j+\beta - \beta^{k/2})^{2-k}}$. This is larger than $f$, and therefore we have a contradiction on the objective value of $A$'s schedule.
\end{proof}

Now we are finally ready to prove the theorem.

\begin{theorem}
If the schedule $A$ has cost at most $f$ then there exists a valid solution to the 3-partition instance.
\end{theorem}
\begin{proof}
By Lemma \ref{lem:threesmstlk} and Lemma \ref{lem:numsm} we know that $|T_i \setminus U_i|= 3i$.  This implies that $3$ jobs are completed between the end of $I_i$ and the end of $I_{i+1}$ for $i$ from $0$ to $m-1$ and $3$ jobs are completed after $I_{m-1}$. Let $P_i$ be the three jobs completed between the end of $I_i$ and the end of $I_{i+1}$ and $P_m$ be the remaining $3$ jobs. Lemma \ref{lem:agesmstlk} and Lemma \ref{lem:exactagesmstlk} state that $\sum_{l \in T_i \setminus U_i} b_l = Bm-Bi$ for $i$ from $1$ to $m-1$.  This implies that $\sum_{j \in P_i} b_j = B$.  Thus, $P_i$ contains exactly $3$ jobs such that their total processing time is exactly $B$ for $i$ from $1$ to $m$.  Hence, $P_1, \ldots, P_m$ corresponds to a solution to the 3-partition problem.
\end{proof}

\subsection{$3$-partition $\rightarrow$  $k$-norm of stretch:}  Now we show that there if there is a valid solution to the $3$-partition then there is a solution to the $k$-norm with an objective at most $f$.   Consider a valid solution to the $3$-partition problem $P_1, P_2, \ldots, P_m$.  Consider the following solution to $k$-norm problem instance.  Each job of Type 2 and 3 is scheduled as soon as it arrives. Now consider Type 1 jobs.  The three Type 1 jobs corresponding to integers in $P_i$ are scheduled during $[(i-1)(B+\alpha), iB+(i-1)\alpha)$.   Note that these jobs can be exactly scheduled during this interval.  Let $\mathcal{A}$ denote this schedule.

Now we bound the cost of Type 1 jobs in the schedule $\mathcal{A}$ by separating the intervals where the jobs could possibly be waiting.

\begin{lemma}
\label{lem:type1blacksmstlk}
For the schedule $\mathcal{A}$ the total cost Type 1 jobs accumulate during closed times is less than $\sum_{l \in T_1}  \frac{1}{ \Delta_s^k} (\beta - \lambda b_l)^k + (3m-3i) \frac{1}{ \Delta_s^k} \Bigg ( \frac{k\alpha}{(s_i + \beta - \beta^{k/2} )^{1-k}} - \frac{k(1-k)(\beta^{k/2})^2}{2(s_i+\beta - \beta^{k/2})^{2-k}} \Bigg) +  \frac{1}{ \Delta_s^k} \frac{k(1-k)(\beta^{k/2})\lambda (mB-iB)}{(s_i+\beta - \beta^{k/2})^{2-k}}$.
\end{lemma}
\begin{proof}
Consider an interval $I_i$ where $1 \leq i \leq m-1$.  Let $U^{\mathcal{A}}_i$ be the set of Type 1 jobs that are unsatisfied in the schedule $\mathcal{A}$ during $I_i$.  The total accumulated cost during $I_i$ for these jobs can be bounded as follows knowing that $|U^{\mathcal{A}}_i| = 3m-3i$ and $\sum_{l \in U^{\mathcal{A}}_i } b_l = mB - iB $.

 \begin{eqnarray*}
&&\sum_{l \in U_i} \frac{1}{b_l^k} \left( (s_i+\beta+\alpha-\lambda b_l)^k - (s_i+\beta-\lambda b_l)^k\right) \\
&\leq& \sum_{l \in U_i} \frac{1}{b_l^k}\Bigg( \Bigg ( (s_i+ \beta - \beta^{k/2})^k+ \frac{k(\beta^{k/2}+\alpha-\lambda b_l)}{(s_i + \beta - \beta^{k/2} )^{1-k}} \Bigg) - \Bigg ( (s_i+ \beta - \beta^{k/2})^k+ \frac{k(\beta^{k/2}-\lambda b_l)}{(s_i + \beta - \beta^{k/2} )^{1-k}}\\
&& \;\;\;\;  - \frac{k(1-k)(\beta^{k/2}-\lambda b_l)^2}{2(s_i+\beta - \beta^{k/2})^{2-k}} \Bigg)\Bigg )   \;\;\;\; \mbox{[Fact \ref{fact:taylor}, $b_l \leq B$ for all $l$ and $2(\beta^{k/2}+ \alpha) \leq \beta - \beta^{k/2}$]}\\
&=& \sum_{l \in U_i} \frac{1}{b_l^k} \Bigg ( \frac{k\alpha}{(s_i + \beta - \beta^{k/2} )^{1-k}} - \frac{k(1-k)(\beta^{k/2}-\lambda b_l)^2}{2(s_i+\beta - \beta^{k/2})^{2-k}} \Bigg)\\
&=& \sum_{l \in U_i} \frac{1}{ \Delta_s^k} \Bigg ( \frac{k\alpha}{(s_i + \beta - \beta^{k/2} )^{1-k}} - \frac{k(1-k)(\beta^{k/2}-\lambda b_l)^2}{2(s_i+\beta - \beta^{k/2})^{2-k}} \Bigg) \;\;\;\; \mbox{[$b_l \geq \Delta_s$ for all $l$]}\\
&=& |U_i| \frac{1}{ \Delta_s^k}  \Bigg ( \frac{k\alpha}{(s_i + \beta - \beta^{k/2} )^{1-k}} - \frac{k(1-k)(\beta^{k/2})^2}{2(s_i+\beta - \beta^{k/2})^{2-k}} \Bigg) + \sum_{l \in U_i} \frac{1}{ \Delta_s^k}  \frac{k(1-k)(2((\beta^{k/2})\lambda b_l)-(\lambda b_l)^2)}{2(s_i+\beta - \beta^{k/2})^{2-k}} \\
&\geq& (3m-3i) \frac{1}{ \Delta_s^k} \Bigg ( \frac{k\alpha}{(s_i + \beta - \beta^{k/2} )^{1-k}} - \frac{k(1-k)(\beta^{k/2})^2}{2(s_i+\beta - \beta^{k/2})^{2-k}} \Bigg) +  \frac{1}{ \Delta_s^k} \frac{k(1-k)(\beta^{k/2})\lambda (mB-iB)}{(s_i+\beta - \beta^{k/2})^{2-k}} \\
&& \;\;\;\; \mbox{[$|U^{\mathcal{A}}_i| = mB-iB$ and $\sum_{l \in U^{\mathcal{A}}_i} b_l = mB-iB$ ]}
\end{eqnarray*}

Now consider the interval $I_0$.  No Type 1 jobs is completed by the end of $I_0$, so the total cost accumulated during this interval is $\sum_{l \in T_1} \frac{1}{b_l^k} (\beta - \lambda b_l)^k \leq \sum_{l \in T_1}  \frac{1}{ \Delta_s^k} (\beta - \lambda b_l)^k$.
\end{proof}

\begin{lemma}
\label{lem:type1nonblacksmstlk}
For the schedule $\mathcal{A}$ the total cost Type 1 jobs accumulate during open times is less than $\frac{3m^2kB}{\Delta_s^k( \beta -  \lambda B  )^{1-k}}$.
\end{lemma}
\begin{proof}
Fix any Type 1 job $l$. The most a Type 1 job will accumulate during a maximal contiguous open time interval is during the $B$ time steps time $0$.  This because the objective function is concave and every maximal contiguous open time interval has length $B$.  Knowing that the arrival time of any Type 1 job $l$ is before time $(\beta -\lambda \beta B)$ and $b_l \geq \Delta_s$, the most job $l$ can accumulate in a single maximal contiguous open time interval is the following.

 \begin{eqnarray*}
&&  \frac{1}{ \Delta_s^k} (\beta -  \lambda B +B)^k -  \frac{1}{ \Delta_s^k} (\beta -  \lambda B )^k \\
&\leq&  \frac{kB}{\Delta^k_s( \beta -  \lambda B  )^{1-k}} \;\;\;\; \mbox{[Fact \ref{fact:taylor} and $2B \leq \beta$]}\\
\end{eqnarray*}

There are $3m$ jobs total and $m$ open maximal contiguous time intervals.  Thus, the total cost the schedule $\mathcal{A}$ can accumulate during open times is less than $\frac{3m^2kB}{\Delta_s^k( \beta -  \lambda B  )^{1-k}}$.
\end{proof}

Now we can bound the total cost of $\mathcal{A}$.  Using Fact \ref{fact:cost23smstlk} the total cost for Type 2 and 3 jobs in $\mathcal{A}$ is $f_{2,3}$.  This and lemmas \ref{lem:type1blacksmstlk} and \ref{lem:type1nonblacksmstlk} imply that the total cost of $\mathcal{A}$ is less than $f_{2,3} + \frac{3m^2kB}{\Delta^k_s( \beta -  \lambda B  )^{1-k}} + \sum_{l \in T_1}\frac{1}{\Delta^k_s} (\beta - \lambda b_l)^k + (3m-3i)\frac{1}{\Delta^k_s}\Bigg ( \frac{k\alpha}{(s_i + \beta - \beta^{k/2} )^{1-k}} - \frac{k(1-k)(\beta^{k/2})^2}{2(s_i+\beta - \beta^{k/2})^{2-k}} \Bigg) + \frac{1}{\Delta^k_s}\frac{k(1-k)(\beta^{k/2})\lambda (mB-iB)}{(s_i+\beta - \beta^{k/2})^{2-k}} = f$.

\begin{theorem}
If there exists a valid solution to the 3-Partition instance then there is a schedule for the $k$-norm of stretch problem instance with an objective value at most $f$.
\end{theorem}

\end{document}